\documentclass[a4paper,12pt]{article}
\usepackage{amsmath}
\usepackage{amssymb}
\usepackage{amsthm}
\usepackage{latexsym}
\usepackage{mathrsfs}
\usepackage[dvips]{graphicx}
\usepackage{axodraw4j}
\usepackage{wick}
\usepackage{pstricks}
\usepackage{color}
\usepackage{diagrams}
\def\f(#1){{\mathop{f}^{(#1)}}}
\def\m(#1){{\mathop{m}^{(#1)}}}
\def\C(#1){{\mathop{C}^{(#1)}}}
\def\p(#1){{\mathop{p}^{(#1)}}}

\def\ben{\begin{equation}}
\def\een{\end{equation}}
\def\bena{\begin{eqnarray}}
\def\eena{\end{eqnarray}}
\def\non{\nonumber}

\def\d{{\rm d}}

\def\C{{\cal C}}

\def\mr{{\mathbb R}}

\def\G{{\cal G}}

\def\P{{\rm P}}

\newcommand{\myid}{{\bf 1}}

\newcommand{\mn}{{\mathbb N}}
\newcommand{\mz}{{\mathbb Z}}
\newcommand{\iho}{{\rm i}}
\newcommand{\mc}{{\mathbb C}}

\renewcommand{\a}{{\mathbf b}}

\newcommand{\e}{{\rm e}}

\theoremstyle{definition}

\newtheorem{thm}{Theorem}

\begin{document}

\title{Perturbative Quantum Field Theory via Vertex Algebras}

\author{
Stefan Hollands$^{1}$\thanks{\tt HollandsS@Cardiff.ac.uk}\:,
%%%
Heiner Olbermann$^{1}$\thanks{\tt OlbermannH@Cardiff.ac.uk}\:,
\\ \\
%%%
{\it ${}^{1}$School of Mathematics,
     Cardiff University, UK} \medskip \\
%%%
}
\date{29 June 2009}
%%%%
\maketitle

\begin{abstract}
In this paper, we explain how perturbative quantum field theory can be
formulated in terms of (a version of) vertex algebras. Our starting point is the
Wilson-Zimmermann operator product expansion (OPE). Following ideas of a previous paper~[arXiv:0802.2198],
we consider a consistency (essentially associativity) condition satisfied by the
coefficients in this expansion. We observe that the information in the OPE coefficients can
be repackaged straightforwardly into ``vertex operators'' and that the consistency
condition then has essentially the same form
as the key condition in the theory of vertex algebras. We develop a general theory
of perturbations of the algebras that we encounter, similar in nature to the Hochschild
cohomology describing the deformation theory of ordinary algebras. The main part of the paper is devoted to the question how
one can calculate the perturbations corresponding to a given interaction Lagrangian (such as $\lambda \varphi^4$) in practice,
using the consistency condition and the corresponding non-linear field equation.
We derive graphical rules, which display the
vertex operators (i.e., OPE coefficients) in terms of certain multiple series of hypergeometric type.
\end{abstract}

%%%%%%%%%%%%%%%%%%%%%%%%%%%%%%%%%%%%%%%%%%%%%%%%%%%%%%%%%%%%%%%%%%%

%\draft
\sloppy

\section{Introduction}
\label{sec1}

It is becoming increasingly clear that Quantum Field Theory (QFT) is not only
a very successful theoretical formalism to describe a variety of physical
situations such as elementary particle collisions or critical phenomena, but also a highly interesting
and complex mathematical structure. This manifests itself not least in the range of
different mathematical disciplines that come to play an important role in this theory, such as
functional analysis, combinatorics, complex analysis, algebra, geometry, measure theory etc.

Not surprisingly, there exist correspondingly many approaches to QFT, such as the path-integral
approach, the approach via diagrammatic/perturbative expansions, algebraic approaches,
stochastic quantization, axiomatic approaches, and more. These approaches are, at some level,
known to be equivalent, but each one of them seems to have different weaknesses and strengths. In a recent paper \cite{Hollands:2008ee} (see also~\cite{holwald} for a related proposal),
one of us proposed a new approach of an algebraic nature that is based on elevating the Kadanoff-Wilson-Zimmermann
{\em operator product expansion} (OPE) \cite{Wilson:1969zs,Wilson:1972ee} to the status of a fundamental structure. The operator product expansion
is traditionally viewed as a property of the Schwinger functions of
the composite fields of the theory. It can be stated as saying that
\ben\label{OPE}
\Big\langle \mathcal{O}_a(x) \mathcal{O}_b(0) \, \prod_{i} \mathcal{O}_{d_i}(y_i) \Big\rangle \sim
\sum_c C^c_{ab}(x) \, \Big\langle \mathcal{O}_c(0) \,  \prod_i \mathcal{O}_{d_i}(y_i) \Big\rangle \, .
\een
Here, $x\in \mr^D$ is a point in Euclidean space not equal to $0$, the
$y_i$ are distinct points with $|y_i| > |x|$,
and $a,b,c$ etc. are indices labeling the composite fields of the theory, which in turn are monomials in the basic
field and its derivatives. The OPE-coefficients $C_{ab}^c$ are distributions that are independent of the ``spectator fields''
$\mathcal{O}_{d_i}(y_i)$,  whereas the brackets indicate the
Schwinger functions of the (Euclidean) QFT under consideration. They may be constructed e.g. by
giving precise mathematical sense to a functional integral of the type
\ben
\label{pathintegral}
\Big\langle \prod_i \mathcal{O}_{a_i}(x_i)  \Big\rangle = \lim_{\Lambda \to \infty}
\int
%_{\S'(\mr^D)}
\prod_i \mathcal{O}_{a_i}(x_i) \, \exp \left[ - S_{\Lambda}(\varphi) \right] \, \d \mu_\Lambda (\varphi) \, .
\een
either in perturbation theory~\cite{Keller:1990ej,kellerkj}, or non-perturbatively, see
e.g.~\cite{Brydges:1993vb,Jaffebook,Rivasseau:1991ub,Magnen:2007uy}. In the above formula, $\d \mu_\Lambda(\varphi)$ is
a Gaussian measure associated with the free part of a classical action, and $S_\Lambda$ is
the full action including all counterterms, depending on the cutoff scale $\Lambda$. The OPE coefficients
characterize the short-distance properties ($|x| \to 0$) of the Schwinger functions. The formula is
normally understood as an asymptotic expansion in this limit, with coefficients $C_{ab}^c$ becoming
more and more smooth in their argument as the dimension of the operator indicated by $\mathcal{O}_c$ increases.

The proposal of~\cite{Hollands:2008ee,holwald} was to view the OPE coefficients not as a secondary structure, but as the defining
structure of the theory---some kind of ``structure functions'' of a suitable algebraic structure governing the
field theory. In order to display this algebraic structure, we looked at Schwinger
functions with a triple insertion $\mathcal{O}_a(x) \mathcal{O}_b(y) \mathcal{O}_c(0)$, assuming that $0<|x-y|<|y|<|x|$. It is then
plausible that one obtains equivalent results when performing successive OPE's in the following two alternative ways
indicated by the different positioning of the parenthesis in $(\mathcal{O}_a(x) \mathcal{O}_b(y)) \mathcal{O}_c(0)
= \mathcal{O}_a(x) (\mathcal{O}_b(y) \mathcal{O}_c(0))$. For example, the left side means that we perform an OPE of the fields
$\mathcal{O}_b(y)$ with the field $\mathcal{O}_c(0)$, and a subsequent OPE of the result with $\mathcal{O}_a(x)$. The equality sign between
both sides means that the results should be the same when inserted in a Schwinger function as above.
In terms of the OPE-coefficients this means that we expect the relation
\ben\label{consistency}
\sum_d C_{ab}^d(x-y) C_{dc}^e(y) = \sum_d C_{bc}^d(y) C_{ad}^e(x) \, .
\een
We called this the ``consistency'' or ``associativity'' condition in~\cite{Hollands:2008ee}.
We observe that this relation becomes much more transparent if we introduce the following notation:
first, we introduce an abstract infinite dimensional complex vector space, $V$, whose basis elements $a \in V$
are in one-to-one correspondence with the composite fields $\mathcal{O}_a$ of the theory. Thus,
by a slight abuse of notation, we use the labels $a,b,c$ both as labels of the fields,
and simultaneously for the basis elements of $V$. We also define the corresponding
(algebraic) dual vector space $V^*$. We then define a ``vertex operator'' $Y(a, x): V \to V$ as the endomorphism
of $V$ whose matrix elements are given by
\ben\label{ydef}
\langle c | Y(a, x) | b \rangle := C_{ab}^c(x) \,\, ,
\een
where we use the usual physicist's notation for basis vectors of $V$ as ``kets'' $|a\rangle$ and the corresponding
dual basis vectors of $V^*$ as ``bras'' $\langle b|$.
With this notation, the consistency condition reads simply
\ben
\label{consistency1}
Y(a,x)Y(b,y) = Y(Y(a, x-y)b, y) \, ,
\een
which holds again when $|x|>|y|>|x-y|>0$.
In this form, our condition is almost identical in appearance to
(one form of) the key relation in the theory of vertex operator algebras
\cite{kac,Borcherds:1983sq,Frenkel:1988xz,gaberdiel}, hence the name ``vertex operator'' for $Y$.
We explain the relationship of our approach to vertex algebras
in somewhat more detail in sec.~\ref{sec:5} of this paper, but
we emphasize from the outset that our setup is intended to be
much more general than that usually encountered in
this theory; for example, we do not assume that the theory is conformally invariant, and we admit
general dimensions $D$, even though we expect non-trivial examples only in $D \le 4$.
As a consequence, the contents of our consistency condition and that in vertex algebras
are actually rather different despite the similar appearance.
An alternative treatment of the free field vertex algebra in $D$ dimensions can already be found
in~\cite{Nikolov:2003df} and~\cite{Borcherds:1997cx}.

The main purpose of the paper is to outline how the standard constructions in QFT, such as
perturbative expansion, BRST cohomology etc. can be formulated in terms of the vertex operators $Y$,
and what new viewpoint one thereby gets for these constructions. As we will see, if we combine the
consistency condition with a perturbative expansion, and with the non-linear field
equations corresponding to the classical action $S_\Lambda$ in the path integral above,
then we obtain in a new scheme for doing perturbative calculations. In more detail, the
contents of the paper are as follows:

\begin{itemize}
\item In sec.~\ref{gen_set} we begin by explaining the general setup. In particular, we explain in
abstract terms how ordinary perturbation theory fits into the framework of vertex algebras. As we
show translating constructions of~\cite{Hollands:2008ee} into the framework of vertex algebras, one can define a cohomology of ``Hochschild-type'' associated with a $D$-dimensional
vertex algebra (in our sense), and perturbations may be characterized in terms of this cohomology.
We also explain how the framework may be generalized to also include the BRST-construction for
gauge theories.

\item In sec.~\ref{free_field}, we explain how the vertex algebra of the $D$-dimensional free
massless field can be obtained from the Schwinger functions, and we thereby obtain a concrete
(albeit somewhat trivial) example for our setting. This section also serves as the
starting point to the later sections.

\item In sec.~\ref{sec:3}, we then construct the deformations (``perturbations'') of the
free field vertex algebra corresponding to an interaction term such as $\lambda \varphi^4$
in the classical Lagrangian. We show how one can obtain a closed form expression for
the $i$-th order deformation of the vertex operators based on the repeated use of the consistency
condition and the (non-linear) field equation [see eqs.~\eqref{amplitude2} and~\eqref{altern} for
the final form of these expressions]. The case $D=2$ is in many ways simpler than
the case $D>2$, so we discuss it separately. However, in both cases, the final result is
expressed in terms of sums that are associated with certain tree graphs and associated
loop graphs. The resulting sums and integrals involve Gegenbauer/Legendre functions
and are reminiscent of the expressions one obtains using the ``Gegenbauer polynomial $x$-space technique'' of
\cite{Chetyrkin:1980pr} for multi-loop Feynman diagrams.

\item In sec.~\ref{sec:4}, we discuss an alternative representation of the deformed
vertex operators in terms of certain multiple infinite sums. These sums have a similarity with
the hypergeometric series, and we believe that this opens up an interesting connection between
our vertex algebras and
a class of special functions.

\item In sec.~\ref{sec:5} we discuss in some detail the similarities and differences between
our notion of vertex algebra and notions that were previously given in the context of
2-dimensional CFT's.

\item Various results and definitions related to Legendre functions in $D$-dimensions needed in
the main text are given in app.~\ref{app:A} and~\ref{app:C}.

\end{itemize}

This paper is about the mathematical structure of QFT, but it is by itself not completely
mathematically rigorous. We analyze the structure and exploit the consequences of the
consistency condition in a mathematically rigorous way, but we do not justify the
consistency condition itself (apart from free field theory) and rather treat it as a hypothesis.
In this sense, our approach can be thought of as some sort of ``bootstrap''.
In order to justify the consistency condition, one can e.g. start from the OPE of the Schwinger
functions, and show that performing the expansion of a triple product of fields lead to
the same result, as we described.
We will come back to this issue in another paper~\cite{kophololb}.
\\

{\em Notations and conventions:} $x,y$ etc. denote points in $\mr^D$, with scalar product
$x \cdot y = \sum_\mu x_\mu y_\mu$, and norm $r^2 = |x|^2 = x \cdot x$. We also use the
notation $\hat x := x/r \in S^{D-1}$ for the angular part if $x \neq 0$.
$\P(z, \nu, D)$ denote the Legendre functions, see appendix~\ref{app:A}, and $\psi = \Gamma'/\Gamma$ is the
Psi-function. The surface
area of the $(D-1)$-dimensional sphere is abbreviated $\sigma_D = \frac{2\pi^{D-2}}{\Gamma(D/2)}$ and $K_D =
\sqrt{D-2}$.
The natural numbers $\mathbb N$ include $0$,  $\mathbb N=\{0,1,2,\dots\}$.
\section{General setup}
\label{gen_set}

In the previous section, we explained the basic idea that connects QFT as formulated in terms
of Schwinger functions with vertex operators $Y(a,x)$ satisfying a consistency condition.
Let us isolate the properties that we expect these vertex operators to have, reflecting the
general known properties of the Schwinger functions:
\begin{enumerate}
\item First, we state formally that there is an identity operator in field
theory. This is a distinguished element $\myid \in V$. The OPE of $\myid$ with any
other field is trivial and the vertex operator associated with $\myid$
is hence given by $Y(\myid, x) = id$, where $id$ is the identity on $V$.
\item We have already said that the OPE coefficients are in general expected to be
distributions on $\mr^D$. In Euclidean space, the situation is actually better and
the coefficients are analytic except for the origin. Thus, if we denote by $\mathcal{A}(\mr^D \setminus \{0\})$
the set of such analytic functions then we expect that $Y(a, \, . \,) \in
\mathcal{A}(\mr^D \setminus \{0\}) \otimes {\rm End}(V)$.
\item The Schwinger functions have an obvious invariance property under the Euclidean
group $SO(D)$ (or its covering group, if there are spinor fields in the theory). These
invariance properties are inherited by the OPE coefficients, and we hence have corresponding
invariance properties of the vertex operators: $Y(a,g\cdot x)=R(g)Y(R(g)^{-1}a,x)R(g)^{-1}$, $g\in SO(D)$,
for some representation $R$ of $SO(D)$ on $V$. Note, however, that there is no similar covariance under translations,
as the vertex operator in effect depend on the choice of the origin $0 \in \mr^D$ as the reference point in the OPE.
\item The key condition is the consistency condition, which we have already motivated and which we
repeat:
\ben
Y(a,x)Y(b,y) = Y(Y(a, x-y)b, y) \, ,
\een
for all $a,b \in V$ and all $x,y \in \mr^D$ subject to $0<|x-y|<|y|<|x|$. Note that there is
an implicit statement about the convergence of an infinite sum made here: If we apply e.g.
the right side of the equation to a vector $c \in V$, then we effectively state that
$Y(b,y)c$ is in the domain of $Y(a,x)$. When written in a standard basis of composite
fields, the vector $d = Y(b,y)c$ typically is an infinite linear combination of such fields even
if $b,c$ are from a basis of composite fields. Thus, $Y(a,x)d$ will also be an infinite sum when written in
the basis of composite fields, and we require, in effect, that this sum converges. As we discuss in more detail
in sec.~\ref{sec:5}, there is no such issue in the usual formulation of vertex algebras, where all sums are either
formal, or finite.
\end{enumerate}

The Schwinger functions have other properties that one can readily translate into properties of the
vertex operators, such as (anti-) symmetry properties under the exchange of fields,
scaling properties, hermiticity properties, etc. To keep the discussion transparent at this point, we will not
go into this here, but state some of these conditions as we go along.

In renormalized perturbation theory, one is naturally led to consider field-redefinitions, for
example when considering changes in the renormalization scheme. A field redefinition is simply a linear
transformation which maps a quantum field to a linear combination of quantum fields, and thereby gives a corresponding
transformation of the Schwinger functions. Such field redefinitions change in an evident
way also the OPE coefficients of the theory. In terms of the vertex operators, a field redefinition is
simply an invertible complex linear map $Z: V \to V$ which commutes with $R(g)$ for all $g \in SO(D)$, and which
satisfies $Z \myid = \myid$. The OPE coefficients associated with the transformed fields give rise to a correspondingly
transformed vertex operator, given by
\ben\label{fieldredef}
Y'(a, x) = Z \, Y(x, Z a) \, Z^{-1}
\een
This vertex operator satisfies the same properties as above, and will be considered
``equivalent''.

An important construction in QFT is to consider perturbations of a given theory. The idea is usually to start
from a free field theory with explicitly known Schwinger functions/OPE and to construct
perturbations of these quantities, order by order in a ``small'' parameter $\lambda$ characterizing
the size of the perturbation. In the path integral, one expands ${\rm exp}(-S_\Lambda)$ in terms of
this parameter, and then performs the Gaussian integral for each term in this expansion. Such a procedure
will also give rise to a corresponding expansion of the OPE coefficients, see e.g. \cite{Keller:1992by}, and thereby
to an expansion of the vertex operators of the form
\ben\label{ypert}
Y(a,x) = \sum_{i = 0}^\infty \lambda^i Y_i(a, x) \, .
\een
The zeroth order term $Y_0(a, x)$ of this---in general only formal---expansion corresponds to the
free field theory around which one expands. The higher terms represent the perturbations. Because
the full $Y(a,x)$ must satisfy the properties 1)---4) stated above for all $\lambda$ (or at least in the
sense of formal perturbation series), the expansion coefficients $Y_i(a,x)$ must satisfy analogous
properties. The most prominent one is the consistency condition, which at $i$-th order reads
\ben\label{consistency2}
\sum_{j=0}^i Y_j(a,x) Y_{i-j}(b,y) = \sum_{j=0}^i Y_j(Y_{i-j}(a,x-y)b,y) \, .
\een
It is satisfied at order $i=0$, because one can prove it for a free field theory. As already
pointed out in~\cite{Hollands:2008ee}, there is an interesting cohomological interpretation of eq.~\eqref{consistency2},
and we now briefly explain how this comes about. Let $n\ge 0$ and define $\Omega^{n+1}(V)$ to be a
space of linear maps
\ben
f_n(x_1, \dots, x_n): V \otimes \dots \otimes V \to {\rm End}(V) \,
\een
that are defined for certain configurations $(x_1, \dots, x_n) \in \mr^{nD}$ of mutually
different points, and where there are $n$ tensor copies of $V$. More precisely,
$f_n$'s should be real analytic functions of $x_1, \dots, x_n \in \mr^{nD}$ in the open domain
\ben
D_n = \left\{(x_1,...,x_n)\in\mathbb{R}^{nD}:r_{1,i-1}<r_{i-1,i}<r_{i-2,i}<...<r_{1,i}\right\}\,,
\een
where $r_{i,j}=|x_i-x_j|$, taking values in ${\rm End}(V^{\otimes n}, {\rm End}(V))$.
Next, we define from $Y_0$ a linear operator $b: \Omega^n(V) \to \Omega^{n+1}(V)$ by the formula
\bena
&&(bf_n)(x_1, \dots, x_{n+1}; a_1, \dots, a_{n+1}) := Y_0(a_1, x_1) f_n(x_2, \dots, x_{n+1};
a_2, \dots, a_{n+1}) \non\\
&+& \sum_{i=1}^n (-1)^i f_n(x_1, \dots, \widehat x_i, \dots, x_{n+1}; a_1, \dots,
Y_0(a_{i}, x_i-x_{i+1})a_{i+1}, \dots a_{n+1}) \non\\
&+& (-1)^{n+1} f_n(x_1, \dots, x_n; a_1, \dots, a_n)Y_0(a_{n+1}, x_{n+1}) \, .
\eena
This linear operator is presumably not defined on all such $f_n$, but only on
a certain domain of definition, but we ignore this issue here for simplicity. We formally calculate~\cite{Hollands:2008ee}
using the consistency condition for $Y_0$ that
\ben
b(bf_n) = 0 \, ,
\een
i.e., $b$ is a coboundary operator. The domains $D_n$ are needed in order to be able to apply the consistency
condition. We denote by
\ben
H^n(V, Y_0) := \{ {\rm ker} \, b: \Omega^n \to \Omega^{n+1} \}/\{{\rm ran} \, b : \Omega^{n-1} \to \Omega^n \}
\een
the $n$-th cohomology ring of the chain complex $(\oplus_n \Omega^n(V), b)$.

The connection to the consistency condition
eq.~\eqref{consistency2} at $i$-th order in perturbation theory
now arises as follows: At $i=1$, the equation can be restated
as simply saying that
\ben
bY_1 = 0 \, .
\een
If $Y_1$ arises from $Y_0$ by merely a field redefinition~\eqref{fieldredef} with
$Z = \sum_{i=1}^n z_i \lambda^i$, then this means
\ben
Y_1 = bz_1 \, .
\een
Consequently, the non-trivial first order perturbations may be viewed as elements in $H^2(V, Y_0)$, and
hence such perturbations only exist if this space is non-trivial.
Continuing in this way, the condition~\eqref{consistency2} at $i$-th order can be stated as saying that
\ben\label{inhom}
bY_i = w_i \,\,\, ,
\een
where $w_i \in \Omega^3(V)$ is defined by the terms in eq.~\eqref{consistency} with $j\neq 0, j\neq i$.
It can be checked inductively~\cite{Hollands:2008ee} that $bw_i = 0$, so $[w_i]$ defines a class in $H^3(V, Y_0)$.
Evidently, the $i$-th order perturbation $Y_i$ exists, if, in fact, $w_i \in {\rm ran} \, b$, i.e.
if it defines the zero class in $H^3(V, Y_0)$. The freedom of choosing different solutions to
eq.~\eqref{inhom} corresponds to elements in $H^2(V, Y_0)$. In this way, we get an interpretation
of perturbation theory in terms of the cohomology rings $H^n(V, Y_0)$.

There is no difficulty to extend this framework to more complicated situations where one has
additional symmetries that one would like to preserve in the deformation process. Suppose e.g. that the free theory
is a free $U(1)^N$ gauge theory with associated ghosts. Then in the free field theory we have
an associated free BRST-operator $s_0$
with the property $s_0^2 = 0$ and $s_0 \circ gh = (gh - id) \circ s_0$, where $gh$ is the ``ghost number'', i.e. a linear operator on $V$ with integer spectrum
that provides a corresponding grading. This acts on the composite
fields of the theory, and
hence, as a linear transformation $s_0: V \to V$ satisfying $\gamma s_0 + s_0 \gamma = 0$,
where $\gamma: V \to V$ is a $\mz_2$-grading of $V$ that corresponds to the
Bose/Fermi character of the fields. The concrete form of $s_0$ in terms of the gauge field and ghost/auxiliary fields
is given in many textbooks, see e.g.~\cite{ZinnJustin}. To explain our general scheme it is only important to know that the
BRST-invariance of the Schwinger functions of the free
gauge theory gives rise to a corresponding invariance of the OPE, namely, the OPE of BRST-invariant
fields again contains only such fields. In terms of the
vertex operators, this is expressed by the relation
\ben\label{s0cons}
s_0 Y_0(a, x) + Y_0(\gamma a, x) s_0 + Y_0(s_0 a, x) = 0 \, .
\een
We would now like to ask whether it is possible to find a deformation of $Y_0$ as above to a $Y$, and a corresponding
deformation of $s_0$ to an $s$
\ben
s = \sum_{i=0}^\infty s_i \lambda^i \, ,
\een
such that $s^2 = 0$, i.e.
\ben\label{s2}
0 = \sum_{j=0}^i s_j s_{i-j} \, ,
\een
and such that eq.~\eqref{s0cons} continues to hold for $Y,s$, that is, at $i$-th order,
\ben\label{s0cons1}
\sum_{j=0}^i s_j Y_{i-j}(a, x) + Y_j(\gamma a, x) s_{i-j} + Y_j(s_{i-j} a, x) = 0 \, .
\een
The problem of finding simultaneously the $s_i$ and $Y_i$ subject to eqs.~\eqref{s0cons1},~\eqref{s2} and~\eqref{consistency2}
is again of a cohomological nature. To set up the cohomology ring
in question, we extend the action of $s_0$ from $V$ to $\Omega^n(V)$ by setting
\bena
(Bf_n)(x_1, \dots, x_n; a_1, \dots, a_n) &:=& s_0 f_n(x_1, \dots, x_n; a_1, \dots, a_n) \\
& +& \sum_{j=1}^n \gamma f_n(x_1, \dots, x_n; a_1, \dots,a_{j-1},\gamma s_0 a_j,\gamma a_{j+1}, \dots,\gamma a_n)\non
%&& + f_n(x_1, \dots, x_n; \gamma a_1, \dots, \gamma a_n) s_0
\eena
Then one verifies~\cite{Hollands:2008ee} that $B^2 = 0$ using $s_0^2 = 0$ and that $bB+Bb = 0$ using eq.~\eqref{s0cons}, i.e.
$B$ is another coboundary operator which is compatible with $b$.
In the language of differential complexes, $(b,B)$ gives rise to a double complex $(\oplus_{n,g} \Omega^{n,g}(V), b,B)$,
whose cohomology rings are denoted $H^{n,g}(V, Y_0, s_0)$. Here $n$ is as above, and $g$ corresponds to the
grading of $V$ (and correspondingly the spaces $\Omega^n(V) = \oplus_g \Omega^{n,g}(V)$) by the ghost number $gh$. We have the following commutative diagram:
\begin{diagram}
&\rTo^{B} &  H^{0,0}(V, Y_0, s_0)  &\rTo^{B}  &H^{0,1} (V, Y_0, s_0)    &\rTo^{B} & H^{0,2}(V, Y_0, s_0) &\rTo^{B} &\\
& & \dTo_{b} & &\dTo_{b}&&\dTo_{b}&\\
&\rTo^{B} & H^{1,0} (V, Y_0, s_0) &\rTo^{B}  &H^{1,1}  (V, Y_0, s_0)   &\rTo^{B} & H^{1,2} (V, Y_0, s_0) &\rTo^{B} &\\
& & \dDots   &&\dDots &&\dDots  &\\
\end{diagram}
As is standard in this situation, we may form the total complex whose cohomology rings are given by
\ben
H^m(V, Y_0|s_0) = \bigoplus_{n+g=m} H^{n,g}(V, Y_0, s_0) \, ,
\label{cohomsums}
\een
and whose total coboundary operator is $B+b$, i.e. $(B+b)^2 = 0$.
The direct sum in eq.~\eqref{cohomsums} consists of the cohomology groups lined up on the diagonal
running from $H^{m,0}(V, Y_0, s_0)$ to $H^{0,m}(V, Y_0, s_0)$ in the above diagram.\\
If we form $\alpha_i = (s_i, Y_i, 0, 0, \dots)$, then the
combined condition for the first order perturbation $\alpha_1$ arising from the first order
consistency condition and the first order BRST-invariance can be written as
\ben
(b+B)\alpha_1 = 0 \, ,
\een
while the condition for the first order perturbations to be due to a first order field redefinition
$\zeta_i = (z_i, 0, 0, \dots)$ is written as
\ben
\alpha_1 = (B+b) \zeta_1 \, .
\een
Thus, first order perturbations of the BRST operator and vertex operators are given by a class $[\alpha_1] \in H^2(V, Y_0|s_0)$.
Likewise, the conditions for the $i$-th order perturbations $\alpha_i$ can be written as
\ben
(B+b) \alpha_i = \beta_i \, ,
\een
where $\beta_i$ is calculable from the $\alpha_j, j \le i-1$, and where one can compute $(B+b) \beta_i = 0$. Thus,
the potential $i$-th order obstruction is the class $[\beta_i] \in H^3(V, s_0|Y_0)$. The details of this
analysis are completely analogous to~\cite{Hollands:2008ee}.

\section{The free field OPE vertex algebra}
\label{free_field}
In this section, we illustrate our abstract framework
for the OPE in a simple example. Our example is
the free quantum field theory obeying the linear field equation
\ben
\label{ffe}
\Delta \varphi = 0 \, .
\een
The space $V$ of fields in this theory may be taken to be the unital, free, commutative ring
generated by the identity $\varphi$ and its derivatives. In other words, the elements of $V$ are in one-to-one
correspondence with monomials in $\partial_{\mu_1} \dots \partial_{\mu_k} \varphi$,
and $\partial_\mu, \mu=1,...,D$ are the derivations that act as if they were
ordinary partial derivatives. To implement the field equation,
we simply set to zero any expressions
containing a factor the form
$\delta^{\mu_i \mu_j} \partial_{\mu_1} \dots \partial_{\mu_k} \varphi$, i.e., monomials that would vanish if
$\varphi$ was an actual field satisfying the field equation. Because monomials containing a trace of
$\partial_{\mu_1} \dots \partial_{\mu_k} \varphi$ are set to zero, $V$ is spanned by all trace-free monomials.
Thus, if we denote by curly brackets $t_{\{\mu_1 \dots \mu_k\}}$ the trace-free part of a symmetric tensor,
then a basis of $V$ is given by $\myid$, together with the set of monomials of the form
$\prod \partial_{\{\mu_1} \cdots \partial_{\mu_k\}} \varphi$.

It is convenient for latter purposes to choose a particular basis. For this, we consider the space
of harmonic polynomials in $D$ real variables homogeneous of degree $l$, i.e. the set of all
polynomials $h(x)$ in $D$ variables $(x \in \mr^D)$ with complex coefficients satisfying
$h(tx) = t^l h(x)$, and $\Delta h(x) = 0$. Some relevant facts about such polynomials
are collected in appendix~\ref{app:A}. We denote by $h_{l,m}, m= 1, \dots, N(D,l)$ a basis of
degree $l$ harmonic polynomials. (The number of linearly independent polynomials of this kind, $N(D,l)$, can be
found in appendix~\ref{app:A}.) We normalize this basis so that
\footnote{With this normalization,
the harmonic polynomials restricted to $S^{D-1}$ are the $D-1$-dimensional spherical harmonics.}
\ben
\label{harmnorm}
\int_{S^{D-1}} \overline h_{l,m}(\hat x) h_{l',m'}(\hat x) \, \d\Omega(\hat x) = \delta_{l,l'} \delta_{m,m'}
\een
where $\d \Omega$ is the standard integration element on the sphere, and $\hat x = x/|x|$.
A basis of $V$ is then given by $\myid$,  together with the elements
\ben\label{vadef}
a =  \prod_{l,m} \frac{1}{\sqrt{a_{l,m}!}}
\left( c_l^{-1}\bar h_{l,m}(\partial) \varphi \right)^{a_{l,m}} \, \quad \, ,
\een
where $a = \{ a_{l,m} \in \mn \mid l \ge 0, m=1,...,N(l,D) \}$ is identified at the
same time with a vector in $V$ and a multi-index of
non-negative integers, only finitely many of which are non-zero, and $c_l$ is given in appendix~\ref{Y0op}.
%For later
%convenience, we also set
%\ben
%c_l = \frac{2^l \, \Gamma(l+1) \Gamma(l+D/2-1)}{\Gamma(D/2-1)}  \, .
%\een
The Schwinger functions of the model are well-known. For $2s$ factors of the basic field they
are given by ($D>2$)
\ben
\Big \langle \prod_{i=1}^{2s} \varphi(x_i) \Big\rangle = \sum_{\sigma\in \text{Sym}(2s)}\prod_{j=1}^s
|x_{\sigma(2j-1)}-x_{\sigma(2j)}|^{2-D}
%\non\\
%\Big \langle \prod_{i=1}^{2s+1} \varphi(x_i) \Big\rangle &=& 0
\label{Schwinger}
\een
for non-coincident $x_i \in \mr^D$. For $2s+1$ factors of the basic field the Schwinger function is zero.
Partial derivatives can be taken
on both sides of the equation to get the Schwinger functions of all $\varphi$-linear fields.
The Schwinger functions of composite operators can be obtained in the same manner,
writing a composite field as a product of $\varphi$-linear fields with the same argument $x\in\mathbb{R}^D$
and omitting permutations $\sigma$ with $x_{\sigma(2j-1)}=x_{\sigma(2j)}$ for some $j\in\{1,\dots,s\}$
in eq.~\eqref{Schwinger}. For example, if none of the composite field has derivatives, we have
\ben
\label{Schwinger2}
\Big \langle \prod_{v=1}^{n} \varphi^{p_v} (x_v) \Big\rangle = \sum_{{\rm graphs} \, G} \prod_{e = (vw) \in G}
|x_{v}-x_{w}|^{2-D} \, .
\een
Here, the sum is over all graphs with coordination numbers $\{p_v\}$, and the product is over all
edges $e = (vw)$ of the graph. For composite fields with derivatives, one has a similar formula.
From the Schwinger functions, one gets the OPE coefficients according to eq.~\eqref{OPE},
and from the OPE coefficients, one gets the $Y_0(a, x)$ according to eq.~\eqref{ydef}. In
$D=2$ dimensions, the propagators are replaced by $\ln |x_v - x_w|$.
We now present the results of these calculations, leaving the details to appendix~\ref{Y0op}.
The subscript ``0'' reminds us that we are dealing with a free field in this section.

In order to present explicitly the $Y_0$'s, it is convenient to
view $V$ as a ``Fock-space,'' with $a_{l,m}$ (see eq.~\eqref{vadef}) interpreted as the ``occupation number''
of the ``mode'' labeled by $l,m$, and with $\myid$ playing the role of "Fock-vacuum"
denoted $|0\rangle$ (vanishing
occupation number). On this Fock-space, one can then define
creation and annihilation operators $\a_{l,m} , \a_{l,m}^+: V \to V$, see appendix~\ref{Y0op}.
They satisfy the standard commutation relations
\ben
\left[ \a_{l,m}^{}, \a_{l',m'}^{+} \right] =
\delta_{l,l'}\delta_{m,m'} \,\, id \, , \quad
\left[ \a_{l,m}^+, \a_{l',m'}^+ \right] =
\left[ \a_{l,m}^{}, \a_{l',m'}^{} \right] = 0 \label{commutation}
\een
where $id$ is the identity operator on $ V$.
In this language, the basis
elements of $V$ are written as
\ben\label{veca}
a = \prod_{l,m} \frac{(\a_{l,m}^+)^{a_{l,m}}}{\sqrt{a_{l,m}!}}
\, \, |0 \rangle \, .
\een
We now give the formula for $Y_0(\varphi, x)$ corresponding to the basic field. For $D>2$, this is given by
\begin{multline}\label{freephiv}
Y_0(\varphi, x) = K_D \, r^{-(D-2)/2} \,
 \sum_{l=0}^{\infty} \, \sum_{m=1}^{N(l,D)} \frac{1}{\sqrt{\omega(D,l)}} \times \\
\Big[ r^{l+(D-2)/2} h_{l, m}(\hat x) \, \a_{l,m}^{+}
+ r^{-l-(D-2)/2} \overline{ h_{l, m}(\hat x) } \, \a_{l,m}^{} \Big] \, ,
%&=& r^{-(D-2)/2} \sum_{k=-\infty}^{+\infty} h_{k,m}(x)\, \a_{k,m} \, ,
\end{multline}
where $K_D=\sqrt{D-2}$, and the "frequency" $\omega(l,D)$ is given by
$2l+D-2$, see appendix~\ref{Y0op}.
For a general $a \in V$ of the form eq.~\eqref{vadef}, the vertex operators $Y_0(a,x): V \to V$ are
\ben
Y_0(a,x) =
\,\,
: \prod_{l,m}
 \frac{1}{(a_{l,m}!)^{1/2}} \left\{c_l^{-1} h_{l,m}(\partial) Y_0 (\varphi, x) \right\}^{a_{l,m}} :
\,\,
 \,  .
\label{freevertex}
\een
Here, double dots $: \dots :$ mean
"normal ordering", i.e., all creation operators are to the right of all annihilation operators.

It can be checked explicitly that  the $Y_0$ satisfy the properties 1)---4).
The representation $R: SO(D) \to {\rm End}(V)$ in 3) is characterized by
\ben
R(g) \, \a_{l,m}^+ \, R(g)^{-1} = \sum_{m'=1}^{N(D,l)} \mathcal{D}^l_{m,m'}(g) \a_{l,m'}^+ \, ,
\een
where $\mathcal{D}^l$ in turn is the representation\footnote{In $D=3$, $l$ corresponds to the
angular momentum quantum number and $\mathcal{D}^l$ is the corresponding irreducible representation. For
$D>3$, the representation $\mathcal{D}^l$ is reducible.} of $SO(D)$ on the space of degree $l$ harmonic
polynomials, i.e. $\sum_{m'} \mathcal{D}_{m,m'}^l(g) h_{l,m'}^{}(x) = h_{l,m}^{}(g \cdot x)$. In order
to demonstrate the consistency condition 4), one must make use of the
identities related to the harmonic polynomials given
in appendix \ref{app:A}. We are not going to go through this lengthy calculation here.

\section{Perturbative OPE vertex algebras}
\label{sec:3}
In the previous section we have described the (well-known) OPE of the
free, massless bosonic field in $D$ dimensions. The new viewpoint was to
understand this as a vertex algebra. Of course, free field theories are only of limited interest,
and we therefore now look at perturbations of the free theory. Normally, perturbations are characterized
by an interaction Lagrangian, which, together with appropriate counterterms, is inserted into the path
integral~\eqref{pathintegral} in order to obtain the perturbation series of the Schwinger functions.
To make this work, one first considers a regulated path integral, with a regulated
interaction, $S_\Lambda$, and one then removes the regulator $\Lambda$. This procedure
is explained in many textbooks, see e.g.~\cite{ZinnJustin}, and it leads also to the
definition of the OPE coefficients, $C_{ab}^c$, see e.g.~\cite{Keller:1992by} for a derivation using
the Wilson-Wegner-Polchinski RG-flow equations.

As we have explained, in this paper, we want to pursue an approach wherein the OPE is
elevated to the status of a fundamental relation, and we should therefore also have a method
to calculate the perturbations of the OPE coefficients---or equivalently the $Y$'s---directly,
without recourse to the Schwinger functions. In principle, we have outlined how this works in
sec.~\ref{gen_set}. But for this we would need to understand more explicitly the---rather abstractly
defined---cohomology rings $H^2(V, Y_0)$ and $H^3(V, Y_0)$. We have so far been unable to
do so. However, one would think that perturbations should also be describable in terms of
an interaction Lagrangian, in the same way as in the usual approach, because one is talking about
the same theory after all.

We now explain one way how to proceed. Let us assume that the
``bare interaction'' (in the usual QFT parlance) is given by a polynomial
$\lambda P(\varphi) = \lambda \sum \frac{c_p}{p!} \varphi^p$. In order to get a well-defined perturbative
definition of the Schwinger functions, we also assume that the interaction is renormalizable, i.e.
${\rm deg}(P) \frac{D-2}{2} \le D$. It is a well-known fact in standard perturbation theory that the
Schwinger functions of the theory may then be defined so that
\ben
\Big\langle [\Delta \varphi(x) - \lambda P'(\varphi(x)) \, ] \mathcal{O}_a(0) \,
\prod_i \mathcal{O}_{d_i}(y_i)  \Big\rangle = 0 \, .
\een
Here, $\mathcal{O}_a, \mathcal{O}_{d_i}$ are arbitrary composite fields, and
the arguments satisfy $|y_i| > |x| > 0$. \footnote{
Otherwise, there would also be ``contact terms'' in the above equation.}
The expression in brackets $[\dots]$ is of course the non-linear field equation.
Let us now apply the OPE to the expression $[\Delta \varphi(x) - \lambda P'(\varphi(x)) \, ] \mathcal{O}_a(0)$
in the above Schwinger function. Then we get a relation between
the OPE coefficients involving $\Delta \varphi$ and those involving $P'(\varphi)$. In terms of the vertex
operators this relation is
\ben
\label{vertexfieldeq}
\Delta Y(\varphi, x) = \lambda \, Y(P'(\varphi), x) \, ,
\een
where we are now viewing $P'(\varphi)$ as an element in $V$, the abstract vector space
whose elements are in one-to-one correspondence with the composite fields.

Next, we expand the vertex operators in a (formal) perturbation series in $\lambda$ as in eq.~\eqref{ypert},
and this immediately leads to the relation
\ben\label{mastereq}
\Delta Y_i(\varphi, x) = Y_{i-1}(P'(\varphi), x) \, .
\een
The evident strategy is now to try to design an iterative scheme from this equation, by
calculating the order $i$ vertex operator on the left in terms of the lower order $i-1$
vertex operator on the right, starting inductively with $i=1$, as all vertex operators
of order $i-1=0$ have been given in the previous section. However, such a procedure runs into the
immediate difficulty that the right side involves the vertex operator associated with the
composite field $P'(\varphi)$, whereas the left side only gives the vertex operator
of the basic field $\varphi$. We must therefore introduce a second induction loop which
constructs, iteratively, the vertex operators of an arbitrary $a \in V$ from those of
$\varphi$. It is the essential strength of our approach that this is possible, using the
consistency condition in perturbative form, see eq.~\eqref{consistency2}.

Thus, suppose that we are given $Y_i(\varphi, x)$, and all other vertex operators up
to order $1, \dots, i-1$. How to find $Y_i(\varphi^2, x), Y_i(\varphi^3, x),$ etc.?
Consider points such that $0< |x-y| < |y|< |x| $ and the following special
case of the $i$-th order consistency condition:
\ben
\sum_{j=0}^i Y_j(\varphi, x) Y_{i-j}(\varphi, y) = \sum_{j=0}^i Y_{i-j}(Y_j(\varphi, x-y)\varphi, y) \, .
\een
Under the hypothesis that this condition indeed holds for the---yet to be constructed---terms in this
equation that are not known at this point, we can draw the following conclusion. Let us look at the
term with $j=0$ on the right side of the equation.
%which is $Y_i(Y_0(x-y, \varphi)\varphi, y)$.
Using the known form of the $Y_0$'s (from the free theory), we have
$Y_0(\varphi,x-y)\varphi = |x-y|^{-(D-2)} \, \myid + \varphi^2$,
plus terms that are smooth in $x-y$ and vanish for $x=y$. Using $Y_i(\myid, y) = 0$ for $i>0$,
we hence arrive at the relation
\ben\label{recursion}
Y_i(\varphi^2, y) = \sum_{j=0}^i Y_j(\varphi, x) Y_{i-j}(\varphi, y) -
\sum_{j=1}^i Y_{i-j}(Y_j(\varphi, x-y)\varphi, y) + \dots \, .
\een
The dots represent terms that vanish in the limit as $|x-y| \to 0$. The singular terms with the $-$-signs may be
thought of as some sort of ``counterterms'', which cancel off the divergence of the first term on the right side in
the limit.
The key thing to note is now that all terms on the right side are known, by induction. Thus, taking the limit, we
obtain the desired vertex operator $Y_i(\varphi^2, x)$, and, by iterating this procedure in an obvious way, all other
$i$-th order vertex operators $Y_i(\varphi^3, x), Y_i(\varphi^4, x), \dots$.

In summary, our iterative scheme is set up in such a way that, at order $i$, we have to perform one inversion
of the Laplace operator in eq.~\eqref{mastereq} to get $Y_i(\varphi, x)$, and then we subsequently have to construct
all other vertex operators at order $i$ via the consistency condition. Unfortunately, it is not evident from what we
have said that the vertex operators constructed in this way will satisfy the
consistency condition to all orders. Furthermore,
there is certainly the freedom to add to the $i$-th order solution to the
inhomogeneous Laplace equation $Y_i(\varphi, x)$ a
solution to the homogeneous equation. It would seem that both issues are connected, and that one must impose the
validity of the $i$-th order consistency condition in order to (partly) fix this ambiguity. One would then, furthermore,
expect that the ambiguity is equivalent to the usual sort of renormalization ambiguity, which in our framework is given
by eq.~\eqref{fieldredef} (with $Z = \sum \lambda^i z_i$) and a change in the coupling constant $\lambda$ by a (formal)
diffeomorphism.

We will not address these issues in the present paper but rather focus on the kind of mathematical expressions that one
obtains following the above iterative scheme. We divide our discussion into several parts. First, in the next subsection
we discuss how to take the inverse of the Laplace operator in a way that is suitable in our setting. Then in
subsec.~\ref{graphrules},
we will discuss how to organize the terms that appear in the iteration process by a graphical notation involving trees.
In subsec.~\ref{renormalization}, we outline how the ``counterterms'' may be incorporated into the graphical notation.
In sec.~\ref{sec:4}, we find another representation of the vertex operators in terms of infinite sums of hypergeometric type.

\subsection{Inverting the Laplace operator}
\label{subsec:3.1}
In our inductive scheme, we have to invert the Laplace operator at each step of the induction.
According to the general setup, the vertex operators are functions $Y(a, \, . \,) \in {\mathcal A}(\mr^D \setminus \{0\})
\otimes {\rm End}(V)$, so we should invert on this function space. However, in perturbation theory, the
space of functions is actually much more restricted. At zeroth order, the vertex operators are in fact
infinite sums of products of creation/anhihilation operators, harmonic polynomials $h_{l,m}(\hat x)$ and
powers $r^k$, where $l \in \mn, k \in {\mathbb Z}, 0 < m \le N(D,l)$, see eqs.~\eqref{freephiv} and~\eqref{freevertex}. Unfortunately, such functions are not stable under
the inverse of the Laplace operator---we also get factors of $\ln r$. Inverting again, we get factors of $\ln^2 r$,
and so on. To incorporate the logarithms, we are thus led to introduce the following spaces of functions for
$i \in \mn$:
\ben
\mathcal{E}_i=\text{span}\left\{r^k \, \ln^j r \, h_{l,m}(\hat x):\,\,
k\in \mathbb{Z},\, j,l\in \mn, \,j\leq i, \, 1 \le m \le N(l,D) \right\}.
\een
Evidently, the union $\cup_j \mathcal{E}_j$ is a filtered ring ($\mathcal{E}_i \mathcal{E}_j \subset \mathcal{E}_{i+j}$).
Any right inverse of the Laplace operator, $G$,
takes us between $\mathcal{E}_i \to \mathcal{E}_{i+1}$ for all $i$.
Our inductive scheme will hence give us $i$-th order vertex operator in the space
\ben
Y_i(a, \, . \,) \in \mathcal{E}_i \otimes {\rm End}(V) \, .
\een
In order to give an explicit formula for $G$ we introduce a residue integral
trick for computing right inverses of the Laplacian which we are going to use extensively. We first define
\bena
G\left(r^k\,h_{l,m}(\hat x)\right) &=& G\left(\frac{1}{2\pi \iho} \oint_C \frac{\d\delta}{\delta}
r^{k+\delta}  h_{l,m}(\hat x)\right)\non\\
%&=&\frac{1}{2\pi i}\oint_C \frac{d\delta}{\delta} \, G\left(
%r^{k+\delta}  \, h_{l,m}(\hat x)\right)\non\\
&:=& \frac{1}{2\pi \iho} \oint_C \frac{\d\delta}{\delta}
\frac{r^{k+2+\delta}}{(k+2+\delta)(k+D+\delta)-l(l+D-2)}  \, h_{l,m}(\hat x) \, .
\label{residuetrick}
\eena
This defines $G$ as an operator on $\mathcal{E}_0 \to \mathcal{E}_1$. To extend the trick \eqref{residuetrick}
to all of $\mathcal{E} = \cup_j \mathcal{E}_j$, assume that we are given
$f(x)\in\mathcal{E}_j$ as a $j$-fold residue integral of the form
\bena
f(x) =\prod_{i=1}^j\left(\frac{1}{2\pi \iho} \oint_{C_i}\frac{r^{\delta_i}\d\delta_i}{\delta_i}\right)
\, F(\delta_1,\dots,\delta_j)\,
r^k \, h_{l,m}(\hat x)\,\,  ,
\label{generalresidue}
\eena
where $F(\delta_1,\dots,\delta_j)$ does not depend on $x$.
It is always possible to represent $f\in\mathcal{E}_j$ as a linear combination of expressions
of the form \eqref{generalresidue}, and in fact, this is the form we will encounter.
We define the Laplace inverse of $f(x)$ by
\bena
G f(x) &:=&
\prod_{i=1}^{j+1}\left(\frac{1}{2\pi \iho} \oint_{C_i}\frac{r^{\delta_i}\d\delta_i}{\delta_i}\right)
%\oint_{C_{j+1}}\frac{d\delta_{j+1}}{\delta_{j+1}}\,
%\left(\frac{r}{L_{j+1,l,m}}\right)^{\delta_{j+1}}
F(\delta_1,\dots,\delta_j)\non\\
&& \times \, \left[ (k+2+\sum_{i=1}^{j+1}\delta_i)(k+D+\sum_{i=1}^{j+1}\delta_i)-
l(l+D-2)\right]^{-1} r^{k+2}\, h_{l,m}(\hat x).
\label{Gdef}
\eena
We remark that the order of the residue integrals is not arbitrary here; the
integral over $\delta_{j+1}$ has to be performed first.

The above formula is not the only possible choice for $G$. In fact,
any other choice $G'$ of the right inverse compatible with
$\Delta G= id_{\mathcal{E}_i}, G: \mathcal{E}_i \to \mathcal{E}_{i+1}$ can
be parametrized  by constants $A_{j,l,m}, B_{j,l,m}$,
$j,l\in\mathbb{N}, m\in\{1,...,N(l,D)\}$:
\bena
G' \,[r^k \, \ln^j r \, h_{l,m}(\hat x)] & = &
G \, [r^k \, \ln^j r \, h_{l,m}(\hat x)] +\non\\
 & & \delta^{k+2}_l \, A_{j,l,m} \, r^l \, h_{l,m}(\hat x)+
\delta^{k+2}_{-l-D+2} \, B_{j,l,m} \, r^{-l-D+2} \, h_{l,m}(\hat x)\,\, ,
\label{ambig}
\eena
where $\delta^a_b$ is the Kronecker delta.
At each iteration step, we are free to choose one set of constants $A_{j,l,m}, B_{j,l,m}$.\\
This freedom is partly restricted by the consistency condition. One expects that the remaining
freedom corresponds to the renormalization ambiguities in the conventional framework. As already said, we will
not prove here that our particular choice for $G$ in eq.~\eqref{Gdef} actually leads to a
set of vertex operators that fulfill the consistency condition.

\subsection{Graphical rules for computing vertex operators}
\label{graphrules}

We now want to consider in more detail the iterative scheme for calculating
vertex operators in perturbation theory described at the beginning of this section, based on our
recipe for inverting $\Delta$ that we have just described.

\subsubsection{The case $D=2$:}

We will explain this first for the OPE vertex algebra in $D=2$, with
interaction $\lambda \, P(\varphi) = \lambda \, \sum \frac{c_p}{p!} \varphi^p$.
What makes the construction in this theory simple is that---as we will see---the
``counterterms'' in eq.~\eqref{recursion} have no effect. This is
directly related to fact that the field theory is ``super-renormalizable'' in field theory parlance.
As usual in perturbation theory, it is of no concern that the
interaction polynomial is not semi-bounded, even though this would render impossible
the non-perturbative construction of the Schwinger functions.

\begin{figure}[htp]
\vspace{10mm}
\unitlength=0.5pt
\SetScale{0.5}
\begin{center}
\fcolorbox{white}{white}{
  \begin{picture}(498,98) (47,-31)
    \SetWidth{1.0}
    \SetColor{Black}
    \Line(160,66)(160,34)
    \Line(160,34)(96,2)
    \Line(96,2)(48,-30)
    \Line(96,2)(96,-30)
    \Line(96,2)(144,-30)
    \Line(160,34)(160,-30)
    \Line(160,34)(272,-30)
    \Vertex(160,34){4}
    \Vertex(96,2){4}
    \Line(432,66)(432,34)
    \Line(432,34)(432,-30)
    \Vertex(432,34){4}
    \Vertex(496,2){4}
    \Line(496,2)(448,-30)
    \Line(496,2)(496,-30)
    \Line(496,2)(544,-30)
    \Line(432,34)(320,-30)
    \Line(432,34)(496,2)
  \end{picture}
}
\end{center}

\caption{Two trees representing the terms $G\left(\left(G:Y_0(\varphi,x)^3:\right)Y_0(\varphi,x)^2\right)$
and $G\left(Y_0(\varphi,x)^2\left(G:Y_0(\varphi,x)^3:\right)\right)$ respectively. Even though the trees
are related by a reflection, they have to be counted as different.
Both make a contribution to the vertex operator $Y_2(\varphi,x)$ in the theory
with interaction $\lambda P(\varphi)=\lambda \varphi^4/4!$. }
\label{fig:tree_ex}
\end{figure}
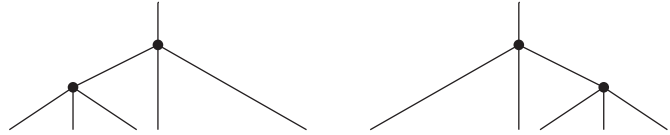

In $D=2$, the angular part and the $N(l,D) = 2$ harmonic polynomials at order $l>0$ are
\ben
\hat x = \e^{\iho\alpha} \, , \quad h_{l,\pm}(\hat x) = \frac{1}{\sqrt{2\pi}} \e^{\pm \iho l \alpha}
\een
At zeroth order, the vertex operator linear in $\varphi$ is given by
\ben
Y_0(\varphi, x) = \a_0 \, \ln r + \a_0^+ +
 \sum_{l=1}^{\infty} \sum_{m=\pm 1} \, \frac{1}{\sqrt{2l}}
\Big[ r^{l} \e^{\iho ml\alpha} \, \a_{l,m}^{+}
+ r^{-l} \e^{-\iho ml\alpha} \, \a_{l,m}^{} \Big] \, .
%\non\\
%&=& \frac{1}{2\pi\iho}  \oint_C \frac{\d \delta}{\delta}
%\Big[ \sum{l \ge 0, \, m=\pm 1} r^{l+\delta} \e^{\iho ml\alpha} \, \a_{l,m}^{+}
%+ r^{-l+\delta} \e^{-\iho ml\alpha} \, \a_{l,m}^{} \Big]
%\, .
\een
When determining the vertex operator $Y_{i+1}(\varphi, x)$ at order $i+1$,
we have to calculate $Y_{i}(\varphi^2,x)$, see eq.~\eqref{recursion}, then $Y_i(\varphi^3, x)$ and
so on, and then apply the inverse $G$ of $\Delta$ as in eq.~\eqref{mastereq}. Eq.~\eqref{recursion}
contains what we called ``counterterms'', which are the terms with the $-$-sign in front. However, it is not difficult
to derive that, in $D$ spacetime dimensions, and for a polynomial interaction, the scaling
degree (i.e., the power of $r$ of the most singular term apart from log's) of an $i$-th order vertex operator is
\ben
{\rm sd} \, \langle c| Y_i(a, x) |b \rangle \leq \dim(a) + \dim(b) - \dim(c) + i[D - \tfrac{D-2}{2} \, {\rm deg}(P)] \, ,
\een
where $\dim$ is a grading of $V$ by the engineering dimension of the fields and is defined for
a field of the form~\eqref{vadef} by
\ben
\dim(a) = \sum_{l, m} a_{l,m}(l + \tfrac{D-2}{2}) \, ,
\een
i.e., each factor of $\varphi$ counts as having dimension $\frac{D-2}{2}$, and
each derivative as $1$, which gives $l$ for the $l$-th order
differential operator $\bar h_{l,m}(\partial)$.
It follows that, in $D=2$ dimension with a polynomial interaction, the
scaling degree of all the terms $Y_j(\varphi, x-y)\varphi$ is strictly
positive for $j>0$. Hence, we may drop the counterterms in perform the limit
$|x-y| \to 0$ just on the first term. Thus, in $D=2$ and $i>0$, \eqref{recursion} reads
\ben
\label{simplerecursion}
Y_i(\varphi^2,x)=\sum_{j=0}^{i} Y_j(\varphi,x)Y_{j-i}(\varphi,x) \, ,
\een
and an analogous equation can easily be derived for the vertex operators
$Y_i(\varphi^p, x)$. The only difference is that we now have $p$ terms on the right
side and a corresponding $p-1$-fold summation.
We may thus form $Y_i(P'(\varphi), x)$, and hence, using the
field equation~\eqref{vertexfieldeq}, we get $Y_{i+1}(\varphi, x)$. Explicitly, this is $(i>0)$
\ben
Y_{i+1}(\varphi,x)= \sum_p \frac{c_p}{p!} \,\, G \!\! \sum_{j_1+...+j_p=i}
Y_{j_1}(\varphi,x) \cdots Y_{j_p}(\varphi,x) \, .
\label{itsimple}
\een
For $i=0$, the formula is instead [compare eq.~\eqref{freevertex}]
\ben
Y_{1}(\varphi,x)= \sum_p \frac{c_p}{p!} \,\, G \!
:Y_{0}(\varphi,x)^p: \, .
\label{it0simple}
\een
Each term in the sum over $p$ has an obvious graphical representation
by a vertex with $p$ incoming lines, see fig.~\ref{fig:tree_ex}.

%%%
%% Heiner modified 04/06/09
%%%
Adopting such a graphical notation
immediately helps one to see that if we
iterate eq.~\eqref{itsimple} starting from $i=0$ to arbitrary $i$, then the resulting expression
will be organized in terms of trees whose vertices have coordination number $p=1,2,\dots,{\rm deg}(P)$,
where $P$ was the polynomial characterizing the interaction.
%The lines of the tree are oriented, where we take the orientation to be pointing away from the
%root, downwards in the tree.
We use the notation $v\prec w$ to indicate that a vertex or leaf $v$ can be reached from the vertex
$w$ following the tree downwards. By $v\preceq w$, we mean $(v\prec w\text{ or } v=w)$.
As we have to heed the order of the creation and annihilation operators in
the term represented by a tree, we have to count as different
trees that are related to each other by a reflection or similar symmetry operation,
see again fig.~\ref{fig:tree_ex}.

In the remainder of the subsection, we want to describe in somewhat more detail what
the mathematical expression is for each such tree. Let $T$ be a tree on $i$ vertices, and
let $Y_i(T, \varphi, x)$ be the contribution to $Y_i(\varphi, x)$ coming from that tree.
Thus,
\ben\label{yit}
Y_i(\varphi, x) = \sum_{{\rm trees \,} T {\rm \, on \,} 1,...,i} \frac{\prod c_{p_v}}{\prod p_v!} Y_i(T, \varphi, x) \, ,
\een
where $p_v$ is the coordination number of the vertex $v$. If we start from the leaves of the tree, then
to each leaf $j$ ending in a vertex $v$ we have to associate a pair $(l_{j}, m_{j}) \in {\mathbb Z}_+ \times \{ \pm 1\}$
and one of the following factors
\ben\label{leavefactors}
\frac{1}{\sqrt{2l_{j}}} r^{l_{j}} \e^{\iho m_{j} l_{j}\alpha} \, \a_{l_{j},m_{j}}^{+} \quad \text{or} \quad
\frac{1}{\sqrt{2l_{j}}} r^{-l_{j}} \e^{-\iho m_{j} l_{j}\alpha} \, \a_{l_{j},m_{j}} \, .
\een
Or, if $l_j = 0$, we have to associate one of the factors $\a_0^+$ or
$\oint \frac{\d \delta_{j}}{\delta_{j}} \, r^{\delta_{j}} \a_0$, using the
residue trick to generate the logarithm. The creation/annihilation operators of the leaves associated
with the same vertex must be normal ordered, by eq.~\eqref{it0simple}. It is handy to distinguish the
two cases in eq.~\eqref{leavefactors} by calling the lines associated with the first type of
factor ``incoming'', while calling ``outgoing'' the other one. For each tree, we will
have to sum over all possible orientations of the leaves, because this
will correspond to different terms contributing to $Y_i(\varphi,x)$.\\
Let us now consider a vertex $v$ that has no further vertices attached to it downwards in the tree;
i.e. if we follow a line downwards starting from $v$, we arrive at a leaf.
At $v$, we have to multiply the factors in eq.~\eqref{leavefactors} associated to the leaves attached to $v$,
put this product into normal order
and then apply the inverse Laplacian $G$.
It is efficient to take care of the phase factors  by introducing, for each
vertex, an auxiliary integration variable $0 \le \beta \le 2\pi$, and to use the trivial identity
\bena\label{orthogonal}
&& \prod_{j\, {\rm outgoing}} \e^{\iho m_{j} l_{j} \alpha}  \prod_{j \, {\rm incoming}} \e^{-\iho m_{j} l_{j} \alpha}\\
&=& 2 \sum_{l=1}^\infty \int_{0}^{2\pi} \frac{\d \beta}{2\pi} \, \cos[l(\alpha-\beta)] \, \prod_{j\, {\rm outgoing}}
\e^{\iho m_{j} l_{j} \beta}  \prod_{j \, {\rm incoming}} \e^{-\iho m_{j} l_{j} \beta} + \text{($l=0$ term)} \non \, \, ,
\eena
which follows immediately using that the functions $\frac{1}{\sqrt{2\pi}}\e^{\pm \iho l\alpha}$ form an orthonormal
basis on $[0,2\pi]$.
The inverse of the Laplace operator has to be applied to an expression of the form
$r^{\tilde\nu} \cos(l(\alpha-\beta))$, where $r^{\tilde\nu}$ results from collecting the powers of $r$ of the
factors~\eqref{leavefactors} associated with the leaves of $v$. The power is thus
\ben
{\tilde\nu} = \sum_{j \, {\rm outgoing}} l_{j} - \sum_{j \, {\rm incoming}} l_j \, ,
\een
where we have assumed that none of the $l_j$'s is zero. Now we introduce $\delta \in \mc \setminus \mz$
and apply our residue trick from section \ref{subsec:3.1} to evaluate $G\left(r^{\tilde\nu} \cos(l(\alpha-\beta))\right)$.
(For each incoming line with $l_j=0$, we must replace
the corresponding term with another such $\delta_j$.) The result is a contour integral with integrand
$2 \frac{\cos(l(\alpha-\beta))}{\nu^2-l^2} r^{\nu+2}$, with $\nu=\tilde\nu+\delta$.
(This holds for $l\neq 0$, for $l=0$ we have $\frac{1}{\nu^2} r^{\nu+2}$ .) Having introduced the additional
integration parameter $\beta$ now pays off, as we can carry out the sum over $l$ using
the following formula\footnote{This may be viewed as a degenerate case of the
Dougall identity, see appendix~\ref{app:A}. To prove the identity, consider the contour integral
$$\oint_C\frac{\cos\left(z(\alpha-\pi)\right)\d z}{z(\nu-z)\sin\pi z}$$
where $C$ is a circle around the origin with radius $\pi(M+1/2),\,M\in\mathbb N$.
For $M\rightarrow \infty$, the integral vanishes and we obtain eq.~\eqref{cosid} via
the residue theorem.} for $\nu \in \mc \setminus {\mathbb Z}$:
\ben\label{cosid}
\frac{1}{2\nu^2} + \sum_{l=1}^\infty \frac{\cos(l\alpha)}{\nu^2-l^2} = \frac{\pi}{\sin(\nu\pi)}
\frac{\cos(\nu \alpha-\nu \pi)}{2\nu} \, .
\een
We can repeat this procedure
for each of the remaining vertices of the tree, moving upwards in the tree. At each new vertex $v$
we introduce a new integration variable $\beta_v$, and a new summation variable $l_{v}$---which
is summed over using the above cosine identity---as well as a
variable $\nu_{v}$ defined similarly as above.
More precisely, when we use the residue trick to perform the inverse of the Laplace operator,
we must first introduce for each vertex $v$ in the tree a new variable
$\delta_{v} \in \mc \setminus \mathbb Z$, and the residue in this variable then has to be taken in the end. If we do
all this, we hence arrive at the following {\em graphical rules} for calculating $Y_i(T, \varphi, x)$,
and hence $Y_i(\varphi, x)$:

\begin{enumerate}
\item Draw the tree $T$ with $i$ vertices. Label the vertices by an index, $v$, and the lines by
pairs of indices $(vw)$. The vertex $v$ has coordination number $p_v$. The leaves also carry
indices.
\item With each leaf $j$ adjacent to a vertex $v$, associate a pair
$(l_{j}, m_{j}) \in {\mathbb Z}_+ \times \{ \pm 1\}$ and one of the following factors
\ben\label{leavefactors2}
\frac{1}{\sqrt{2l_{j}}} \e^{\iho m_{j} l_{j}\beta_v} \, \a_{l_{j},m_{j}}^{+} \quad \text{or} \quad
\frac{1}{\sqrt{2l_{j}}} \e^{-\iho m_{j} l_{j}\beta_v} \, \a_{l_{j},m_{j}} \, .
\een
The first factor is chosen if the line associated with the leaf is oriented upwards,
and the second if it is oriented downwards.
For the zero modes ($l_j = 0$), we have to associate one of the factors $\a_0^+$ or
$\oint \frac{\d \delta_{j}}{\delta_{j}} \, r^{\delta_{j}} \a_0$, where $\delta_j \in \mc \setminus \mz$,
again depending on the orientation. The creation/annihilation operators of the leaves
connected to the same vertex are to be normal ordered.

\item With each vertex $v$ associate a parameter $\delta_v \in \mc \setminus \mz$, and
a parameter $\beta_v \in [0,2\pi]$.

\item With each vertex $v$ we associate $\nu_{v} \in \mc \setminus \mz$ defined by
\ben\label{nukdef}
\nu_{v} = \sum_{{\rm in \, leaves} \, j \prec v} l_j -
\sum_{{\rm out \, leaves} \, j \prec v} l_j + \sum_{{\rm vertices} \, w \preceq v} (2+\delta_w) \, .
\een
The ``2'' results from the inversion of the Laplacian, which
at each inversion step (i.e., each vertex) raises the power of the radial coordinate by 2. The
$\delta_w$ arises from the residue trick for the Laplace inversion at each vertex $w$ below $v$.

\item With the root, associate the parameter $\alpha \in [0,2\pi]$, and
the factor $r^{\nu_{\rm root}}$ collecting all the factors of $r$, where $x=r\e^{\iho\ alpha}$.
The number $\nu_{\rm root} \in \mc$
is defined as in eq.~\eqref{nukdef}, but with the vertex $v$ replaced by the root, so that the
sums contain contributions from all leaves of $T$.

\item With each line $(vw)$ (not connecting a leaf) associate a factor $$\frac{\pi}{\sin (\pi \nu_w)}
\frac{\cos[(\beta_v-\beta_w) \nu_w - \pi \nu_w]}{ \nu_w}.$$ This results from the application of
the cosine identity~\eqref{cosid}.

\item Perform the sum over all $l_j, m_j$ and zero modes. Furthermore, perform  the integrals
$$\prod_{{\rm vertices} \, v} \int_0^{2\pi} \d \beta_v
\quad \text{and} \quad
\prod_{{\rm vertices} \, v} \frac{1}{2\pi \iho}  \oint_{C_v} \frac{\d \delta_v}{\delta_v}\, . $$
Finally, take the sum over all possible orientations of the leaves.
\end{enumerate}

\begin{figure}[htp]
\medskip
\unitlength=0.5pt
\SetScale{0.5}
\begin{center}
\fcolorbox{white}{white}{
  \begin{picture}(288,172) (207,-31)
    \SetWidth{1.0}
    \SetColor{Black}
    \Line(336,140)(336,108)
    \Line(336,108)(272,44)
    \Line[arrow,arrowpos=0.5,arrowlength=5,arrowwidth=2,arrowinset=0.2](272,44)(208,-20)
    \Line[arrow,arrowpos=0.5,arrowlength=5,arrowwidth=2,arrowinset=0.2](272,44)(336,-20)
    \Line[arrow,arrowpos=0.5,arrowlength=5,arrowwidth=2,arrowinset=0.2,flip](336,108)(464,-20)
    \Vertex(272,44){4}
    \Vertex(336,108){4}
    \Text(204,-46)[lb]{\small{\Black{$j_1$}}}
    \Text(332,-46)[lb]{\small{\Black{$j_2$}}}
    \Text(460,-46)[lb]{\small{\Black{$j_3$}}}
    \Text(250,44)[lb]{\small{\Black{$v_2$}}}
    \Text(310,108)[lb]{\small{\Black{$v_1$}}}
  \end{picture}
}
\end{center}

\caption{Tree contributing to $Y_2(\varphi,x)$, with fixed orientation of the leaves.}
\label{fig:tree_ex2}
\end{figure}
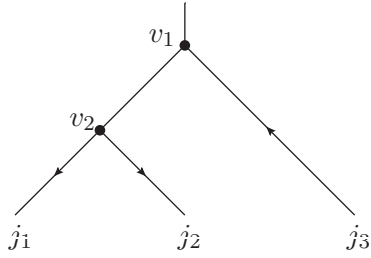

The last step requires some further comment. Our first comment concerns the choice of the integration
contours $C_v$ in the residue integral. They can be chosen as small circles
around the origin with radii $|\delta_v|$ chosen so that for any subset $\mathcal V$
of the vertices in the tree, and for any
values of the phases of the $\delta_v, v\in\mathcal V$, we have $\sum_{v\in\mathcal V}\delta_v \in\mathbb{C}\setminus \mathbb{Z}$.
We need this in order for our residue trick to work; the exponents $\nu_v$ have to be non-integer.
The second remark concerns
the convergence of the multiple summation over the counters $l_j$. To illustrate the general point,
we consider the example tree with a fixed orientation of the leaves given in fig.~\ref{fig:tree_ex2}.

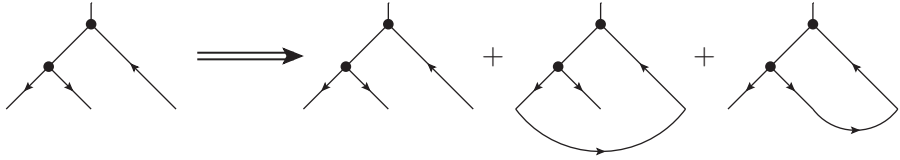
\begin{figure}[htp]
\vspace{10mm}
\unitlength=0.5pt
\SetScale{0.5}
\begin{center}
\fcolorbox{white}{white}{
  \begin{picture}(674,116) (79,-31)
    \SetWidth{1.0}
    \SetColor{Black}
    \Line(144,84)(144,68)
    \Line(144,68)(112,36)
    \Line[arrow,arrowpos=0.5,arrowlength=5,arrowwidth=2,arrowinset=0.2,flip](80,4)(112,36)
    \Line[arrow,arrowpos=0.5,arrowlength=5,arrowwidth=2,arrowinset=0.2,flip](144,4)(112,36)
    \Line[arrow,arrowpos=0.5,arrowlength=5,arrowwidth=2,arrowinset=0.2,flip](144,68)(208,4)
    \Vertex(112,36){4}
    \Vertex(144,68){4}
    \Line[arrow,arrowpos=0.5,arrowlength=5,arrowwidth=2,arrowinset=0.2,flip](368,68)(432,4)
    \Vertex(336,36){4}
    \Line(368,68)(336,36)
    \Line[arrow,arrowpos=0.5,arrowlength=5,arrowwidth=2,arrowinset=0.2,flip](304,4)(336,36)
    \Line[arrow,arrowpos=0.5,arrowlength=5,arrowwidth=2,arrowinset=0.2,flip](368,4)(336,36)
    \Vertex(368,68){4}
    \Line(368,84)(368,68)
    \Line[arrow,arrowpos=0.5,arrowlength=5,arrowwidth=2,arrowinset=0.2,flip](528,68)(592,4)
    \Line[arrow,arrowpos=0.5,arrowlength=5,arrowwidth=2,arrowinset=0.2,flip](464,4)(496,36)
    \Line(528,68)(496,36)
    \Line(528,84)(528,68)
    \Vertex(528,68){4}
    \Vertex(496,36){4}
    \Line[arrow,arrowpos=0.5,arrowlength=5,arrowwidth=2,arrowinset=0.2,flip](528,4)(496,36)
    \Line(688,84)(688,68)
    \Vertex(688,68){4}
    \Line[arrow,arrowpos=0.5,arrowlength=5,arrowwidth=2,arrowinset=0.2,flip](688,68)(752,4)
    \Line(688,68)(656,36)
    \Line[arrow,arrowpos=0.5,arrowlength=5,arrowwidth=2,arrowinset=0.2,flip](624,4)(656,36)
    \Line[arrow,arrowpos=0.5,arrowlength=5,arrowwidth=2,arrowinset=0.2,flip](688,4)(656,36)
    \Vertex(656,36){4}
    \Arc[arrow,arrowpos=0.5,arrowlength=5,arrowwidth=2,arrowinset=0.2](720,28)(40,-143.13,-36.87)
    \Arc[arrow,arrowpos=0.5,arrowlength=5,arrowwidth=2,arrowinset=0.2](528,52)(80,-143.13,-36.87)
    \SetWidth{1.5}
    \Line[arrow,arrowpos=1,arrowlength=14,arrowwidth=5.6,arrowinset=0.2,double,sep=5](224,44)(296,44)
    \Text(440,36)[lb]{\small{\Black{$+$}}}
    \Text(600,36)[lb]{\small{\Black{$+$}}}
  \end{picture}
}
\end{center}
\caption{Application of Wick's theorem to a tree. On the left side of the arrow, we have the tree
from fig.~\ref{fig:tree_ex2}. On the right side, we have the representation of the normal ordered
expression for it. Contractions are represented by an additional line.}
\label{fig:ex_wick}
\end{figure}

We apply the above rules and obtain the following expression for fig.~\ref{fig:tree_ex2}:
\bena
%G\left(\left(G:Y_0(\varphi,x)^2:\right)Y_0(\varphi,x)\right) \non\\
&&\oint_{C_1}\oint_{C_2}\frac{\d\delta_{v_1}}{\delta_{v_1}}\frac{\d\delta_{v_2}}{\delta_{v_2}}
\int_0^{2\pi}\int_0^{2\pi} \d \beta_{v_1}\d \beta_{v_2}\sum_{l_1,l_2,l_3}\sum_{m_1,m_2,m_3}\non\\
&\times&
\frac{\pi\cos\left[(\beta_1-\beta_2-\pi)\nu_2\right]}
{\nu_2\sin\pi\nu_2}\frac{\pi\cos\left[(\alpha-\beta_1-\pi)\nu_1\right]}{\nu_1\sin\pi\nu_1}\non\\
&\times& e^{\iho\ beta_{2}(-m_1l_1-m_2l_2)}e^{\iho\ beta_{1}m_3l_3}
r^{\nu_1}\a^+_{l_{1},m_1}\a^+_{l_{2},m_2}\a_{l_{3},m_3}
\eena
with $\nu_2=-l_1-l_2+2+\delta_2,\,\,\nu_1=-l_1-l_2+l_3+4+\delta_1+\delta_2$,
which makes a contribution to the vertex operator $Y_2(\varphi,x)$ in the theory with interaction
$\lambda P(\varphi)=\lambda\varphi^3/3!$. We are going to normal-order
the product of creation and annihilation operators $\a^+_{l_{1},m_1}\a^+_{l_{2},m_2}\a_{l_{3},m_3}$,
which will make it easier to calculate matrix elements of the vertex operator.
(Recall that the vertex operators were related to the OPE coefficients by the formula
$\langle c| Y(a, x) |b \rangle = C^c_{ab}(x)$.)
The only terms for which normal ordering has non-trivial effects are those where either $(l_{1},m_{1})=(l_{3},m_{3})$
or $(l_{2},m_{2})=(l_{3},m_{3})$. These terms can be visualized by joining
the leaves $j_1,j_3$ or $j_2,j_3$ respectively by a new line, see fig.~\ref{fig:ex_wick}.

We see that, in general, we will ``close loops'' in the tree when we apply
Wick's theorem (see eq.~\eqref{Wick}) to the products of creation and annihilation operators.
Thus, when normal ordering the operators in the expression $Y_i(T, \varphi, x)$,
we get contractions between the creation and annihilation operators between the
individual leaves (not attached to the same vertex, as these already are normal ordered).

Let $j$ and $k$ be leaves. Let $j$ be outgoing and standing to the left of $k$,
which shall be incoming. Their contraction gives $\delta_{l_j, l_k} \delta_{m_j,m_k}$.
We combine this with the other factors in eq.~\eqref{leavefactors2} and carry out the sum over
$m_j$, getting
$\frac{\cos[l_{j}(\beta_v-\beta_w)]}{l_{j}}$, when $l_j \neq 0$, where
$v$ is the vertex adjacent to $j$ and $w$ the vertex adjacent to $k$.
%The plus/minus sign depends on the orientations of the leaves.
When $l_{j} = 0$, we get instead $\frac{1}{\delta_j}$.
We will represent each such new factor by a new line joining the respective leaves.
If we apply this systematically, we are thus led to a wider class of graphical objects
that are obtained from our tree $T$ by joining an arbitrary number of leaves, but never joining
two leaves from the same vertex. These new graphs, which we call $G$,
are not trees any more, but also contain loops.
The orientation of the loops is always from the ``left'' to the ``right'', pointing from
a leaf representing a creation operator to a leaf representing
an annihilation operator.
In the language of graphs, $T$ is a ``spanning
tree'' for each $G$, and we collect these graphs in the set
\ben
\G(T) = \{ \text{graphs $G$} \mid \text{$T$ a spanning tree for $G$} \}.
\een
Our graphs $G$ contain three different kinds of lines, or edges $e$: The edges $e \in T$ that
were already present in the tree $T$, the edges $e \in G \setminus T$ that
were created by joining two leaves, and the leaves $e$ that were not joined (the ``external lines'').
The edges in the first category carry momenta $\nu_{e} \in \mc$ that are
determined by the momentum conservation rule \eqref{nukdef}. The edges in the second
category carry ``loop'' momentum $l_{e} \in \mz$, and
the leaves $e$ carry ``external'' quantum numbers $l_{e}, m_{e}$. The
collection of loop and external momenta together is
the same as the assignment $l_{e}$ above. We denote by $Y_i(G, \varphi, x)$
the contribution to $Y_i(\varphi, x)$ from such an individual loop graph $G$ where the
sum over all possible orientations of the external lines is understood. In other words,
\ben\label{yig}
Y_i(\varphi, x) = \sum_{{\rm trees \,} T {\rm \, on \,} 1,...,i} \frac{\prod c_{p_k}}{\prod p_k!}
\sum_{G \in {\mathcal G}(T)}  Y_i(G, \varphi, x) \, .
\een
The contribution
$Y_i(G, \varphi, x)$ from an individual graph $G \in \G(T)$, is given in turn by
\bena
Y_i(G, \varphi, x) &=& \sum_{\substack{\rm leaf\\ \rm orientations}}\,\sum_{l_e,m_e, \,e \,{\rm leaf}}\,
\sum_{l_e,\, e \in G \setminus T }
\left( \prod_{{\rm vertices} \, v} \int_{C_v} \frac{\d \delta_v}{\delta_v} \int_0^{2\pi} \d \beta_v \right) \non\\
&\times&
\prod_{e \in T} \frac{\pi}{\sin\pi(l_e+\delta_e)} \, \frac{\cos[(l_e+\delta_e)(\beta_e-\pi)]}{l_e+\delta_e}\,
%\left(
%\begin{matrix}
%\nu_{ij}+D-3\\
%\nu_{ij}
%\end{matrix}
%\right)
\prod_{e \in G \setminus T}
%\left(
%\begin{matrix}
%h_{ij}+D-3\\
%h_{ij}
%\end{matrix}
%\right)
\frac{\cos(\beta_e l_e)}{l_e} \non\\
&\times& \exp \left( \ln r \left\{ \sum_{{\rm in \, leaves} \, e} l_e - \sum_{{\rm out \, leaves} \, e} l_e + \sum_{{\rm vertices} \, v} (2+\delta_v) \right\} \right) \non\\
&\times& : \prod_{{\rm leaves} \, e}  \a_{l_e,m_e}^\pm \frac{\e^{\pm \iho m_e l_e \beta_e}}{\sqrt{2l_e}} : \, .
\label{amplitude}
\eena
This formula requires several comments. First, we have used now the notation ``$e$'' for the edges of
$e$, which include those present already in $T$, and those that close the loops, i.e. in $G \setminus T$.
To each of the latter, we have associated an integer $l_e \in \mathbb N$, which is summed over.
Each of the lines $e$ already present in $T$ also carries an $l_e \in \mathbb Z$, which is determined by
the ``conservation rule''
\ben
\label{lconservation}
2 =
\sum_{e \, {\rm outgoing}} l_e
%\,+ \sum_{e=(wv),\,w\succ v} l_e
-\sum_{e \, {\rm incoming}} l_e
%-\sum_{e=(vw),\,w\prec v} l_e\,.
\een
at every vertex $v$, where the incoming lines $e$ are either incoming leaves or lines below $v$ (in the sense of
the relation $\prec$), and where the outgoing lines $e$ are either outgoing leaves or the line above $v$ (in the sense of
the relation $\succ$). Hence, the numbers $l_{e}, e \in T$ are determined by the $L$ loop
momenta $l_{e}, e \in G \setminus T$ and external momenta (leaves) $l_e$ via the
momentum conservation rule at the vertices of $G$. That momentum conservation
rule comes from eq.~\eqref{nukdef}. We have also introduced $\delta_e$ as the sum
$\sum_v \delta_v$ of all those $\delta_v \in \mc \setminus \mathbb Z$ that are associated with
a vertex ``below'' the line $e$ in the original tree, and this accounts for the corresponding
term in eq.~\eqref{nukdef}. Finally, also the leaves (i.e. uncontracted lines) carry indices $l_e,m_e$
that are summed, and we have set $\beta_e = \beta_v - \beta_w$ if $e = (vw)$, with
$\beta_v = \alpha$ if $v$ is the root, and $\beta_e=\beta_v$ if $e$ is a leaf associated to the vertex $v$.
As always, $x$ is related to $r$ and $\alpha$ by
$x = r\e^{\iho \alpha}$.

\begin{figure}[htp]
\vspace{10mm}
\unitlength=0.5pt
\SetScale{0.5}

%\begin{center}
\fcolorbox{white}{white}{
  \begin{picture}(272,456) (59,-19)
    \SetWidth{1.0}
    \SetColor{Black}
    \Line(112,360)(112,417)
    \SetWidth{0.0}
    \Vertex(112,280){5.657}
    \SetWidth{1.0}
    \Line[arrow,arrowpos=0.5,arrowlength=5,arrowwidth=2,arrowinset=0.2](112,160)(112,205)
    \Line[arrow,arrowpos=0.5,arrowlength=5,arrowwidth=2,arrowinset=0.2](112,120)(112,75)
    \Text(128,176)[lb]{\small{\Black{$j$}}}
    \Text(128,88)[lb]{\small{\Black{$j$}}}
    \Text(296,384)[lb]{\normalsize{\Black{$\rightarrow\quad\frac{1}{\sin(\pi\nu_w)} \, \frac{\cos[(\nu_w)(\beta_e-\pi)]}{2\nu_w}$}}}
    \Text(296,288)[lb]{\normalsize{\Black{$\rightarrow\quad  r^{2+\delta_v} \quad\text{(and impose ``momentum conservation'' eq.~ \eqref{nukdef})}$}}}
    \Text(296,184)[lb]{\normalsize{\Black{$\rightarrow\quad  \frac{1}{\sqrt{2l_{j}}}r^{l_j} \e^{\iho m_{j} l_{j}\beta_v} \, \a_{l_{j},m_{j}}^{+} $}}}
    \Text(296,96)[lb]{\normalsize{\Black{$\rightarrow\quad \frac{1}{\sqrt{2l_{j}}} r^{-l_j}\e^{-\iho m_{j} l_{j}\beta_v} \, \a_{l_{j},m_{j}}$}}}
    \Text(120,288)[lb]{\small{\Black{$v$}}}
    \Line[dash,dashsize=10](112,328)(112,280)
    \Line[dash,dashsize=10](112,280)(64,248)
    \Line[dash,dashsize=10](112,280)(112,248)
    \Line[dash,dashsize=10](112,280)(160,248)
    \SetWidth{0.0}
    \Vertex(72,24){5.657}
    \Vertex(152,24){5.657}
    \SetWidth{1.0}
    \Arc[arrow,arrowpos=0.5,arrowlength=5,arrowwidth=2,arrowinset=0.2](112,45.333)(45.333,-151.928,-28.072)
    \Text(56,32)[lb]{\small{\Black{$v$}}}
    \Text(160,32)[lb]{\small{\Black{$w$}}}
    \Text(296,24)[lb]{\normalsize{\Black{$\rightarrow \quad \frac{\cos(\beta_e l_e)}{l_e}$}}}
    \Text(112,-24)[lb]{\small{\Black{$l_e$}}}
    \SetWidth{0.0}
    \Vertex(112,120){5.657}
    \Vertex(112,208){5.657}
    \Text(96,120)[lb]{\small{\Black{$v$}}}
    \Text(96,208)[lb]{\small{\Black{$v$}}}
    \Vertex(112,360){5.657}
    \Vertex(112,416){5.657}
    \Text(88,360)[lb]{\small{\Black{$w$}}}
    \Text(88,416)[lb]{\small{\Black{$v$}}}
  \end{picture}
}
%\end{center}

\caption{The rules for the ``amplitude'' of a graph. The appropriate summations and integrals have to be
understood. Slightly deviating from the earlier statement of the rules, we do not associate all
powers of $r$ with the root but all leaves and all vertices carry some power of $r$. }
\label{fig:2D_rules}
\end{figure}
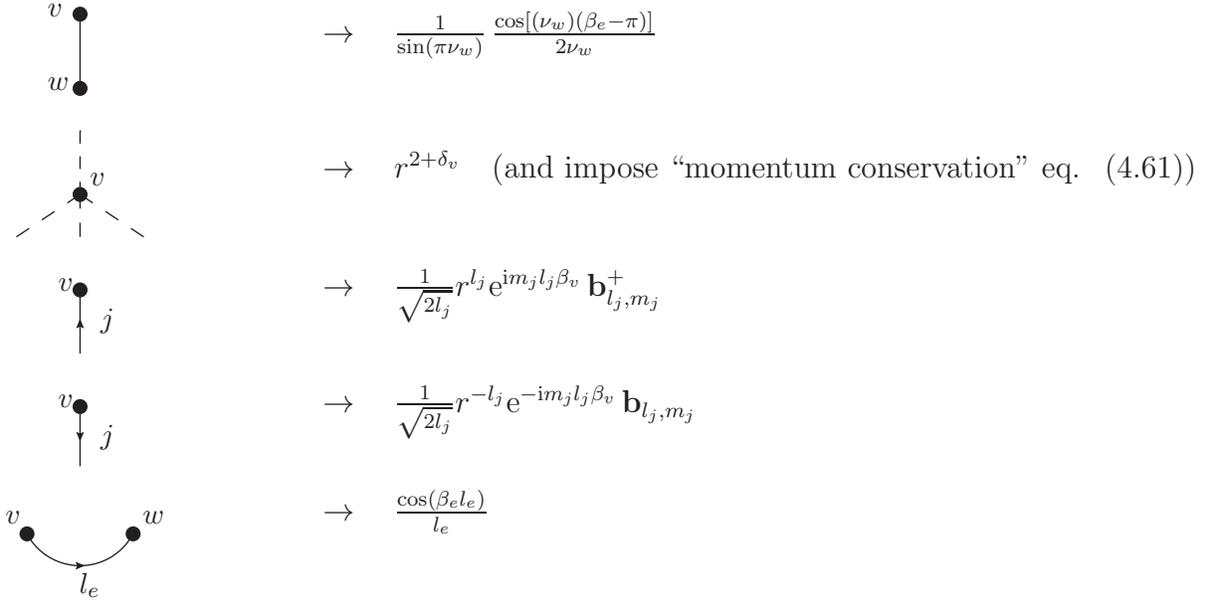

The formula for $Y_i(G, \varphi, x)$ given above looks quite complicated. To see that the sums
converge, we now introduce the following trick that, in effect, replaces the sums by integrals.
The basic trick is to implement the conservation rule~\eqref{lconservation} by the integral
\ben\label{constrick}
\int_{0}^{2\pi} \frac{\d t_v}{2\pi} \exp \left( \iho t_v\left\{ 2+\sum_{e = (vw): v \succ w} l_e - \sum_{e = (vw): v \prec w} l_e \right\} \right)  \,\,\, ,
\een
with one new integration variable $t_v$ per vertex $v$ because momentum conservation holds at
each vertex. Once the momentum conservation rule has been
implemented (at the expense of the new integrals), we can now sum over {\em all} $l_e$ independently,
and not just the $l_e$ associated with the loop lines $e \in G \setminus T$. We {\em then} do the new integrals
afterwards. The advantage is that the sums can now be performed explicitly. We hence  trade the summations
over $l_e, e \in G \setminus T$ for the integrals over the $t_v$, where $v$ runs through the vertices, as we now
explain in more detail.

The infinite summations that we have to deal with occur at each line $e \in G$ that is not
an external line (leaf), and they are performed with the aid of the formula
\ben\label{gauss}
\sum_{l=0}^\infty \frac{\cos[(l+\delta)\beta] \, h^l}{l+\delta} =
\frac{1}{2\delta} \Bigg(\e^{\iho \delta \beta}\,  {}_2 F_1( \delta, 1; 1+\delta; h\e^{\iho \beta}) +
\e^{-\iho \delta \beta} {}_2 F_1( \delta, 1; 1+\delta; h\e^{-\iho \beta}) \Bigg) \, ,
\een
which holds for $\delta \in \mc \setminus \mz$, and $|h| < 1$, and which
follows straightforwardly from the definition of the Gauss hypergeometric series, see appendix~\ref{app:A}.
In the above complicated formula~\eqref{amplitude}
for $Y_i(G, \varphi, x)$, we now apply the formula at each internal line $e\in T$, with
the choice $l=l_e$, $\delta = \delta_e$,
$\alpha = \beta_e$, and $h = \e^{\iho t_e}$, where $t_e = t_v - t_w$ for the line $e=(vw)$.
For the loop lines ($e \in G \setminus T$), we have the formula
\ben
\label{loopsum}
\sum_{l\in\mathbb{N}}\frac{\cos(l\alpha)}{l}h^l=-\ln\sqrt{1+h^2-2h\cos\alpha}\,.
\een
However, we cannot apply these formulae straightforwardly, for two reasons. First, they only hold for $|h| < 1$, while our choice would correspond to $|h|=1$, which is on the boundary of the
disk of convergence of the Gauss hypergeometric series. Secondly, the series in $Y_i(G, \varphi, x)$
have to go over both positive and negative values of $l_e$ if $e$ is neither a leaf nor a loop line,
whereas eq.~\eqref{gauss} only goes over
non-negative values. The remedy to this difficulty is as follows. We first split each sum
over $l_e \in \mz$ into positive and negative values of $l_e$. Then, for the positive values
of $l_e$, we replace $t_e$ by $t_e + \iho 0$ (i.e., we add a small positive imaginary part),
and for negative values of $l_e$ we replace $t_e$ by $t_e - \iho 0$. This then justifies the
exchange of summation and integration over the $t_v$ {\em before} we take the small
imaginary part to zero, and the remaining integration over the $t_v$ must then be understood in
the sense of distributions.

The trick affects our formula~\eqref{amplitude} for $Y_i(G, \varphi, x)$ as follows. Instead of the sum/integral
in the first line of eq.~\eqref{amplitude}, we now have
\ben
\sum_{\substack{\rm leaf\\ \rm orientations}}\sum_{l_e,m_e,\,e \, {\rm leaf}} \left( \prod_{{\rm vertices} \, v} \int_{C_v} \frac{\d \delta_v}{\delta_v} \int_0^{2\pi} \e^{2\iho t_v} \d t_v \int_0^{2\pi} \d \beta_v \right) \, ,
\een
i.e., we have gotten rid of the---potentially dangerous---summation over the loop momenta, and replaced
these by additional integrations, which are easier to control as we will see in a moment. Furthermore,
the term in the first product $(e \in T)$ in the second line becomes
\bena
\label{g2sum}
g_2(\delta_{e}, \cos \beta_e, t_e) &:=&
\frac{\pi}{\sin\pi\delta_e}\Big(\sum_{l=0}^\infty (-1)^{l}\e^{\iho( t_e+\i0)l}
\frac{\cos[(l+\delta_e)\beta_e] }{l+\delta_e}\non\\
\,\,\,\,\,&&\,\,\,\,\,\,\,\,\,\,\,\,\,\,+\sum_{l=-\infty}^{-1} (-1)^{l}\e^{\iho( t_e-\i0)l}\frac{\cos[(l+\delta_e)\beta_e] }{l+\delta_e}\Big)\,.
\eena
Using eq.~\eqref{gauss}, we get
\bena
\label{g2def}
&&g_2(\delta_e, \cos \beta_e, t_e) = \\
&&
\frac{\e^{+\iho \delta_e \beta_e}}{2\delta_e \sin \pi\delta_e}\, \Bigg( {}_2 F_1( \delta_e, 1; 1+\delta_e; -\e^{\iho (+\beta_e + t_e + \iho 0)}) -
{}_2 F_1( -\delta_e, 1; 1-\delta_e; -\e^{\iho( -\beta_e - t_e + \iho 0)}) -1\Bigg) \non \\
&+&
\frac{\e^{-\iho \delta_e \beta_e}}{2\delta_e \sin \pi\delta_e}\, \Bigg( {}_2 F_1(\delta_e, 1; 1+\delta_e; -\e^{\iho (-\beta_e + t_e + \iho 0)}) -
{}_2 F_1(-\delta_e, 1; 1-\delta_e; -\e^{\iho( +\beta_e - t_e + \iho 0)}) -1\Bigg)  \, . \non
\non
\eena
The term in the second product $(e \in G\setminus T)$ is as in eq.~\eqref{loopsum}.
The remaining parts of the formula \eqref{amplitude} are unchanged. The total
effect of these manipulations is reflected in eq.~\eqref{amplitude2}, setting
$D=2$ and $\epsilon_v = 0$ there.

A possible divergence can now only come from the $\d \beta_v \d t_v$ integrations,
and the danger can only come from configurations near $\beta_e = \pm t_e+\pi$, where
the argument of the hypergeometric function tends to 1. It is at this
stage that having integrals instead of sums pays off, because we can now use
the well-known expansion formula for the Gauss hypergeometric function near $x=1$:
\ben
{}_2 F_1(\delta, 1; 1+\delta; x) =  2\delta \, [ \gamma_{\rm E} - \psi(\delta) ] - 1
- \delta \, \ln(1-x)  + o(1-x) \, .
\een
When we apply this to the ${}_2F_1$-factors in the terms above, we find that
\ben
g_2(\delta_e, \cos \beta_e, t_e)
\sim -\frac{\e^{\iho \delta_e \beta_e}}{2 \sin(\delta_e \beta_e)} \, \ln |\beta_e - t_e -\pi| +
\frac{\e^{-\iho \delta_e \beta_e}}{2 \sin(\delta_e \beta_e)} \, \ln |\beta_e + t_e -\pi| \, ,
\een
near $\beta_e = \pm t_e + \pi$,
where we have also used standard identities such as $\ln(x+\iho 0) = \ln |x| + \iho \pi \theta(x)$.
The contour integrals over the parameters $\delta_e$ are harmless as long as we choose the contours
such that ${\rm dist}(\delta_e, \mz) > {\rm const.}$
%%%% HO 07/06/09
These estimates suffice to show that any matrix element $\langle c| Y_i(\varphi, x)| b\rangle$ is
convergent. To see this, we recall that $Y_i(\varphi, x)$ was the sum over all trees $T$
on $i$ elements of the quantities $Y_i(T, \varphi, x)$, which in turn was the sum
over all $G \in {\mathcal G}(T)$ of the quantities $Y_i(G, \varphi, x)$, for which we gave
a formula above. Now, when forming a matrix element $\langle c| Y_i(G, \varphi, x)| b \rangle$,
we get a sum of terms of the form
\ben
\Big \langle c \Big| \,\, : \prod_{{\rm leaves} \, e}\a_{l_e,m_e}^\pm \frac{\e^{\pm \iho m_e l_e \beta_e}}
{\sqrt{2l_e}} : \,\, \Big| b \Big \rangle\,.
\label{cr_an_prod}
\een
We write  $\langle c |=(c!)^{-1/2}\langle 0 |\prod \left(\a_{l,m}\right)^{c_{l,m}}$ and
$|b\rangle=(b!)^{-1/2}\prod \left(\a^+_{l,m}\right)^{b_{l,m}}|0 \rangle$.
Furthermore, for a fixed assignment of orientations and indices $l_e,m_e$ to each of the (uncontracted) leaves
$e\in G$, we introduce the multiindices $a^+,a^-$ by
\bena
a^+_{l,m}&=& \# \left\{e\in G: e \,{\rm incoming},(l_e,m_e)=(l,m)\right\}\non\\
a^-_{l,m}&=&\# \left\{e\in G: e \,{\rm outgoing},(l_e,m_e)=(l,m)\right\}\,\,.
\eena
In order for the matrix element~\eqref{cr_an_prod} not
to vanish, all creation operators have to be contracted with annihilation operators
of the same indices.
%, i.e. we have to have multiindices $b_a,b_c,c_a,c_b$ fulfilling
%\bena
%b=%b_a+b_c,\,\,\,c=c_a+c_b\non\\
%a^-=b_a,\,\,\,a^+=c_a,\,\,\, b_c=c_b\,\,.
%\label{pairing}
%\eena
Remember that we are considering the contribution of a graph $G$ with a fixed number of leaves. In the
above notation, that means $|a^+|+|a^-|$ is fixed.
We also assume that the multiindices $b,c$ are fixed. This means there is maximally one choice for $a^+,a^-$
so that the matrix element~\eqref{cr_an_prod} does not vanish. In this case the latter is equal to
\ben
(\ref{cr_an_prod}) =
\prod_{e \,{\rm incoming}}\frac{\e^{\iho m_e l_e \beta_e}}{\sqrt{2l_e}}
\prod_{e \,{\rm outgoing}}\frac{\e^{-\iho m_e l_e \beta_e}}{\sqrt{2l_e}}\,.
\een
The sum over the orientations and indices of the leaves in eq.~\eqref{amplitude} thus reduces
to the sum over all graphs with $|a^+|$ incoming leaves carrying indices from $a^+$ and
$|a^-|$ outgoing leaves carrying indices from $a^-$. This sum is obviously finite.\\
\\
Thus, the only source of divergent behavior are the
integrals over the $\beta_v, t_v$. By our estimates above, these are bounded from above by
(a constant times) the integral of the form
\ben
I_G = \prod_{{\rm vertices} \, v \, {\rm in} \, G} \left( \int_0^{2\pi} \d \beta_v \int_0^{2\pi} \d t_v \right)
\prod_{e \in G} \ln|t_e \pm \beta_e-\pi| \, .
\een
Such integrals are standard ``loop integrals'' in quantum field theory (in position space), and
they are well-known to be finite. Hence we conclude that our formula for $Y_i(\varphi, x)$ gives
a finite answer, despite the infinite sums.\\

\subsubsection{The case $D>2$}
%%%%%%%%% HO modified 29/05/09
\label{sec:4.2.2}
Let us now consider the field theory with interaction $\lambda \, P(\varphi) =
\lambda \,  \frac{1}{4!} \varphi^4$, in dimension $D>2$. The main difference to $D=2$ is that
eq.~\eqref{simplerecursion}, and hence eq.~\eqref{itsimple}, no longer holds, as the ``counterterms''
in eq.~\eqref{recursion} now do not have a vanishing limit as $|x-y| \to 0$. This is closely related to the
fact that, for $D=2$, we were dealing with the special case of a super-renormalizable theory whose divergences
are of a particularly benevolent nature. In dimensions $D>2$, we must work with a version of eq.~\eqref{recursion} that also incorporates the counterterms, and this will lead to correspondingly more complicated graphical rules.\\

Our aim is to express the vertex operator $Y_{i}(\varphi,x)$ as a function of the
already known $0$-th order vertex operators, and then find a graphical representation for this
expression.\\
We list the equations that we need to decompose $Y_{i}(\varphi,x)$ into 0-th order operators.
Eq. \eqref{mastereq} reads
\ben
\label{invfield}
Y_{i+1}(\varphi,y)=\frac{1}{3!}\,G\,Y_{i}(\varphi^3,y)\,.
\een
Moreover, we need the equations \eqref{recursion} and
\ben
\label{recursion2}
Y_i(\varphi^3, y) = \sum_{j=0}^i Y_j(\varphi, x) Y_{i-j}(\varphi^2, y) -
\sum_{j=1}^i Y_{i-j}(Y_j(\varphi, x-y)\varphi^2, y) -\frac{1}{|x-y|^{D-2}}Y_i(\varphi,y)+ \dots \, ,
\een
where the dots stand again for terms vanishing in the limit $x-y\rightarrow 0$.\\
We do not perform the limit $x-y\rightarrow 0$ for the moment, so
each time we use either of the eqs.~\eqref{recursion} or \eqref{recursion2}, we have
to introduce a new variable from $\mathbb{R}^D$. We can choose this new variable ($x$ above) to be
$(1+\epsilon)$ times the old variable ($y$ above, i.e. $x=(1+\epsilon)y$),
where a new regulator $\epsilon_v>0$ has to be introduced each time we apply
either of eqs.~\eqref{recursion} or \eqref{recursion2}. \\
The result of repeatedly applying eqs.~\eqref{invfield}, \eqref{recursion} and \eqref{recursion2}
is a sum of products of nested $0$-th order vertex operators whose arguments from $\mathbb{R}^D$ depend
on the initial variable from $\mathbb{R}^D$ and the $\epsilon_v$'s.
We now focus on a special partial sum,
the sum of ``tree-like'' summands. We call a summand tree-like if, when tracing back its path through
the iteration process, at each iteration step it does neither belong to the counterterms nor to the
smooth terms represented by dots.
Another way to put this is to say that we are only interested in the terms that we would have obtained
by dropping the counterterms and dots in eqs.~\eqref{recursion} and \eqref{recursion2}.

A formal definition
of the tree-like summands can be given as follows. Let $T$ be a tree such that all vertices have
coordination number four. With each vertex or leaf $v$ of the
tree associate a regulator $\epsilon_v$. If we have three such trees $T_1, T_2, T_3$ with number
vertices $i_1, i_2, i_3$, then we can join
these at their root to form a new tree, $T = \cup_w T_w$ with $i$ vertices. The three vertices now attached to the new root
(corresponding to the $3$ subtrees $T_1, T_2, T_3$) carry new regulators denoted e.g. $\epsilon_1, \epsilon_2, \epsilon_3$.
The recursive definition of the tree-like contributions to the $i$-th order vertex operators is then
\ben
Y_i(T, \varphi, x) = \frac{1}{3!} \sum_{\substack{i_1+i_2+i_3=i \\ \cup_w T_w = T}} \,\, \prod_{w=1}^3 Y_{i_w}(T_w, \varphi, x_w) \, ,
\een
where $x_w = (1+\epsilon_w)x$, and where the dependency of the quantities $Y_i(T, \varphi, x)$ etc. on the
regulators $\{\epsilon_v\}$ has been suppressed to lighten the notation.
The vertex operators $Y_i(\varphi, x)$ are given by a sum of these tree-like terms [see eq.~\eqref{yit}], plus counterterms. Moreover,
it will turn out that these tree-like summands can be thought of as building blocks for the (more complicated) counterterms.
We will not discuss these here, but outline their construction in the next section.  The regulators $\{\epsilon_v\}$
 cannot be sent to zero before taking the sum of all contributions to the vertex operator,
including the counterterms.\\

We would now like to find a closed form expression
for the tree-like terms, applying the same kind of reasoning as for $D=2$. We have
to take into account some differences. The first difference is that the trigonometric functions in
the expression for the free vertex operator are now replaced by the harmonic polynomials, see eq.~\eqref{harmnorm}.
The analogue of relation~\eqref{orthogonal} is
\bena\label{orthogonal1}
&& \prod_{j\, {\rm outgoing}} h_{l_{j},m_j}(\hat x)  \prod_{j \, {\rm incoming}} \overline h_{l_{j},m_j}(\hat x) \\
&=& \sum_{l=0}^\infty \frac{2l+D-2}{\sigma_D} \int_{S^{D-1}} \d\Omega(\hat y) \, \P(D,l, \hat x \cdot \hat y)
\, \prod_{j\, {\rm outgoing}} h_{l_{j},m_j}(\hat y)  \prod_{j \, {\rm incoming}} \overline h_{l_j,m_j}(\hat y) \, \, ,\non
\eena
using this time the orthogonality of the harmonic polynomials, as well as the formula for the
Legendre polynomials $\P(z,l,D)$ in $D$ dimensions given in appendix~\ref{app:A}.
We proceed as in the case of $D=2$ dimensions.
Let us now consider a vertex $v$ that has no further vertices attached to it downwards in the tree.
Let the leaves attached to $v$ be
labeled by $j$. We collect corresponding factors of $r$, and the harmonic polynomials. The harmonic polynomials
are multiplied by the formula just given, while the factors of $r$ work out as $r^{{\tilde\nu_v}}$,
where ${\tilde\nu_v}$ is now given by
\ben
\label{nuv}
{\tilde\nu_v} = \sum_{j \, {\rm incoming}} l_j  - \sum_{j \, {\rm outgoing}} (l_{j}+D-2) \,.
\een
Thus, in total we get a factor of
$(2 l + D -2)r^{\tilde\nu_v} \, \P( \hat x \cdot \hat y,l,D)$. We have to apply the
inverse of the Laplacian to this expression using the residue trick, introducing
$\delta_v$ and $\nu_v={\tilde\nu_v}+\delta_v$. When we do this,
we get a contour integral with integrand
\ben
\frac{(2 l + D -2) \P(\hat x \cdot \hat y, l,D)r^{\nu+2}}{ \nu_v(\nu_v+D-2) - l(l+D-2)}
\, .
\een
The sum over $l$ is now readily performed\footnote{This key observation is due to~\cite{Holland}.} using
the {\em generalized Dougall's identity}, see appendix~\ref{app:C}, which states that
for any $\nu \in \mc \setminus \mz$ and $-1 \le z \le +1$ and $D \ge 3$, we have the identity
\ben
\label{dougalltrick}
\sum_{l=0}^\infty
\frac{(2l+D-2) \,  \P(z; l, D)}{\nu(\nu+D-2) - l(l+D-2)}
= \frac{\pi}{\sin \pi \nu} \, \P(-z, \nu, D) \, .
\een

From the vertices connected to the leaves, we then work our way upwards, repeating for
each new vertex the same procedure. We will then end up with a similar set of graphical rules.
The main difference to $D=2$ is that we have to use Legendre functions instead of the trigonometric functions,
and  that we must take into account the
regulators $\epsilon_v$. As one can see, this means that we have to introduce an extra factor
\bena
&&\left((1+\epsilon_j)\prod_{\substack{{\rm vertices}\, v\succeq j}}(1+\epsilon_v)\right)^{l_j}\non\\
&\text{or } &\left((1+\epsilon_j)\prod_{\substack{{\rm vertices}\, v\succeq j}}(1+\epsilon_v)\right)^{-l_j-D+2}
\eena
for each leaf $j$, depending on whether it is incoming or outgoing, to get the correct rules for $D>2$.

\medskip
\noindent
To summarize, we have the following {\em graphical rules} for calculating $Y_i(T, \varphi, x)$,
the regularized contribution of a tree $T$ to the vertex operator $Y_i(\varphi,x)$ for the theory
with interaction $\lambda P(\varphi)=\lambda \varphi^4/4!$ in $D>2$ (a general interaction polynomial
is completely analogous):
\begin{enumerate}
\item[1')]
Draw all trees with $i$ vertices with incidence number $4$. Label the vertices by an index $v$
and the lines by pairs of indices $(vw)$. The leaves also carry indices.

\item[2')] With each vertex $v$ associate a parameter $\delta_v \in \mc \setminus \mz$,
a parameter $\hat y_v \in S^{D-1}$ and a regulator $\epsilon_v>0$.

\item[3')] With each leaf $j$ adjacent to a vertex $v$, associate a pair
$(l_{j}, m_{}) \in {\mathbb N} \times \{1,...,N(D,l) \}$ and one of the following factors
\bena
           & \frac{K_D}{\sqrt{\omega(D,l_j)}}\, h_{l_j,m_j}(\hat{y}_v) \,\a^{+}_{l_j,m_j}\non\\
\text{or } & \frac{K_D}{\sqrt{\omega(D,l_j)}}\, \overline{h_{l_j,m_j}}(\hat{y}_v) \, \a_{l_j,m_j}
\label{0summand}
\eena
The first factor is chosen if the line associated with the leaf is oriented upwards,
and the second if it is oriented downwards. The creation/annihilation operators of the leaves
connected to the same vertex are to be normal ordered.
Also, write down the regularizing factor
\bena
&\left((1+\epsilon_j)\prod_{\substack{{\rm vertices}\, v\succeq j}}(1+\epsilon_v)\right)^{l_j}\non\\
\text{or } &\left((1+\epsilon_j)\prod_{\substack{{\rm vertices}\, v\succeq j}}(1+\epsilon_v)\right)^{-l_j-D+2}
\eena

\item[4')] With each vertex $v$ we associate $\nu_{v} \in \mc \setminus \mz$ defined by
\ben
\nu_{v} = \sum_{{\rm in \, leaves} \, j \prec v} l_j - \sum_{{\rm out \, leaves} \, j \prec v} (l_j + D-2) +
\sum_{{\rm vertices} \, w \preceq v} (2+\delta_w) \, .
\een
The ``2'' results from the inversion of the Laplacian, which
at each inversion step (i.e., each vertex) raises the power of the radial coordinate by 2. The
$\delta_w$ arises from the residue trick for the Laplace inversion at each vertex $w$ below $v$.

\item[5')] With the root, associate the parameter $\hat x \in S^{D-1}$, and
the factor $r^{\nu_{\rm root}}$, where $x=r\hat x$. The number $\nu_{\rm root} \in \mc$
is defined as in eq.~\eqref{nukdef}, but with the vertex $v$ replaced by the root, so that the
sums contain contributions from all leaves of $T$.

\item[6')] With each line $(vw)$ connecting vertices $v,w$ associate a factor
$$\frac{\pi}{\sin \pi \nu_{w}} \, \P(-\hat y_v \cdot \hat y_w; \nu_{w}, D).$$ This results from the application of
the Dougall formula.

%\item[7')]
%\ben
%\label{regular}
%\left(\prod_{\substack{{\rm vertices\, and}\\ {\rm leaves}\, w\succeq v}}(1+\epsilon_w)\right)^{\nu_v}\,.
%\een

\item[7')] Perform the sum over all $l_j, m_j$. Furthermore, perform  the integrals
$$\prod_{{\rm vertices} \, v} \int_{S^{D-1}} \d \Omega(\hat y_v)
\quad \text{and} \quad
\prod_{{\rm vertices} \, v} \frac{1}{2\pi \iho}  \oint_{C_v} \frac{\d \delta_v}{\delta_v}\, . $$
Finally, take the sum over all possible orientations of the leaves.
\end{enumerate}

If we proceed as in the case $D=2$ and perform the sum of all tree-like terms as in
eqs.~\eqref{yig},~\eqref{yit} we do not get the complete vertex operator
$Y_i(\varphi,x)$, so in this sense our rules are not complete.
To get the complete vertex operator, we should also incorporate the counterterms in
$D>2$ (see next section). The sum over graphs~\eqref{yig}
depends on the regulators $\epsilon_v$, and is divergent for
$\epsilon_v\rightarrow 0$. The expectation is that these divergences are canceled by counterterms.
We will briefly expose this idea in the next subsection, without explaining it in full.

Let us write down explicitly
the contributions $Y_i(G, \varphi, x)$ from loop graphs $G$ as in eq.~\eqref{yig}. In $D$ dimensions,
the formula is
\bena
Y_i(G, \varphi, x) &=& \sum_{\substack{{\rm leaf}\\{\rm orientations}}} \sum_{l_e,m_e, \,e \,{\rm leaf}}\,
\sum_{l_e,\, e \in G \setminus T }
\left(\prod_{{\rm vertices}\, v}\frac{1}{2\pi\iho} \int_{C_v} \frac{\d \delta_v}{\delta_v} \int_{S^{D-1}} \d \Omega(\hat y_v) \right) \label{amplitude1}\\
&\times&
\prod_{e \in T} \frac{\pi}{\sin\pi(l_e+\delta_e)} \, \P(- \hat y_v \cdot \hat y_w, l_e+\delta_e, D) \,
%\left(
%\begin{matrix}
%\nu_{ij}+D-3\\
%\nu_{ij}
%\end{matrix}
%\right)
\prod_{e \in G \setminus T}
%\left(
%\begin{matrix}
%h_{ij}+D-3\\
%h_{ij}
%\end{matrix}
%\right)
\e^{-\theta_e l_e}\P(\hat y_v \cdot \hat y_w, l_e, D) \non\\
&\times& \exp \left( \ln r \left\{ \sum_{{\rm in \, leaves} \, e} l_e - \sum_{{\rm out \, leaves} \, e} (l_e+D-2) + \sum_{{\rm vertices} \, v} (2+\delta_v) \right\} \right) \non\\
&\times& \prod_{{\rm in \, leaves}\, e}\left(\prod_{v\succ e}(1+\epsilon_v)\right)^{l_e}
\prod_{{\rm out \, leaves}\, e}\left(\prod_{ v\succ e}(1+\epsilon_v)\right)^{-l_e-D+2}\non\\
&\times&: \prod_{{\rm leaves}\,e} \sqrt{\frac{D-2}{2l_e+D-2}}   h_{l_e, m_e}(\hat y_e) \a_{l_e,m_e}^\pm : \, . \non
\eena
In the last line, we have set $\hat y_e=\hat y_v$ for an uncontracted leaf $e$ associated to some vertex $v$.
The $\theta_e $ are defined as follows:
If $v$ is a vertex, let
\ben
\theta_v = \ln \prod_{w: w \succeq v} (1+\epsilon_w) \, ,
\een
and if $e = (vw) \in G \setminus T$, let $\theta_e = \theta_v - \theta_w$.
%for edges $e = (vw) \in G \setminus T$ (i.e. the loop edges) and are given by
%$\theta_v = \ln \prod_{w: w \preceq v} (1+ \epsilon_v)$.
The regulators are always chosen so that
$\theta_e >0$, but evidently, $\theta_e \to 0$ as the regulators are removed.

%As before, $e=(v w )$ denotes a line of the graph $G$ connecting the vertices $v$ and $w$.\\

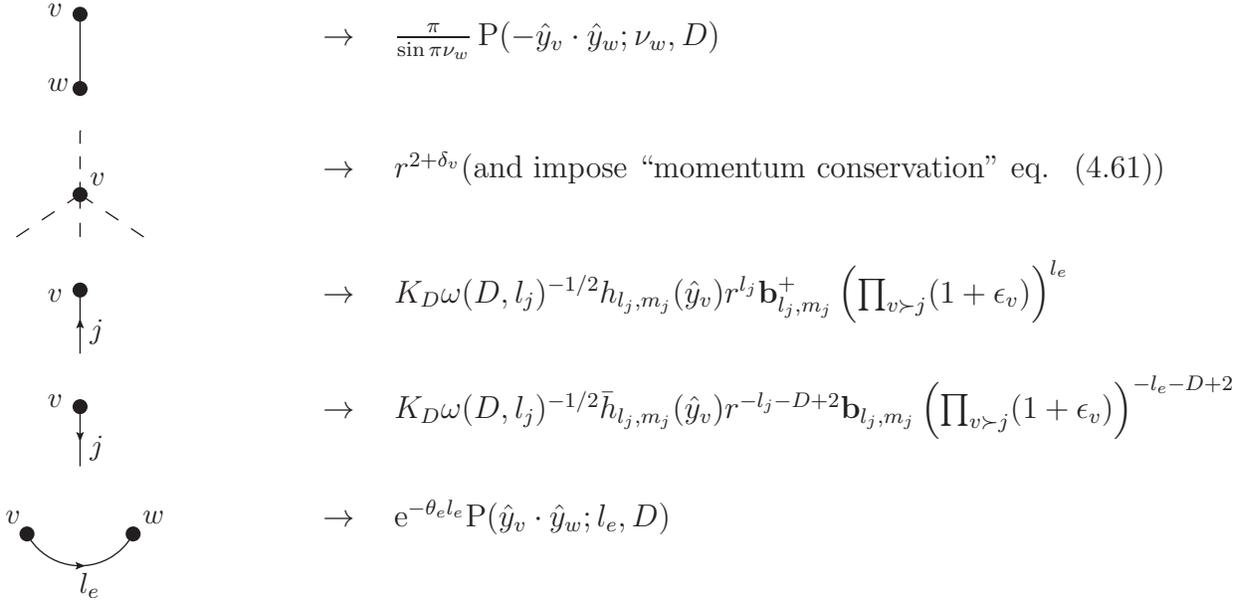
\begin{figure}[htp]
\vspace{10mm}
\unitlength=0.5pt
\SetScale{0.5}

%\hspace{1cm}
%\begin{center}
\fcolorbox{white}{white}{
  \begin{picture}(272,456) (59,-19)
    \SetWidth{1.0}
    \SetColor{Black}
    \Line(112,360)(112,417)
    \SetWidth{0.0}
    \Vertex(112,280){5.657}
    \SetWidth{1.0}
    \Line[arrow,arrowpos=0.5,arrowlength=5,arrowwidth=2,arrowinset=0.2](112,160)(112,205)
    \Line[arrow,arrowpos=0.5,arrowlength=5,arrowwidth=2,arrowinset=0.2](112,120)(112,75)
    \Text(120,168)[lb]{\small{\Black{$j$}}}
    \Text(120,80)[lb]{\small{\Black{$j$}}}
    \Text(296,384)[lb]{\normalsize{\Black{$\rightarrow\quad \frac{\pi}{\sin \pi \nu_{w}} \, \P(-\hat y_v \cdot \hat y_w; \nu_{w}, D)$}}}
    \Text(296,288)[lb]{\normalsize{\Black{$\rightarrow\quad r^{2+\delta_v} \text{(and impose ``momentum conservation'' eq.~ \eqref{nukdef})}$}}}
    \Text(296,184)[lb]{\normalsize{\Black{$\rightarrow\quad K_D\omega(D,l_j)^{-1/2} h_{l_j,m_j}(\hat{y}_v) r^{l_j} \a_{l_j,m_j}^{+}\left(\prod_{ v\succ j}(1+\epsilon_v)\right)^{l_e}$}}}
    \Text(296,96)[lb]{\normalsize{\Black{$\rightarrow\quad  K_D\omega(D,l_j)^{-1/2}{ \bar h_{l_j,m_j}(\hat{y}_v)} r^{-l_j-D+2} \a_{l_j,m_j}\left(\prod_{ v\succ j}(1+\epsilon_v)\right)^{-l_e-D+2}$}}}
    \Text(120,288)[lb]{\small{\Black{$v$}}}
    \Line[dash,dashsize=10](112,328)(112,280)
    \Line[dash,dashsize=10](112,280)(64,248)
    \Line[dash,dashsize=10](112,280)(112,248)
    \Line[dash,dashsize=10](112,280)(160,248)
    \SetWidth{0.0}
    \Vertex(72,24){5.657}
    \Vertex(152,24){5.657}
    \SetWidth{1.0}
    \Arc[arrow,arrowpos=0.5,arrowlength=5,arrowwidth=2,arrowinset=0.2](112,45.333)(45.333,-151.928,-28.072)
    \Text(56,32)[lb]{\small{\Black{$v$}}}
    \Text(160,32)[lb]{\small{\Black{$w$}}}
    \Text(296,24)[lb]{\normalsize{\Black{$\rightarrow\quad \e^{-\theta_e l_e}\P(\hat y_v \cdot \hat y_w; l_e, D)$}}}
    \Text(112,-24)[lb]{\small{\Black{$l_e$}}}
    \Text(88,120)[lb]{\small{\Black{$v$}}}
    \SetWidth{0.0}
    \Vertex(112,120){5.657}
    \Vertex(112,208){5.657}
    \Text(88,200)[lb]{\small{\Black{$v$}}}
    \Vertex(112,360){5.657}
    \Vertex(112,416){5.657}
    \Text(88,416)[lb]{\small{\Black{$v$}}}
    \Text(88,360)[lb]{\small{\Black{$w$}}}
  \end{picture}
}
%\end{center}
\caption{The rules for  the ``amplitude'' of a graph in dimension $D\geq 3$.
The appropriate summations and integrals have to be
understood. As in fig.~\ref{fig:2D_rules},
we associate the powers of $r$ with the leaves and vertices instead of
with the root.}
\label{fig:3D_rules}
\end{figure}

It is possible to make the source of divergences more transparent by introducing the same trick as
in $D=2$ to replace the summations over the angular momentum numbers $l_e$ by integrals.
%From now on, we will neglect
%the regularizing factors \eqref{regular}; this will lead to divergent expressions.
%We want to investigate the nature of these divergences.\\
At each vertex, we have the momentum conservation rule eq.~\eqref{lconservation}. As in $D=2$,
we can realize this rule by introducing an integral similar to~\eqref{constrick}
at each such vertex. We can then perform the summations over the $l_e$ as we have done in eq.~\eqref{g2def}.
The overall result of these manipulations is the following. Instead of the sum/integral
in eq.~\eqref{amplitude1}, we now have
\ben
\sum_{\substack{{\rm leaf}\\{\rm orientations}}} \sum_{l_e,m_e, \,e \,{\rm leaf}}\left( \prod_{{\rm vertices} \, v} \int_{C_v} \frac{\d \delta_v}{\delta_v} \int_{S^1 \times S^{D-1}} \e^{2 \iho t_v} \, \d t_v \wedge \d \Omega(\hat y_v) \right) \, ,
\een
i.e., we have gotten rid of the summation over the loop momenta, and replaced
these by additional integrations over the parameters $t_v$.
The term in the first product $(e \in T)$ in the second line of eq.~\eqref{amplitude1} becomes,
with $\beta_e := \arccos(\hat y_v \cdot \hat y_w)$ for the line $e = (vw)$ connecting the vertices $v$ and $w$:
\bena
\label{gDsum}
g_D(\delta_{e}, \cos \beta_e, t_e) &:=&
\frac{\pi}{\sin\pi\delta_e}\Big(\sum_{l=0}^\infty (-1)^{l}\e^{\iho( t_e+\i0)l}\P(\cos \beta_e,l+\delta_e,D)\non\\
&&\,\,\,\,\,\,\,\,\,+ \,\,\,\,\, \sum_{l=-\infty}^{-1} (-1)^{l}\e^{\iho( t_e-\i0)l}\P(\cos \beta_e,l+\delta_e,D)\Big) \, .
\eena
Putting together eq.~\eqref{g2def} and eq.~\eqref{diffP}, we get for even $D$
\bena\label{dangerous}
&&g_D(\delta_{e}, \cos \beta_e, t_e) = \frac{\e^{-\iho t_e(D/2-1)}}{\Gamma(D/2) \, \sin \pi\delta_e}
\left(\frac{\partial}{2\sin \beta_e \, \partial \beta_e}\right)^{\frac{D-2}{2}} \\
&\times& \Bigg(
\e^{+\iho \delta_{e} \beta_e} \, f(\delta_e, 1+\e^{\iho (+\beta_e + t_e + \iho 0)}) +
\e^{+\iho \delta_{e} \beta_e} \, f(-\delta_e, 1+\e^{\iho( -\beta_e - t_e + \iho 0)})  \non \\
&+&
\,\,\,\, \e^{-\iho \delta_e \beta_e} \,   f(\delta_e, 1+ \e^{\iho (-\beta_e + t_e + \iho 0)}) +
\e^{-\iho \delta_e \beta_e} \,   f(-\delta_e,1+\e^{\iho( +\beta_e - t_e + \iho 0)})  -\delta_e^{-1} \cos (\delta_e \beta_e) \Bigg) \non \, .
\eena
Here, $f$ is the transcendental function given by
\ben\label{fdef1}
f(\delta,x) = \sum_{n=0}^\infty \frac{(\delta)_n}{n!} [\psi(n+1) - \psi(\delta+n) - \ln x] \, x^n \, ,
\een
also see eq.~\eqref{fdef}.
For odd $D$, there is a similar formula obtained with the aid of
theorem~\ref{lem2} of appendix~\ref{app:C}.

The term in the second product $(e \in G\setminus T)$ in eq.~\eqref{amplitude1} is affected by the regulators.
%If $v$ is a vertex, let
%\ben
%\theta_v = \ln \prod_{w: w \succeq v} (1+\epsilon_w) \, ,
%\een
%and if $e = (vw) \in G \setminus T$, let $\theta_e = \theta_v - \theta_w$. The choice of the regulators is
%always made such that $\theta_e > 0$.
For these edges, we must replace the term in the second product over
$(e \in G\setminus T)$ by the expression
\ben
\sum_{l_e \in \mn} \e^{\iho l_e(t_e + \iho \theta_e)} \P(\cos \beta_e, l_e, D) =
\left[(\e^{\iho \beta_e} -\e^{\iho( t_e+\iho \theta_e)})(\e^{-\iho \beta_e}-\e^{\iho( t_e+\iho \theta_e)})\right]^{-(D-2)/2}\,.
\een
There is also an additional factor of $\prod_{v:(vw)=e \in G \setminus T} \e^{\iho t_v(D-2)}$ arising from the
fact that the momentum conservation rule in $D$ dimensions is slightly different.
The remaining parts of the formula~\eqref{amplitude1} are unchanged. The total effect of these manipulations to
eq.~\eqref{amplitude1} is reflected in eq.~\eqref{amplitude2} below.

The divergence that appears when we set $\epsilon_v = 0$ in our graphical rules for the tree-like terms
$Y_i(T, \varphi, x)$ now manifests itself as a pole of the
transcendental function $f(\delta, x)$ at $x=0$, which corresponds to $\beta_e = \pm t_e - \pi$ in the above formula~\eqref{dangerous}. More precisely, using the expansion of the function $f$ above, we can infer that the divergent behavior of
$g_D$ are sums of terms of the form $\sim (t_e \pm \beta_e -\pi \pm \iho 0)^{2-D}$. It can be seen from this that the
most divergent part of the integrals that we have to consider is now given by an expression of the form
\ben
I_G = \prod_{{\rm vertices} \, v \, {\rm in} \, G} \left( \int_{S^{D-1} \times S^1} \d \Omega_v \d t_v \right)
\prod_{e \in T} |t_e \pm \beta_e-\pi|^{2-D} \prod_{e \in G \setminus T} |t_e + \iho \theta_e \pm \beta_e |^{2-D} \, .
\een
In $D>3$, integrals of this type are no longer convergent for $\epsilon_v = 0$ (i.e. $\theta_e = 0$), but
the divergences are very similar in nature to the
divergences found in ordinary Feynman integrals in $x$-space. (But note that the
``$x$'' space has become $S^{D-1} \times S^1$, and that the $\iho 0$ prescriptions are different).
In the usual approach to perturbation theory via Feynman integrals, these
divergences have to be ``renormalized'' by hand. But in our approach the divergences are expected to cancel automatically
when we include all contributions to $Y_i(\varphi, x)$ including the tree like terms {\em and} the counter terms before
taking the regulators $\epsilon_v$ to zero.
We now explain this in a little more detail.

%%%%%%%% HO 29/05/09

\subsection{Renormalization}
\label{renormalization}
In conventional perturbative quantum field theory, renormalization is necessary
to make the terms in the perturbation series well-defined. In our approach the situation is somewhat different,
because the consistency condition tells us in principle right from the start how to obtain well-defined perturbations
of arbitrary order. We think this is a remarkable feature of the present framework. Nevertheless, we have already
borrowed some vocabulary from renormalization such as  ``counterterms'', and the reason for this is
that we also need to perform various limits in our approach which are quite reminiscent of certain operations in
conventional renormalization theory. In fact, the inclusion of the counterterms
into the rules may be thought of as the ``renormalization'' of the tree-like contributions that we have
treated in the last subsection and that diverge when the regulators are sent to $0$. The counterterms cure
these divergences.

To state the ``complete'' set of rules is in principle straightforward, but generates a rather heavy notation.
To keep the discussion reasonably transparent, we therefore only sketch the basic procedure,
and leave a more detailed discussion to a future paper, where we will also outline the relation to
Hopf-algebras similar to those in~\cite{kreimer}.

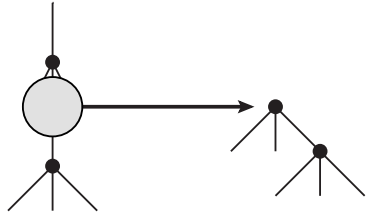
\begin{figure}[htp]
\vspace{10mm}
\unitlength=0.7pt
\SetScale{0.7}

\begin{center}
\fcolorbox{white}{white}{
  \begin{picture}(194,114) (231,-31)
    \SetWidth{1.0}
    \SetColor{Black}
    \Line(256,82)(256,50)
    \Line(256,50)(248,34)
    \Line(256,50)(256,34)
    \Line(256,50)(264,34)
    \Line(256,-6)(232,-30)
    \Line(256,-6)(256,-30)
    \Line(256,-6)(280,-30)
    \SetWidth{2.0}
    \Line[arrow,arrowpos=1,arrowlength=6.667,arrowwidth=2.667,arrowinset=0.2](264,26)(360,26)
    \SetWidth{1.0}
    \Line(376,26)(376,2)
    \Line(376,26)(352,2)
    \Line(376,26)(400,2)
    \Line(400,2)(424,-22)
    \Line(400,2)(376,-22)
    \Line(400,2)(400,-22)
    \Vertex(256,50){4}
    \Vertex(256,-6){4}
    \Vertex(376,26){4}
    \Vertex(400,2){4}
    \Line(256,18)(256,-6)
    \GOval(256,26)(16,16)(0){0.882}

  \end{picture}
}
\end{center}\caption{A decorated tree making a contribution to the vertex operator $Y_4(\varphi,x)$.
More precisely, this tree makes a contribution to
$G\left(\langle \varphi|Y_2(\varphi,\epsilon x)|\varphi^2\rangle Y_1(\varphi,x)\right)$; the tree pointed at by the arrow is a summand to $Y_2(\varphi,\epsilon x)$ and
the tree below the ``blob'' is a summand to $Y_1(\varphi,x)$.}
\label{fig:dec_tree_ex}
\end{figure}

Let us go back to the start of the recursion procedure for evaluating $Y_i(\varphi,x)$.
By eq.~\eqref{invfield},
\ben
Y_i(\varphi,x)=\frac{1}{3!} \, G \, Y_{i-1}(\varphi^3,x)\,.
\een
By eq.~\eqref{recursion2},
\bena
Y_{i-1}(\varphi^3,x)&=&\sum_{j=0}^{i-1}Y_j(\varphi,(1+\epsilon)x)Y_{i-1-j}(\varphi^2, x)\non\\
&-&\sum_{j=1}^{i-1} Y_{i-1-j}(Y_j(\varphi, \epsilon x)\varphi^2, x) -\frac{1}{|\epsilon x|^{D-2}}Y_{i-1}(\varphi,x)+...
\eena
where dots stand for terms vanishing for $\epsilon \rightarrow 0$. The counterterms (the terms with the $-$-sign in front)
can be rewritten as
\bena
\label{counterterms}
&&\sum_{j=1}^{i-1} Y_{i-1-j}(Y_j(\varphi, \epsilon x)\varphi^2, x) +\frac{1}{|\epsilon x|^{D-2}}Y_{i-1}(\varphi,x)\non\\
&=&\sum_{j=1}^{i-1} \quad \sum_{\text{dim}(c)\leq 3(D-2)/2}\langle c,Y_j(\varphi,\epsilon x)\varphi^2\rangle
Y_{i-1-j}(c,x)+\frac{1}{|\epsilon x|^{D-2}}Y_{i-1}(\varphi,x)+...
\eena
where again, dots stand for terms vanishing for $\epsilon \rightarrow 0$.
None of the terms in eq.~\eqref{counterterms} will give rise to tree-like summands
in the final formula for $Y_i(\varphi,x)$.
However, we can now apply the ``incomplete'' rules from the last subsection to the operators
$Y_j(\varphi,\epsilon x),\,Y_{i-1-j}(c,x)$ and $Y_{i-1}(\varphi,x)$ appearing in eq.~\eqref{counterterms},
and thus obtain more terms
that contribute to $Y_i(\varphi,x)$. Graphically, we represent this by a ``decorated tree''
where the trees contributing to the matrix element
$\langle c|Y_j(\varphi,\epsilon x)|\varphi^2\rangle$
are the decorations of a special vertex or blob, to which in turn the roots of the trees contributing
to $Y_{i-1-j}(c,x)$ are attached\footnote{The term
$|\epsilon x|^{2-D}Y_{i-1}(\varphi,x)$ would be represented in the same way by a blob decorated with
the matrix element $|\epsilon x|^{2-D}=\langle\varphi|Y_0(\epsilon x)|\varphi^2\rangle$, to which a tree
contributing to $Y_{i-1}(\varphi,x)$ is attached.}, see fig.~\ref{fig:dec_tree_ex}.

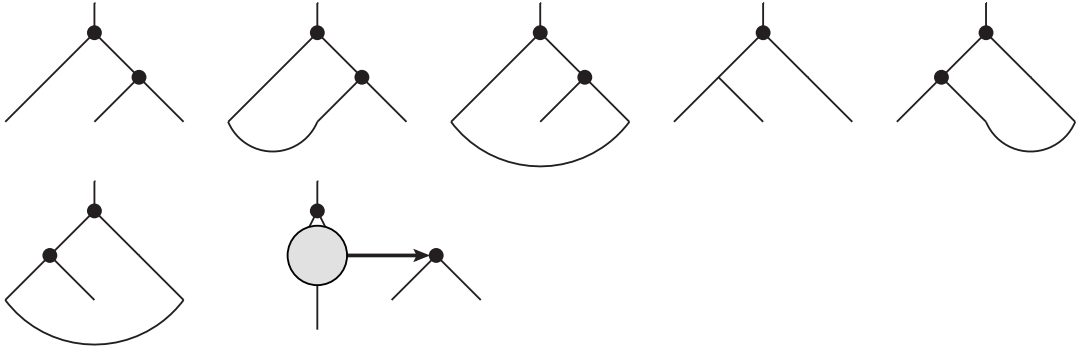
\begin{figure}[htp]
\vspace{10mm}
\unitlength=0.7pt
\SetScale{0.7}

\begin{center}
\fcolorbox{white}{white}{
  \begin{picture}(578,186) (47,-31)
    \SetWidth{1.0}
    \SetColor{Black}
    \Line(96,154)(96,138)
    \Line(96,138)(48,90)
    \Line(96,138)(120,114)
    \Line(120,114)(144,90)
    \Line(120,114)(96,90)
    \Line(216,138)(168,90)
    \Line(216,154)(216,138)
    \Line(240,114)(216,90)
    \Line(216,138)(240,114)
    \Line(240,114)(264,90)
    \Line(336,138)(288,90)
    \Line(336,154)(336,138)
    \Line(336,138)(360,114)
    \Line(360,114)(336,90)
    \Line(360,114)(384,90)
    \Arc(192,100)(26,-157.38,-22.62)
    \Arc(336,126)(60,-143.13,-36.87)
    \Line(456,138)(408,90)
    \Line(432,114)(456,90)
    \Line(456,138)(480,114)
    \Line(480,114)(504,90)
    \Line(456,154)(456,138)
    \Line(552,114)(576,90)
    \Line(576,138)(528,90)
    \Line(576,138)(600,114)
    \Line(600,114)(624,90)
    \Line(576,154)(576,138)
    \Arc(600,100)(26,-157.38,-22.62)
    \Arc(96,30)(60,-143.13,-36.87)
    \Line(96,42)(48,-6)
    \Line(96,58)(96,42)
    \Line(72,18)(96,-6)
    \Line(96,42)(120,18)
    \Line(120,18)(144,-6)
    \Line(216,58)(216,42)
    \Line(216,42)(208,26)
    \Line(216,42)(224,26)
    \Line(216,18)(216,-22)
    \SetWidth{2.0}
    \Line[arrow,arrowpos=1,arrowlength=6.667,arrowwidth=2.667,arrowinset=0.2](224,18)(272,18)
    \SetWidth{1.0}
    \Line(280,18)(304,-6)
    \Line(280,18)(256,-6)
    \Vertex(216,42){4}
    \Vertex(280,18){4}
    \Vertex(72,18){4}
    \Vertex(96,42){4}
    \Vertex(96,138){4}
    \Vertex(120,114){4}
    \Vertex(216,138){4}
    \Vertex(240,114){4}
    \Vertex(360,114){4}
    \Vertex(336,138){4}
    \Vertex(456,138){4}
    \Vertex(576,138){4}
    \Vertex(552,114){4}
    \GOval(216,18)(16,16)(0){0.882}
  \end{picture}
}
\end{center}
\caption{All trees contributing to the operator $Y_2(\varphi,x)$ for $D=4$, $P(\varphi)=\varphi^3/3!$.
In the decorated tree, the tree pointed at by the arrow is contracted with the vectors
$\langle\varphi|,|\varphi\rangle$.}
\label{fig:Y_2}
\end{figure}

Of course we still do not have identified \emph{all} the terms that make a contribution to $Y_i(\varphi,x)$.
We would have to take into account the counterterms at each recursion step in the same manner as above.
But we see that if we do so, we get all summands
in the final regularized formula for $Y_i(\varphi,x)$  represented by (multiply) decorated trees.
To summarize, we have laid out the following idea for calculating a vertex operator $Y_i(\varphi,x)$:
\begin{itemize}
\item Draw a certain set of decorated trees with $i$ vertices.
\item Apply a set of rules similar to those from subsec.~\ref{sec:4.2.2}
to each of these trees, translating them into an ``amplitude'' depending on a
number of regulators $\epsilon_v$.
\item Take the sum of all those amplitudes and take the limits $\epsilon_v\rightarrow 0$
in the appropriate order.
\end{itemize}
The development of rules for drawing
decorated forests mentioned in the first bullet point above and the rules for amplitudes of decorated trees from the
second bullet point is straightforward but cumbersome. They naturally lead to the appearance of a
Hopf algebra structure similar to that found in~\cite{kreimer} for ordinary Feynman diagrams.
We will not state these rules here, leaving this to a future paper. \\
\\
As an example, we consider the operator $Y_2(\varphi,x)$  for the theory with interaction
$\lambda \varphi^3/3!$ in $D=4$.
There are two trees with two 3-valent vertices. We take the sum over all graphs spanned by these trees
and include the one decorated tree that makes a contribution to the sum, see fig.~\ref{fig:Y_2}.

The contribution of the decorated tree is
\bena
&&\sum_{l,m}\left(\frac{1}{2\pi\iho} \right)^2\oint_{C_1}\frac{\d\delta_1}{\delta_1}\oint_{C_2}\frac{\d\delta_2}{\delta_2}
h_{l,m}(\hat x)\omega(4,l)^{-1/2}\a_{l,m}\non\\
&\times&\frac{\epsilon^{\delta_2}}{\delta_2(2+\delta_2)}
\frac{r^{l+\delta_1+\delta_2+2}}{4l+2l(\delta_1+\delta_2)+8+6(\delta_1+\delta_2)+(\delta_1+\delta_2)^2}
\label{Y_2counter}
\eena
Here, we have already performed the (trivial) integrals over the angular variables and we have not
used the trick \eqref{dougalltrick} because the sums over the ``momenta'' of internal lines
are trivial as well.
In the limit $\epsilon\rightarrow 0$, the expression \eqref{Y_2counter} is logarithmically divergent,
as can be seen from evaluating the contour integral in $\delta_2$.
This has to be compared with the divergences in the loop graphs. Consider
the contribution from the second tree in the first line of fig.~\ref{fig:Y_2}:
\bena
&&\sum_{l,m,l_1}\left(\frac{1}{2\pi\iho} \right)^2\oint_{C_1}\frac{\d\delta_1}{\delta_1}\oint_{C_2}\frac{\d\delta_2}{\delta_2}
\int_{S^3}\d\Omega(\hat y_1)\int_{S^3}\d\Omega(\hat y_2)\non\\
&\times& \P(\hat y_1\cdot\hat y_2,l_1,4)
\frac{\pi}{\sin\pi\delta_2}\,\P(\hat y_1\cdot\hat y_2,l-l_1+\delta_2,4)(1+\epsilon)^{l-l_1+\delta_2}\non\\
&\times&\frac{\pi}{\sin\pi(\delta_1+\delta_2)}\,\P(\hat y_1\cdot\hat x,l+2+\delta_1+\delta_2,4)\non\\
&\times& r^{l+2+\delta_1+\delta_2}h_{l,m}(\hat y_2)\omega(4,l)^{-1/2}\a_{l,m}^+
\label{Y_2loop}
\eena
We have only kept track of one regulator $\epsilon$; the other regulators $\epsilon_v$ that one would obtain
by following the graphical rules by the letter give rise to trivial limits $\epsilon_v\rightarrow 0$.\\
\\
The contributions from the other loop graphs are similar. In the loop graphs, the divergences arise
from the loop summations which is the sum over $l_1$ in eq.~\eqref{Y_2loop}. These divergences should be canceled
by the counterterm \eqref{Y_2counter}. In the previous section, we have applied manipulations which
turned the sums into integrals, and the divergences then manifest themselves at the level of the integrals
$I_G$, as we discussed. It is still not an easy matter to see that the counterterms precisely cancel the divergence,
already in this example. See \cite{Holland} for a detailed discussion of examples of that nature. It seems that
a different method is required to make such cancellations manifest in a systematic way.\\
\\
Counterterms also occur in the construction of composite operators. Here, what have been
``tree-like summands'' before become ``forest-like summands''. For an operator $Y_i(a,x)$ these
would have a graphical representation by forests consisting of $|a|$ trees with $i$ vertices overall.
The rules for the contribution to $Y_i(a,x)$ of such a forest are more or less the same as
the rules from section \ref{sec:4.2.2} applied to every tree in the forest. Again, the notation necessary
to state the rules in full, including decorated forests for the counterterms, is quite heavy, and we will
not spell this out here.
%%%%%%%%%%%%%%%%

\section{Vertex algebra and special functions}
\label{sec:4}

In this section, we will give yet another representation of the vertex operator $Y(\varphi, x)$
in the interacting field theory characterized by the interaction
$\lambda \, P(\varphi) = \lambda\sum \frac{c_p}{p!} \varphi^p$. We believe
that this representation is interesting not only because it might be convenient for calculations,
but also because it hints at a deep relation between our vertex algebras and certain special
functions of hypergeometric type.

Our starting point is the representation of $Y(\varphi, x) = \sum_i \lambda^i Y_i(\varphi, x)$ in the
form of eqs.~\eqref{amplitude} and~\eqref{amplitude1}. We further decompose the $i$-th order contribution $Y_i(\varphi, x)$
to the vertex operator into contributions from counterterms (when $D>2$), and from graphs, $G$, as outlined above, and these are
called again $Y_i(G, \varphi, x)$. The main point of this section is to provide an alternative representation for this quantity. Since each $Y_i(G, \varphi, x)$ is an endomorphism in
${\rm End}(V)$, it is sufficient to present a formula for the matrix elements in a basis of $V$.

To present our result in the most economical form, it is convenient to choose a particular
(non-orthogonal) basis of $V$ which is defined as follows. First,
for any $p \in \mr^D, l \in \mn$, we define
$\a_l(p)^+ = \omega(l, D)^{-1/2} \,  \sum  \,  h_{l, m}( p) \, \a_{l, m}^+$.
For $\vec p = (p_1, \dots, p_n) \in \mr^{nD}$ and $\vec l = (l_1, \dots, l_n) \in \mn^n$, we then define
\ben
|\vec p, \vec l \rangle := \prod_{i=1}^n \a_{l_i}(p_i) |0\rangle \, .
\een
We remember that any element in $V$ corresponds to a composite field, i.e., a formal product
of $\varphi$ and its derivatives. For the vectors just defined, this is
\ben
|\vec p, \vec l \rangle = \prod_{i=1}^n  \P(p_i \cdot \partial, l_i, D) \varphi \, .
\een
It is evident from this expression that the vectors $|\vec p, \vec l\rangle$ form an
(overcomplete) basis of $V$. Our aim in this section is to provide an alternative expression for the matrix elements $\langle \vec p_-, \vec l_- | Y_i(G, \varphi, x) | \vec p_+, \vec l_+ \rangle$.

Our starting point is the formula~\eqref{amplitude1} for the $Y_i(G, \varphi, x)$.
The basic idea is to carry out the angular integrals
$\int \d\Omega(\hat y_v)$ first, or rather, to turn these integrals into summations. This is done
by first expanding the Legendre functions using the formula (valid for $\nu \notin \mz$)
\ben
\P(z,\nu,D) = \frac{\sin \pi \nu}{\pi} \, \frac{2^{-(D+1)/2}}{\Gamma(D/2)}
\sum_{n=0}^\infty (-2z)^n \frac{\Gamma(-\nu/2+n/2)\Gamma(\nu/2 + n/2+D/2-1)}{n!}  \, .
\een
which we prove in the appendix~\ref{app:C}. For $\nu = l \in \mz$, we have instead the
well-known formula
\ben
\P(z,l,D) =  \frac{2^{-(D-3)/2}}{\Gamma(D/2)}
\sum_{j=0}^{[l/2]} (-2z)^{l-2j} \frac{\Gamma(l-j+D/2-1)}{j!(l-2j)!}  \, .
\een
If we perform these expansions for all the Legendre functions appearing in eq.~\eqref{amplitude1}, then
we end up with a multiple sum, whose terms contain powers $(\hat y_v \cdot \hat y_w)^{n_e}$,
where $e=(vw)$ is an edge between $v,w$, and where each $n_e$ is the summation counter from the power series expansion of the Legendre functions associated with the edge $e$. To perform the angular integrals, we now further expand each such power using the multinomial formula,
\ben
(\hat y_v \cdot \hat y_w)^{n_e} = \sum_{k_{e,1}+ \dots+ k_{e,D}=n_e}
\frac{n_e!}{\prod_\mu k_{e,\mu}!} \prod_\mu (\hat y_{v,\mu} \hat y_{w,\mu})^{k_{e,\mu}} \, .
\een
Here, and in the following, $\mu$ runs from $1$ to $D$.
After the combined expansions, each term in eq.~\eqref{amplitude1} will now consist of a prefactor times
$\hat y_{v,\mu}$, raised to some power $a_{v,\mu}$. The power is
\ben
a_{v,\mu} = \sum_{e \, {\rm adjacent} \, v} k_{e,\mu} \, ,
\een
where the sum is over all edges $e$ going from the vertex $v$. Thus, the integrals we have to consider
are of the type ($a_i \in \mn$)
\ben
\int_{S^{D-1}} \d \Omega(\hat x) \,  \hat x_{1}^{a_{1}} \cdots \hat x_{D}^{a_D} =
\begin{cases}
2 \frac{\prod_\mu \Gamma\left(\frac{a_\mu+1}{2}\right)}{\Gamma\left(\frac{\sum_\mu a_\mu + D}{2}\right)} & \text{if all $a_i$ even} \, ,
\\
0 & \text{otherwise} \, .
\end{cases}
\label{EulerBeta}
\een
This formula can be viewed as a multi-dimensional generalization of the
standard formula for the Euler Beta-function ($D=2$) and can be proved
e.g. by induction in $D$, expressing $\d\Omega$ in $D$-dimensional polar coordinates. If we combine all the steps we
have described so far, then we end up with the following expression (up to a numerical prefactor) for the vertex operator:
\bena
&&\langle \vec p_-, \vec l_- | Y_i(G, \varphi, x) | \vec p_+, \vec l_+ \rangle =\\
&&
r^{2i} \sum_{\substack{{\rm assignments}\\ \vec l_+,\vec l_-\rightarrow {\rm leaves}}}
\sum_{l_e \in \mathbb{N}: \,  e \in G \setminus T} \,\, \sum_{k_e \in \mn^D: \, e \in T}
\left( \prod_{v \in T} \frac{1}{2\pi \iho}  \int_{C_v} \frac{\d \delta_v}{\delta_v} \right) \non\\
&\times&
\prod_{e \in T}
\frac{\Gamma(-l_e/2-\delta_e/2+|k_{e}|/2) \Gamma(l_e/2+\delta_e+D/2-1+|k_{e}|/2)}{k_{e}!} \,\non\\
&\times& \prod_{v \in T} \frac{\prod_\mu \Gamma((\sum_{e \, {\rm on} \, v} k_{e, \mu}+1)/2)}{\Gamma(
(\sum_{e \, {\rm on} \, v} |k_e|+D)/2)} \,
%\left(
%\begin{matrix}
%\nu_{ij}+D-3\\
%\nu_{ij}
%\end{matrix}
%\right)
\prod_{e \in G \setminus T} \frac{\Gamma(l_e-j_e+D/2-1)}{j_e!(l_e-2j_e)!} \non\\
%\left(
%\begin{matrix}
%h_{ij}+D-3\\
%h_{ij}
%\end{matrir^{}x}
%\right)
&\times&
\prod_{e \, {\rm in}} \frac{\Gamma(l_{+e}-j_e+D/2-1)}{j_e!(l_{+e}-2j_e)!}
\prod_{e \, {\rm out}} \frac{\Gamma(l_{-e}-j_e+D/2-1)}{j_e!(l_{-e}-2j_e)!} \non\\
&\times&
\hat x^{k_0} \, r^{\sum_{e \, {\rm in}} l_{+e} - \sum_{e \, {\rm out}} (l_{-e}-D+2) + \sum_{v \in T} \delta_v } \, (-2)^{\sum_e |k_e|} \, \prod_{e \, {\rm in}} p_{+e}^{k_e} \, \prod_{e \, {\rm out}} p_{-e}^{k_e}\, \non
\prod_{e \in G \setminus T} \e^{-l_e\theta_e}
%\non\\ &\times&
%\non\\
%&\times&\prod_{e \, {\rm in}} \left(\prod_{w\succeq e}(1+\epsilon_w)\right)^{l_w }
%\prod_{e \, {\rm out}}  \left(\prod_{w\succeq e}(1+\epsilon_w)\right)^{-l_w-D+2 }
%\prod_{v \in T} \Big( \prod_{w\preceq v}(1+\epsilon_w)\Big)^{l_w + \delta_w} \, . \non
\label{altern}
\eena
Eq.~\eqref{altern} requires several comments. First, we have the
regularizing factors from eq.~\eqref{amplitude1}. This means that eq.~\eqref{altern} rigourously  makes
sense in arbitrary $D$, but for $D\geq 3$ the expression is divergent if we let $\epsilon_v \to 0$.
We have left out the regularizing factors associated with uncontracted leaves because they remain
regular in the limit $\epsilon_v\to 0 $.
As above, $T$ is a tree on $1, \dots, i$,
and $G$ is a graph from ${\mathcal G}(T)$, the set of all loop graphs for which $T$ is a spanning
tree, and which result from $T$ by joining leaves together and hence forming loops. The edges
are $e$, and the vertices $v$. The leaves on $T$ that were joined to form loops
are hence $e \in G \setminus T$, and with each of them, we have a summation counter $l_e \in \mathbb{N}$.
The $l_e$'s associated with the uncontracted leaves are assigned to the entries in $\vec l_\pm$.
Here, the notation ``$e$ in'' means that $e$ runs through the incoming leaves,
while ``$e$ out'' that it runs through the outgoing leaves. The $l_e$ associated with the incoming leaves is set equal to
an $l_{+e}$ in the vector $\vec l_+$, while the $l_e$ associated with the outgoing leaves is set equal
to an $l_{-e}$ in the vector $\vec l_-$. By the first summation sign in the first line of eq.~\eqref{altern}, we mean
the sum over all possible assignments of the entries of $\vec l_+,\vec l_-$ to the uncontracted
leaves of $G$. A similar notation is used above for the vectors $\vec p_+, \vec p_-$.

The $l_e$ associated with edges that are neither leaves, nor contracted leaves are referred to
as ``$e \in T$'', and if $e=(vw)$ they are also called $l_v$ in the last line.
The value of $l_e$ on such an edge is determined by the momentum conservation rule~\eqref{lconservation}.
Furthermore, with each edge, we have additional summation counters $k_e \in \mn^D$.
 The sum over $k_{e}$ when
$e \in G \setminus T$ is
over those integer vectors for which $2j_e:=l_e - |k_e|$ is even.
For such integer vectors $k = (k_\mu)
\in \mn^D$, and vectors $p = (p_\mu) \in \mr^D$ we use the following multi-index notation:
\ben
|k| = \sum_\mu k_\mu \, , \quad p^k = \prod_\mu p_\mu^{k_\mu} \, , \quad
k! = \prod_\mu k_\mu ! \, .
\een
We write ``$e$ on $v$'' to mean that the sum/product is
running over those edges $e$ going out from the vertex $v$.
The integer vector $k_0 \in \mn^D$ is the counter associated with the root $e = 0$.

Formula~\eqref{altern} is the desired alternative representation of the contribution to the
vertex operator from an individual graph $G$. In $D=2$, the full vertex operator is given by the sum
of the graph contributions~\eqref{altern} with $\epsilon_v=0$, as in eq.~\eqref{yig}. For $D\geq 3$, we
have to keep the regulators $\epsilon_v$ finite. The full vertex operator is the
sum of the contributions from the graphs~\eqref{altern} and the counterterms whose construction was outlined
in sec.~\ref{renormalization}. The limit as $\epsilon_v \to 0$
has to be performed in the end.

The residue integrals $\int \d \delta_v / \delta_v$ can be performed straightforwardly
using the well-known Laurent expansion of the Gamma-function around integer values, which can be
inferred from the standard formula
\ben
\Gamma(1+\delta) = \frac{1}{1+\delta} \e^{\delta(1-\gamma_{\rm E})} {\rm exp}
\left\{
\sum_{n=2}^\infty (-\delta)^n (\zeta_n -1)/n
\right\} \, ,
\een
where $\zeta_n$ are the values of the Riemann Zeta-function\footnote{We expect to
see a connection to the ``$Z$-sums'' described e.g. in~\cite{weinzierl} and their
algebra when performing the residue in eq.~\eqref{altern}.}. Thus, we see that we get a
representation involving only (multiple) sums.

It is worth contrasting the above formula~\eqref{altern} with the integral representations
for the contribution $Y_i(G, \varphi, x)$ that were derived earlier in sec.~\ref{sec:3}. For comparison,
we here repeat these formula, or more precisely, the matrix elements of them. In doing so,
we take the opportunity to replace the integrals $\prod_v \int_{S^{D-1}} \d\Omega(\hat y_v)$ over the
spheres by integrals over the corresponding $SO(D)$-invariants $\hat y_v \cdot \hat y_w$, using a general formula
due to~\cite{fuglede}. More precisely, let us write $\{\vec p_+, \vec p_-\} = \{p_i \mid i=1, \dots, L\}$
for the ``momenta'' associated with the states $|\vec p_\pm, \vec l_\pm \rangle$. We order the vertices so
that the vertex $v=0$ is the root, so that the vertices $v=1, \dots, L$ correspond to the leaves,
and so that the internal vertices are $v = L+1, \dots, L+i+1$. Then, for each pair $(vw)$ of vertices we
introduce a real integration variable $z_{v,w}$, and if $(vw)=e$ represents an edge $e$ of the graph $G$,
we also write $z_{v,w} = z_e$. From these quantities, we also define the following symmetric $D \times D$ matrices:
\ben
Z_{D,k} = \left(
\begin{matrix}
z_{0,0} & \dots & z_{D-2,0} & z_{k,0}\\
\vdots & & \vdots & \vdots \\
z_{0,D-2} & \dots & z_{D-2,D-2} & z_{k,D-2}\\
z_{0,k} & \dots & z_{D-2,k} & z_{k,k}
\end{matrix}
\right) \, .
\een
The formulae derived in sec.~\ref{sec:3} can then be seen to take the form in $D>2$ (up to numerical factors):
\bena
&&\langle \vec p_-, \vec l_- | Y_i(G, \varphi, x) | \vec p_+, \vec l_+ \rangle = \, r^{2i+ \sum_{e \, {\rm in}} l_{+e} - \sum_{e \, {\rm out}} (l_{-e}+D-2)} \non\\
&\times&
\sum_{\substack{{\rm assignments}\\ \vec l_+,\vec l_-\rightarrow {\rm leaves}}} \,
\left( \prod_{L+1 \le v \le L+i+1} \frac{1}{2\pi \iho}  \int_{C_v} \frac{\d \delta_v}{\delta_v} \, r^{\delta_v}
%(1+\epsilon_v)^{h(v) \delta_v}
\right)
%r^{\sum_{L+1 \le v \le L+i+1} \delta_v }
 \non\\
&\times&
\prod_{1 \le v \le L+i+1} \left( \int_0^{2\pi} \e^{2\iho t_v} \d t_v \right)   \,
\, \int_{{\mathscr M}_{D,L+i+1}} \, \prod_{L+1 \le v \le L+i+1} \frac{1}{\sqrt{{\rm  det}(Z_{D,v})}}
\prod_{\substack{0 \le v \le D-2\\ v<w\le L+i+1}} \d z_{w,v} \non\\
&\times& \,\,\,\, \prod_{1 \le v \le L} \delta(\hat x \cdot p_v - z_{0,v})
\,
\prod_{\substack{1 \le v \le D-2\\ v<w\le L}} \delta(p_v \cdot p_w - z_{v,w}) \non\\
&\times&
\,\,\,\,\,\,\, \prod_{e \in T} g_{D}(\delta_e, z_e, t_e) \,
\prod_{\substack{e \in G \setminus T\\ }}
\left[(\e^{\iho \arccos z_e} -\e^{\iho( t_e+\iho \theta_e)})(\e^{-\iho \arccos z_e}-\e^{\iho( t_e+\iho \theta_e)})\right]^{-(D-2)/2} \non\\
\vspace{1cm}
&\times&
%(1 + 2 \e^{\iho (t_e + \iho 0)} z_e + \e^{2\iho (t_e + \iho 0)} )^{-D/2+1} \,
\,\,\,\,\,\,\, \prod_{e \, {\rm in}}^{} \P(z_e, l_{+e}, D) \, \prod_{e \, {\rm out}}^{} \P(z_e, l_{-e}, D) \,
\prod_{1 \le v \le L_+} \e^{\iho t_v(D-2)} \, .
\label{amplitude2}
\eena
Here, $g_D$ are the distributions given explicitly in eq.~\eqref{dangerous}, and the $\delta$'s in
the fourth line are Dirac $\delta$-distributions. In $D=2$, the terms in the last product
are replaced by log's.

Furthermore, ${\mathscr M}_{D,n}$ is the manifold of dimension $(r-1)(n-r/2)$
of all real, positive $n \times n$ matrices $M = (M_{ij})$
of rank $r = min(D, n)$ such that $M_{ii} =1$. Thus, we see that while formula~\eqref{altern}
is given entirely in terms of infinite sums, the above representation~\eqref{amplitude2} is entirely
in terms of convergent integrals. The integral over ${\mathscr M}_{D, L+i+1}$ will become divergent for $D \ge 3$ in the
limit when the regulators are removed,
the divergences coming from the boundary $\partial {\mathscr M}_{D, L+i+1}$.
As was the case for the previous formula, for $D\geq 3$, we would have to calculate the
vertex operator $\langle \vec p_-, \vec l_- | Y_i(\varphi, x) | \vec p_+, \vec l_+ \rangle$ as the sum of all the above terms
$\langle \vec p_-, \vec l_- | Y_i(G, \varphi, x) | \vec p_+, \vec l_+ \rangle$ plus the counterterms, before removing
the regulators. This was outlined in sec.~\ref{renormalization}.

\medskip

It is fair to ask what is the value of having the alternative
representations~\eqref{altern} and~\eqref{amplitude2}. It is not clear that either representation has much of an advantage computationally,
as there is essentially an equivalent number of summations as there are integrations in both
formulae~\eqref{altern} and~\eqref{amplitude2}. However, the alternative representation~\eqref{altern} brings out a striking feature that was far from
obvious when we started the construction of the vertex operators, namely that it can be represented in
terms of (multiple) infinite series of a very special form, with each term being a monomial times
a ratio of Gamma-functions. Because of this feature,
the above series can be viewed as a generalization of the Gauss hypergeometric series, and the particular
form of the series is governed by the graph $G$ under consideration.
The vertex operators $Y_i(a, x)$ may also be defined for general $a \in V$, and their matrix elements
have a similar representation. Furthermore, the vertex operators satisfy the consistency relation~\eqref{consistency2}.
We expect that this relation will give highly non-trivial relations between the above functions
of hypergeometric type. Those relations make this class of multi-variate functions special.
By analyzing the relations that are obtained in more detail, we expect that one can uncover interesting relations between our vertex operator algebras
and the theory of special functions. We think that this is a fruitful direction for further research.

It remains to be understood better how to incorporate the renormalization procedure into this
approach. As we have emphasized before, the above formula only yields a convergent result for $D=2$,
while it is only an incomplete representation of the vertex operator for $D>2$. It appears that
a formula of the above nature including renormalization can also be given, but this involves the use of a Hopf-algebra structure similar to that of~\cite{kreimer}. We will pursue this in another paper as well.

\section{Relation to vertex operator algebras in CFT's}
\label{sec:5}

Vertex operator algebras have been discussed previously in the literature (see e.g.~\cite{Borcherds:1983sq,Frenkel:1988xz,kac})
in the context of conformally invariant QFT's (CFT's), so we explain the
difference between our framework/motivation and those approaches.

First of all, we stress again our framework is in essence a ``repackaging'' of the information contained in the Wilson operator product expansion (OPE), which is a standard tool in QFT. The relation is, as we explained above, $\langle c| Y(a,x) |b \rangle =
C^c_{ab}(x)$. Our new realization is that, when repackaged in this way, the OPE satisfies
properties that can be encoded in some kind of vertex algebra. As we have discussed, this repackaging
allows both for a new conceptual viewpoint of perturbation theory via Hochschild cohomology and
a new way to do calculations in perturbative QFT.

By contrast to the vertex operator algebras that have previously been considered in the context of conformally invariant theories
in $D=2$ dimensions, our approach is intended to work
not just for 2-dimensional CFT's, but for any QFT (in any dimension) whose Schwinger functions have an OPE. Despite this
key difference, there are some evident parallels: As in our approach, one considers endomorphisms $Y(a,x) \in {\rm End}(V)$, where $a \in V$. Also, similar to our approach, the vectors $a$ are interpreted both as states and as fields (``state-field correspondence''), and the $Y(a,x)$ satisfy certain properties that are similar to ours. More precisely,
in the usual approaches, $\mr^2$ is identified with $\mc$, and the vertex operators are formal distributions\footnote{
To be precise, our vertex operators should be written $Y(a, x, \bar x)$, because they are not holomorphic
in $x$ unlike in the CFT context.},  $Y(a, \, . \,) \in {\rm End}(V) \otimes \mc[[x,x^{-1}]]$, where $\mc[[x,x^{-1}]]$ is the ring of formal sums of the form
$\sum_{k \in \mz} a_k x^k$ with complex coefficients. The analogue of the transformation formula under $SO(2)$ is
played by a formula expressing the conformal invariance of the theory, i.e. either under $PSL_2(\mr)$ or
even the Virasoro algebra. A notion of vertex algebra in $D$ dimensions that is in a similar spirit and
is applicable to globally conformally invariant theories has also been introduced, see~\cite{Nikolov:2003df}.
The appearance of the ring of formal Laurent series in the CFT context is connected in an essential way to the fact that,
in conformally invariant theories, the OPE-coefficients only have singularities of a very special form
due to conformal invariance.

Our consistency condition, as stated above in eq.~\eqref{consistency1}, does not make sense as it stands
in the standard CFT-vertex algebra context~\cite{Frenkel:1988xz,kac}, because the left side of the condition would be an element in the ring ${\rm End}(V) \otimes \mc[[x,x^{-1},y,y^{-1}]]$, whereas the right side would be an
element in the ring ${\rm End}(V) \otimes \mc[[(x-y),(x-y)^{-1}]][[y,y^{-1}]]$, and there is no
natural way to identify these rings without some notion of convergence. Such a notion of convergence is
available (and used) in our context, because we deal with the ring of holomorphic functions, but it is not
usually considered in the CFT-context, where one prefers to work with the above rings of formal series. A related problem
with our form of the consistency condition in the CFT-context is that the condition
$0 < |x-y| < |y| < |x|$, which is essential for our form of the consistency condition, does not
even make sense in the ring of formal series considered in the CFT context.

However, in the CFT context, there is another way to formulate a consistency condition which bypasses any convergence
considerations, and which leads to a relation with a superficially similar appearance.
This condition can be stated in various equivalent ways, but a
particularly transparent formulation is that for each $a,b \in V$ there is an $N \in \mn$ such that
\ben\label{lvop}
(x-y)^N [Y(a,x)Y(b,y) - Y(b,y)Y(a,x)]  = 0 \, .
\een
This equation now makes sense in the CFT context as an equation in the ring ${\rm End}(V) \otimes \mc[[x,x^{-1},y,y^{-1}]]$,
but it no longer makes sense in our context: The first term would be defined in the domain $|x|>|y|$, whereas the
second term would be defined in the disjoint domain $|y|>|x|$. Thus, the situation with regards to our consistency condition~\eqref{consistency1} and with~\eqref{lvop} is exactly opposite: Our condition~\eqref{consistency1}
cannot even be formulated in the standard vertex algebra approach in CFT's, while eq.~\eqref{lvop} cannot
even be formulated in our approach. While one can derive from eq.~\eqref{lvop} an equation that
does make sense in the CFT-context, and that is similar in appearance to our consistency condition~\eqref{consistency1},
\ben
(x-y)^N \, Y(a, x-y) Y(b, -y)c = (x-y)^N \, Y(Y(a,x)b,-y)c \, ,
\een
see~\cite[Lemma 4.6]{kac}, one sees that there are notable differences between this and our consistency condition\footnote{In the approach~\cite[Sec. 3.4]{gaberdiel} to CFT's (see also references therein)
the consistency condition is the same as ours. However, he does not impose the restriction $0 < |x-y| < |y| < |x|$ for the validity of the equation. This feature is accidental in CFT and essentially due to the fact that three points $0,x,y \in \mc$ can
be brought into an arbitrary relative position by a conformal transformation, but not by a Euclidean transformation. Hence, also
the underlying reasoning leading to the consistency condition as given in this reference is specific to CFT's and
cannot be repeated in more generic situations.}.

The equation~\eqref{lvop}, which is often called ``locality'' in the CFT-vertex operator algebra context, may also be
restated as saying that the commutator $[Y(a,x),Y(b,y)]$ is a finite sum of derivatives of
``formal delta distributions''\footnote{The formal delta distributions
are defined as the formal series $\delta(x-y) = \sum_{n \in \mz} x^n y^{-n-1}$.} in the sense that
 \ben
[Y(a,x),Y(b,y)] = \sum_{n=0}^{N-1} [\partial^n \delta](x-y) Y(c_n, y) \, , \quad a,b,c_n \in V \, .
\een
Especially when written in the last form one can understand that the ``locality condition'' is intimately related to the fact
that, in 2-dimensional CFT's, the commutator of local operators are supported on the lightcone (in a Minkowski formulation
of the theory)---or equivalently---that the singularities of the OPE coefficients are of the form $(x-y)^{-n}$, where $n \in \mn$
(in a Euclidean formulation of the theory). By contrast, the singularities in a generic QFT do not have to be of this
form, and may e.g. contain log's, as exemplified by the perturbative constructions of this paper. Furthermore, in a generic QFT, it is no longer true that the terms in the OPE  are smooth apart from a finite number of terms, as even terms in the OPE
\ben
\mathcal{O}_a(x) \mathcal{O}_b(0) = \sum_c C^c_{ab}(x) \, \mathcal{O}_c(0)
\een
corresponding to fields $\mathcal{O}_c$ of arbitrarily large dimension still typically contain log's and are thus
not strictly smooth.

For these reasons, we are somewhat pessimistic that a fruitful definition of vertex algebra of the same purely algebraic flavor as described e.g. in~\cite{kac} can also be found for the generic (e.g., perturbative) QFT models. However, we think that this direction deserves further study, and some ideas in this direction have been put forward by~\cite{Borcherds:1997cx}.

\medskip

\noindent
{\bf Acknowledgements:} We would like to thank R. Brunetti, J. Holland, N. Nikolov, and I. Runkel for discussions. We would
also like to thank R. Brunetti for pointing out to us ref.~\cite{fuglede}. The work of H. Olbermann was supported by an EPSRC Doctoral Training Grant.

\appendix

\section{Spherical harmonics and Legendre functions in $D$ dimensions}
\label{app:A}
Polynomials $h(x), x \in \mathbb{R}^D$ which are solutions to the Laplace
equation $\Delta h(x) = 0$ are called
``harmonic polynomials''. Since the Laplace operator $\Delta$ commutes with dilations
$x \mapsto tx$, it follows that any harmonic polynomial can be
decomposed into a sum of homogeneous harmonic polynomials. The harmonic polynomials
satisfying $h(tx) = t^l h(x), l \in \mn$ span a vector subspace of dimension $N(l,D)$
in $\mc[x]$, where
$N(0,D) = 1$ and
\ben
N(l,D) =
\frac{(2l+D-2)(l+D-3)!}{(D-2)!l!} \quad \text{for $l>0$.}
\een
This can be seen for example by noting that the degree $l$ harmonic polynomials $h(x)$ are in
one-to-one correspondence with
totally symmetric, traceless tensors of rank $l$ on $\mr^D$: If
$c_{\mu_1 \dots \mu_l}$ are the components of such a tensor, then
$h(x) = \sum c_{\mu_1 \dots \mu_l} x_{\mu_1} \cdots x_{\mu_l}$ is a harmonic polynomial
of degree $l$, and vice versa. The spherical harmonics in $D$ dimensions are by definition the restrictions of the harmonic polynomials to $S^{D-1}$.

%%HO 05/06/09
In the main text, we consider the a basis $h_{l,m}(x),\, m\in\{1,\dots,N(D,l)\}$ of degree $l$ harmonic polynomials
for each $l \in \mn$. The members of this basis are chosen to satisfy the orthogonality condition eq.~\eqref{harmnorm}. In fact, the $h_{l,m}$ form an orthonormal basis of $L^2(S^{D-1}, \d \Omega)$ when restricted to the sphere. It follows
immediately from the fact that the $h_{l,m}(x)$ are harmonic polynomials that their restrictions $h_{l,m}(\hat x)$ to the
sphere are eigenfunctions of the Laplacian $\hat \Delta$ on $S^{D-1}$ with eigenvalue $-l(l+D-2)$.

For our calculations in appendix~\ref{Y0op}, we need to know in more detail the relation of the harmonic
polynomials $h_{l,m}$ to the traceless symmetric tensors of rank $l$ described above. To state the relevant facts,
we use the familiar multi-index notation, $\alpha=(\alpha_1,\dots,\alpha_D) \in \mn^D$, with
\ben
x^\alpha = \prod_\mu x_\mu^{\alpha_\mu} \, , \quad \partial_\alpha = \prod_\mu \partial_{\mu}^{\alpha_\mu} \, ,
\quad \alpha! = \prod_\mu \alpha_\mu! \, \quad \text{etc.,}
\een
and we write
\ben
h_{l,m}(x)= \sum_\alpha t_{l,m;\alpha} x^\alpha\, .
\een
Combining eq.~\eqref{harmnorm} with theorem~5.14 of \cite{Axler}
 we get
\ben
\sum_\alpha \bar{t}_{l,m;\alpha}t_{l',m';\alpha}\frac{\alpha!}{k_l}=\delta_{l,l'}\delta_{m,m'}.
\een
with $k_l=2^l\Gamma(l+D/2)/\Gamma(D/2)$. This can also easily be proved starting from eq.~\eqref{EulerBeta}.\\
The decomposition of a harmonic function $f$ regular at the origin into harmonic polynomials reads
\ben
f(x)=\sum_{l,m}\left(\int_{S^{D-1}}\d\Omega(\hat x) f(\hat x) \bar h_{l,m}(\hat x)\right)h_{l,m}(x)\,.
\label{harmdecomp}
\een
With
$\partial^\alpha x^\beta|_{x=0}=\delta_{\alpha\beta}\alpha !$
we have
\ben
\bar h_{l,m}(\partial)h_{l',m'}(x)|_{x=0}=
\sum_\alpha \bar t_{l,m;\alpha}t_{l',m';\alpha}\alpha !=\delta_{ll'}\delta_{mm'}k_l
\een
and thus eq.~\eqref{harmdecomp} reads
\ben
\label{harmdecomp2}
f(x)=\sum_{l,m}k_l^{-1} \left(\bar h_{lm}(\partial)f(0)\right) h_{lm}(x).
\een

We also cite theorem 5.20 of \cite{Axler}, which
states that for a harmonic homogeneous polynomial $p$ of degree $l$,
\ben
\label{harmderiv}
p(\partial)\, g(r)=q_l\, r^{2-D}p(x/r^2)
\een
where $r=|x|$,
\ben
g(r)=\begin{cases}r^{2-D} &  \text{for } D=2\\
\ln r & \text{for } D>2\,,\end{cases}
\label{Gr}
\een
and
\ben
q_l=\begin{cases}2^{l-1}\Gamma(l) & \text{for } D=2\\
2^l\Gamma(l+D/2-1)/\Gamma(D/2-1) & \text{for } D>2\,.\end{cases}
\een

\vspace{10mm}

The Legendre polynomials in $D \ge 2$ dimensions are defined as the following invariants under $SO(D)$:
\ben\label{Pdef}
\sum_{m=1}^{N(l,D)}
\overline{h}_{l,m}(\hat x) h_{l,m}(\hat y) =
\frac{2l+D-2}{\sigma_D} \, \P(\hat x \cdot \hat y, l, D) \, .
\een
By construction, the Legendre polynomials $\P(z,l,D)$ are polynomials of degree $l \in {\mathbb N}$.
They are often also called ``Gegenbauer polynomials'' and are denoted alternatively $C^{(D-2)/2}_l(z)$,
with notable differences in the normalization convention throughout the literature.
A generating function is
\ben
\label{expansion}
\frac{1}{D-2} \left(\frac{1}{\sqrt{1-2hz+h^2}} \right)^{D-2} = \sum_{l=0}^\infty \P(z,l,D) h^l \,\, .
\een
This formula holds for $D\ge 3$. For $D=2$, the left side is to be replaced by $-\ln \sqrt{1-2hz+h^2}$.
A generalization of this formula needed in the main text is provided in theorem~\ref{lem2}.
The Legendre polynomials have the symmetry property
$\P(z, l, D) = (-1)^l \P(-z, l, D)$, and satisfy the normalization condition
\ben\label{ncond}
\P(1, l, D) = \frac{(l+D-3)!}{l! (D-2)!} \, .
%\left(
%\begin{matrix}
%l+D-3\\
%l
%\end{matrix}
%\right) \, .
\een
For complex values of the index $\nu \in \mc$ (or $D$), one can define an
analytic continuation by means of the Gauss hypergeometric function
\ben
\P(z, \nu, D) =
%\left(
%\begin{matrix}
%\nu+D-3\\
%\nu
%\end{matrix}
%\right)
\frac{\Gamma(\nu+D-2)}{\Gamma(\nu+1) \, \Gamma(D-1)}
\, {}_2 F_1 \left( -\nu, \nu + D-2, D/2 -1/2, \frac{1-z}{2} \right)  \, .
\een
The Gauss hypergeometric function is
given by the convergent expansion
\ben
{}_2 F_1(a,b;c; x) = \sum_{n=0}^\infty \frac{(a)_n (b)_n}{(c)_n n!} \, x^n \, , \quad
(a)_n = \Gamma(a+n)/\Gamma(a) \, ,
\een
for $|x| < 1$. Note that the above formula has a slight anomaly in $D=2$ dimensions.
Here, it gives $\P(\cos \alpha, \nu, 2) = \cos (\nu \alpha)/2\nu$
in $D=2$ dimensions for $\nu \neq 0$, and this evidently does not have a limit as $\nu \to 0$.
On the other hand, the generating formula definition gives $\P(\cos \alpha, 0, 2) = 1$.
%We will also use the Schlaefli-integral representation
%(see [Whittaker, Watson]):
%\bena
%\P(z,\nu,D)
%%%\frac{2^n\Gamma(\nu+(D-2)/2)\Gamma(\nu+D-2)}{\Gamma(\frac{D-2}{2}\Gamma(2\nu+D-2)}
%%%(1-z^2)^{-\frac{D-3}{2}}\oint_C\frac{(t^2-1)^{\nu+\frac{D-3}{2}}}{(t-z)^{n+1}}dt\non\\
%=2^{3-D-\nu}\,\frac{\sqrt{\pi}\,\Gamma(\nu+D-2)}{\Gamma(\frac{D-2}{2})\Gamma(\nu+\frac{D-1}{2})}\,
%(1-z^2)^{-\frac{D-3}{2}}\,\oint_C\frac{(t^2-1)^{\nu+\frac{D-3}{2}}}{(t-z)^{\nu+1}} \, \d t
%\label{schlaefli}
%\eena
%The contour $C$ is running counter-clockwise around the cut of the integrand between $z$ and $+1$.
The differential equation satisfied by the Legendre functions is
\ben\label{ydiff}
(1-z^2) y'' - (D-1) zy' + \nu(\nu+D-2) y = 0 \, .
\een
%% HO 04/06/09
\section{Theorems for Legendre functions}
\label{app:C}

In the main text, we use certain identities for Legendre functions in
$D$ dimensions that we were not able to find in the literature, and
which we therefore prove here:

\begin{thm}\label{lem1} ({\em Generalized Dougall's formula})
For any $\nu \in \mc \setminus \mz$ and $-1 \le z \le +1$ and $D \ge 3$, we have the identity
\ben
\sum_{l=0}^\infty
\frac{(2l+D-2) \,  \P(z, l, D)}{\nu(\nu+D-2) - l(l+D-2)}
= \frac{\pi}{\sin \pi \nu} \, \P(-z, \nu, D) \, .
\een
\end{thm}
{\it Proof:}
For $D=3$, a proof of the theorem can be given via a contour integral argument, see~\cite{Erdelyi}. We
here give the following conceptually somewhat more transparent proof, valid for arbitrary $D>2$. Let
$\hat \Delta$ be the Laplacian on the sphere $S^{D-1}$. This is an elliptic, second order partial differential operator on a compact manifold with analytic coefficients. Using standard results on the functional
calculus of such operators, we can form the resolvent operator $R_\nu = [\hat \Delta + \nu(\nu+D-2)]^{-1}$
for any $\nu$ such that $\nu(\nu+D-2)$ is not an eigenvalue, i.e. $\nu \notin \mz$. Let $R_\nu(\hat x,
\hat y)$ be the kernel of $R_\nu$, which using general results on the Laplacian on
compact Riemannian manifolds is known to be an analytic function on $S^{D-1} \times S^{D-1}$
apart from coincident points. Near coincident points, one has $R_\nu \sim [d(\hat x, \hat y)]^{-(D-3)/2}$ for
$D > 3$ and $R_{\nu} \sim \ln d(\hat x, \hat y)$ for $D=3$, where $d(\hat x, \hat y) =
\arccos (\hat x \cdot \hat y)$ is the geodesic distance on the sphere. A representation of $R_\nu$
in terms of eigenfunctions of the Laplacian is
\bena
R_\nu(\hat x, \hat y) &=& \sum_{l=0}^\infty \sum_{m=1}^{N(D,l)}
\frac{\overline h_{l,m}(\hat x) h_{l,m}(\hat y)}{\nu(\nu+D-2) - l(l+D-2)} \non\\
&=&
\sigma_D \sum_{l=0}^\infty
\frac{(2l+D-2) \,  \P(\hat x \cdot \hat y, l, D)}{\nu(\nu+D-2) - l(l+D-2)} \, .
\eena
In the second line we have used the definition of the Legendre polynomials.
Hence we see that the kernel $R_\nu$ is, up to a constant, precisely equal
to the left side of the Dougall formula.

By definition, the kernel obeys $[\hat \Delta + \nu(\nu+D-2)] R_\nu = \delta$
in the sense of distributions. However, since $R_\nu$ is evidently invariant under
$SO(D)$-transformations, we may write $R_\nu(\hat x, \hat y) = y(z)$ for
some analytic function of $z = \hat x \cdot \hat y$ when $z \neq 1$. As a consequence
of the differential equation satisfied by $R_\nu$, it can easily be seen that $y$
satisfies the differential equation for the Legendre function of dimension $D$
and degree $\nu$, see eq.~\eqref{ydiff}. Hence we have
\ben
\label{yab}
y(z)= A\, \P(z,\nu,D)+ B\, \P(-z,\nu,D)
\een
for some $A,B\in\mc$ as $\P(z,\nu,D),\P(-z,\nu,D)$ span the solution space of eq.~\eqref{ydiff}.
Furthermore, $\P(-z,\nu,D)$ is singular at $z=1$ and regular at $z=-1$ (see e.g.\cite{Erdelyi}), as is $y(z)$.
By contrast, $\P(z,\nu,D)$ is singular at $z=-1$ and regular at $z=1$. Thus, we must have
$A=0$ in eq.~\eqref{yab}.

%The Legendre differential equation has two linearly independent solutions, but only one
%of them the desired singular behavior of $y$, so we conclude that $y(z,\nu,D) = c(\nu,D) \,
%\P(z, \nu, D)$.
In order to determine the constant $B$, we evaluate
$y(-1)$.
Using $\P(-1,l,D)=(-1)^l \P(1, l, D)$ and the formula~\eqref{ncond}, we find using
various summation identities for the Gamma-function:
\bena
\frac{1}{\sigma_D}\, y(-1)&=& \frac{1}{(D-2)!}
\sum_{l=0}^\infty (-1)^l \left(
\frac{1}{\nu-l} - \frac{1}{\nu+l+D-2}
\right)
\frac{(l+D-3)!}{l!} \non \\
&=&
%%       ???
%% ???   HOW  ???
%%       ???
\frac{\Gamma(-\nu)\Gamma(\nu+D-2)}{\Gamma(D-1)} =
\frac{\Gamma(\nu+D-2)}{\Gamma(\nu+1)\Gamma(D-1)}
%\left(
%\begin{matrix}
%\nu+D-3\\
%\nu
%\end{matrix}
%\right)
\frac{\pi}{\sin \pi \nu} \,\, .
\eena
Comparing this with the normalization of $\P(-z,\nu,D)$ at $z=-1$,
we get the statement of the theorem.
\qed

The next theorem is a generalization of formula~\eqref{expansion}.

\begin{thm}\label{lem2} ({\em Shifted generating functional formula})
For any $\delta \in \mc \setminus \mz$ and $-1 \le z \le +1, |h|<1$ and {\em even} $D \ge 4$, we have the identity
\bena
&& \sum_{l=0}^\infty
h^l \,  \P(z, l+\delta, D) = (2\delta_D)^{-1} \, A_{D}\Bigg\{
(z+\iho\ sqrt{1-z^2})^{\delta_D} \, {}_2 F_1 \Big(\delta_D, 1; \delta_D+1; h(z+\iho\ sqrt{1-z^2}) \Big) \non\\
&&+ (z-\iho\ sqrt{1-z^2})^{\delta_D} \, {}_2 F_1 \Big(\delta_D, 1; 1+\delta_D, h(z-\iho\ sqrt{1-z^2}) \Big)
\Bigg\} \, ,
\eena
where $\delta_D$ and the differential operator
$A_D$ are given by
\bena
A_D &=& \frac{1}{\Gamma(D/2)} \left( \frac{\partial}{2h \, \partial z} \right)^{(D-2)/2}\non\\
\delta_D&=&\delta+(D-2)/2\,.
\eena
For $D=2$ the differential operator is missing.
For {\em odd} $D \ge 5$, we have the formula
\bena
&&\sum_{l=0}^\infty
h^l \,  \P(z, l+\delta, D) = A_{D}
\Bigg\{ \frac{-1}{\sqrt{1+h^2+2hz}} \times \label{gen_func}\\
&&\Bigg(F_1\Big(-\delta_D,\delta_D,1,1;\frac{1-z}{2},\frac{1-t_-}{2}\Big)+
\frac{z-t_-}{2} F_1\Big(1-\delta_D,1+\delta_D,1,2;\frac{1-z}{2},\frac{1-\tau(t_+)}{2}\Big)\Bigg) \Bigg\}\,.\non
\eena
where this time
\bena
A_D &=& \frac{\sqrt{\pi}}{2\Gamma(D/2)} \left( \frac{\partial}{2h \, \partial z} \right)^{(D-3)/2}\non\\
\delta_D&=&\delta+(D-3)/2\,.
\eena
For $D=3$ the differential operator is missing.
$F_1$ is the two-variable generalization of the hypergeometric function defined by
\ben
F_1(a,b,c;d;x,y)=\sum_{m=0}^\infty\sum_{n=0}^\infty
\frac{(a)_{m+n}(b)_m(c)_n}{(d)_{m+n}m!n!}x^m y^n\,,
\een
and we have defined $t_{\pm} =  h^{-1}\left(1\pm\sqrt{1+h^2-2hz}\right)$, and $\tau(t)$
by eq.~\eqref{tautrafo}.
\end{thm}
{\bf Remark}: There is an apparent asymmetry in the formulas for even and odd $D$. One is tempted
to believe that both formulas given for the shifted generating function are valid for all $D$
(when appropriately interpreted), but we have not been able to show this. The first formula may
further be rewritten noting the standard formula
\ben\label{fdef}
f(\delta,x) = (2\delta)^{-1} \, {}_2 F_1(\delta, 1; \delta+1; 1-x) =
\sum_{n=0}^\infty \frac{(\delta)_n}{n!} [\psi(n+1) - \psi(\delta+n) - \ln x] \, x^n \, .
\een
In the main text, we use the theorem to calculate the sums
eq.~\eqref{g2sum} for $D=2$ and eq.~\eqref{gDsum} for $D=3$, where we also make use of the relation
\ben
\P(z,\nu,D)=(-1)^{D-3}\P(z,-\nu-D+2,D)\,.
\een
We then apply the recurrence relation \eqref{diffP} to obtain $g_D(\delta,\cos\beta,t)$ as in
eq.~\eqref{dangerous} (for even $D$).

\medskip

\begin{figure}[htp]
\vspace{10mm}
\unitlength=0.7pt
\SetScale{0.7}
\begin{center}
\fcolorbox{white}{white}{
  \begin{picture}(380,105) (111,-91)
    \SetWidth{1.0}
    \SetColor{Black}
    \Line(112,-71)(272,-71)
    \Vertex(272,-71){2}
    \Vertex(384,-71){2}
    \Vertex(416,-7){2}
    \Bezier(384,-71)(400,-55)(416,-39)(416,-7)%JaxoID:FBez
    \Text(285,-71)[lb]{\small{\Black{$-1$}}}
    \Text(402,-71)[lb]{\small{\Black{$1$}}}
    \Text(424,-7)[lb]{\small{\Black{$z$}}}
    \Arc[dash,dashsize=10,arrow,arrowpos=0.5,arrowlength=5,arrowwidth=2,arrowinset=0.2](400,-39)(51.225,129,489)
    \Text(456,-47)[lb]{\small{\Black{$C^+$}}}
    \Text(416,-39)[lb]{\small{\Black{$\gamma_2$}}}
    \Text(184,-63)[lb]{\small{\Black{$\gamma_1$}}}
  \end{picture}
}
\end{center}
\caption{Cuts chosen to make $(t^2-1)^\delta(t-z)^{-\delta}$ well-defined.}
\label{fig:cuts}
\end{figure}
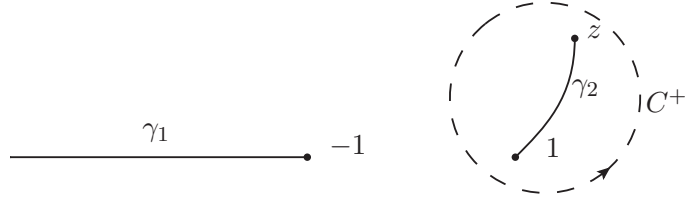

{\it Proof for even $D$:} For $D=2$, the proof of the theorem follows immediately from the definition of
the hypergeometric function. The alternate form~\eqref{fdef} is obtained
from transformation formula~15.3.10 of~\cite{abramowitz} for hypergeometric functions.
For $D$ even and $D>2$, we prove the formula using the recurrence identity
\ben
\label{diffP}
\frac{\d}{\d z} \P(z,\nu,D) = D \, \P(z, \nu-1, D+2)\,.
\een
{\it Proof for odd $D$:} For odd $D$, we proceed using the same recurrence identity, but in order
to be able to do so, we have
to evaluate $\sum_{l=(D-3)/2}^\infty \P(z,l+\delta,3)$, and this requires some extra work.
%%%
%% HO 01/06/09
%%%
We start with the Schlaefli integral formula for Legendre functions \cite{WhWa27},
\ben
P(z,\nu,3)=\frac{1}{2\pi\iho} \oint_{C^+} \frac{(t^2-1)^\nu}{2^\nu(t-z)^{\nu+1}}\d t\,.
\een
To make $(t^2-1)^\nu 2^{-\nu}(t-z)^{-\nu-1}$ single-valued, we have to
introduce two cuts in the complex plane, and we follow \cite{Szmytkowski} choosing these cuts as
the half-line $\gamma_1=(-\infty,-1)$ and a curve $\gamma_2$ joining the points $t=1$ and $t=z$,
parametrized by
\ben
\frac{1+\eta z}{\eta+z} \quad (1\leq \eta <\infty)\,.
\label{cut2}
\een
The contour $C^+$ encircles $\gamma_2$ counterclockwise, see fig.~\ref{fig:cuts}.
In \cite{Szmytkowski}, this particular representation of Legendre functions made it possible
to determine derivatives of Legendre functions with respect to their degree $\nu$.

We set $\delta_D=\delta+(D-3)/2$ and have
\bena
\sum_{l=(D-3)/2}^\infty h^l P(z,l+\delta_D,3)& = &
\sum_{l=0}^\infty h^l P(z,l+\delta_D,3)\non\\
& = &\frac{1}{2^{1+\delta_D}\pi\iho} \oint_{C^+}
\d t\left(\frac{t^2-1}{t-z}\right)^{\delta_D}
\frac{1}{t-z-h(t^2-1)/2}\non\\
& = &\frac{1}{2^{1+\delta_D}\pi\iho} \oint_{C^+}\d t\, \chi(t)\frac{1}{t-z-h(t^2-1)/2}\,,
\label{gen_int}
\eena
where in the first
equation, we have interchanged the order of summation and integration, and in the second
we have set $\chi(t)=(t^2-1)^{\delta_D}(t-z)^{-\delta_D}$ (defined with the same cuts as above).
Our aim will be to express $\chi$ as a sum of two functions $\chi_1,\chi_2$ that
have cuts on $\gamma_1$ and $\gamma_2$ respectively, and are analytic elsewhere. Then
we will be able to carry out the integration in eq.~\eqref{gen_int}.
Assuming $\Re \delta_D<0$, we apply formula 3.1.11 of \cite{Hoe},
\ben
\chi(t)=-\pi^{-1}\int \d w\d \bar w \left(\bar\partial \chi(w,\bar w)\right)(w-t)^{-1}\,.
\label{Hoer}
\een
We introduce a function
$\rho_1$ that equals 1 in a small neighborhood of the cut $(-\infty,1)$ and 0
outside a slightly bigger neighborhood. Also, we introduce $\rho_2$, smooth, equal to 1 in a small
neighborhood of the cut \eqref{cut2} and 0 in a slightly bigger neighborhood,
so that $\text{supp}\rho_1\cap\text{supp}\rho_2=\emptyset$. As $\chi$ is analytic away from the cuts
(i.e. $\bar\partial \chi=0$ on $\mathbb{C}\setminus(\gamma_1\cup \gamma_2)$),
we can modify eq.~\eqref{Hoer} in the following way,
\bena
\chi(t)  & = &\chi_1(t)+\chi_2(t)\,,\non\\
\chi_i(t)& = & -\pi^{-1}\int \d w\d \bar w \left(\bar\partial \chi(w,\bar w)\right)(w-t)^{-1}
\rho_i(w,\bar w)\,,\,\,(i=1,2)\,.
\label{Hoer2}
\eena
Now by theorem 3.1.12 of \cite{Hoe}
\bena
\chi_1(t) & = & \frac{1}{2\pi\iho} \int^{-1}_{-\infty}\d x\left(\chi(x+\iho 0)-\chi(x-\iho 0)\right)(x-t)^{-1}\non\\
       & = & \frac{\sin\delta_D\pi}{\pi}\int^{-1}_{-\infty}\d x \left(\frac{x^2-1}{z-x}\right)^\delta_D
       (x-t)^{-1}\,.
\eena
We substitute $u:=2/(1-x)$ and obtain
\bena
\chi_1(t)& = & -\frac{\sin\delta_D\pi}{\pi}2^{\delta_D}\int^1_0 u^{-1-\delta_D}(1-u)^\delta_D\left(1-\frac{1-z}{2}u\right)^{-\delta_D}
\left(1-\frac{1-t}{2}u\right)^{-1}\non\\
& = &-\frac{\sin\delta_D\pi}{\pi}2^{\delta_D}B(1+\delta_D,-\delta_D)F_1\left(-\delta_D,\delta_D,1,1;(1-z)/2,(1-t)/2\right)
\label{finalg1}
\eena
where in the second equation, we have used formula 3.211 of \cite{gradshteyn}; $F_1$ is the hypergeometric function
of two variables defined above,
and $B$ is the Beta function $B(x,y)=\Gamma(x)\Gamma(y)/\Gamma(x+y)$.
We next determine $\chi_2$,
\bena
\chi_2(t)& = & -\pi^{-1}\int \d w\d\bar w (\partial g)(w-t)^{-1}\rho_2(w,\bar w)\non\\
%     & =\pi^{-1}\int \d w\d\bar w  g\partial\left((w-t)^{-1}\rho_2(w,\bar w)\right)\\
      & = & -\pi^{-1}\int \d\tau\d\bar \tau\frac{\partial w}{\partial \tau}
      \frac{\partial\bar w}{\partial\bar\tau}\left( \frac{\partial\bar\tau}{\partial\bar w}\partial_{\bar\tau}
      \chi(w(\tau))\right)(w(\tau)-t)^{-1}\rho_2(w(\tau),\bar w(\bar\tau))\non\\
      & = & -\frac{1}{2\pi\iho} \int_{-\infty}^{-1}dx \left(\chi(x+\iho 0)-\chi(x-\iho 0)\right)\frac{1-z^2}{(z-x)^2}
      \left(\frac{1-zx}{z-x}-t\right)^{-1}\non\\
      & = & -\frac{\sin\delta_D\pi}{\pi}\int_{-\infty}^{-1}\d x\left(\frac{x^2-1}{z-x}\right)^\delta_D
      \frac{1-z^2}{(z-x)^2}\left(\frac{1-zx}{z-x}-t\right)^{-1}
\label{g_2manip}
\eena
where in the second equation we have performed a change of coordinates,
\ben
\tau(w)=\frac{1-wz}{z-w}\,,
\label{tautrafo}
\een
with inverse $w(\tau)=(1-\tau z)/(z-\tau)$. This coordinate transformation maps $\gamma_1$ on $\gamma_2$ and vice versa.
Again substituting $u=2/(1-x)$, we get
\bena
\chi_2(t)  & = & -\frac{\sin\delta_D\pi}{\pi}\frac{1-z^2}{z-t}2^{-1+\delta_D}\int_0^1\d u\,
        u^{-\delta_D}(1-u)^{\delta_D}\left(\frac{1-z}{2}u\right)^{-1-\delta_D}
        \left(\frac{1-\tau(t)}{2}u\right)^{-1}\non\\
        & = & \frac{\sin\delta_D\pi}{\pi}\frac{1-z^2}{t-z}2^{-1+\delta_D}\non\\
        & \times & B(1+\delta_D,1-\delta_D)F_1(1-\delta_D,1+\delta_D,1,2;(1-z)/2,(1-\tau(t))/2)
\label{finalg2}
\eena
In eq.~\eqref{gen_int}, we replace
\bena
\frac{1}{t-z-h(t^2-1)/2} & = & -\frac{2}{h(t-t_+)(t-t_-)}\non\\
\quad t_{\pm}& = & h^{-1}\left(1\pm\sqrt{1+h^2-2hz}\right).
\eena
For $-1\leq z\leq 1$, $t_+\in\gamma_1$ and $t_-\in\gamma_2$, so $t_-$ lies inside $C^+$ and $t^+$ outside.
We can now calculate the contribution of $\chi_1(t)$ to the integral \eqref{gen_int},
\ben
\frac{1}{2^{1+\delta_D}\pi\iho} \oint_{C^+}\d t\, \chi_1(t)\frac{1}{t-z-h(t^2-1)/2}\,,
\een
which is now a simple residue integral. We obtain
\bena
%\frac{1}{2^{1+\delta_D}\pi\iho} \oint_{C^+}\d t\, \chi_1(t)\frac{1}{t-z-h(t^2-1)/2}
\frac{\sin\delta_D\pi}{\pi}\,\frac{1}{\sqrt{1+h^2+2hz}}\,B(1+\delta_D,-\delta_D)
F_1\left(-\delta_D,\delta_D,1,1;(1-z)/2,(1-t_-)/2\right)\,.
\label{g1contr}
\eena
In a similar manner, we calculate the contribution of $\chi_2$,
\bena
%\frac{1}{2^{1+\delta_D}\pi\iho} \oint_{C^+}\d t\, \chi_2(t)\frac{1}{t-z-h(t^2-1)/2}
\,& \, & \frac{B(1+\delta_D,1-\delta_D)\sin\pi\delta_D(1-z^2)}{4\pi^2 \iho}  \non\\
& \times & \oint_{D^-}\d\tau\bigg(\frac{1-z^2}{(z-\tau)^2}
F_1(1-\delta_D,1+\delta_D,1,2;(1-z)/2,(1-\tau)/2)\frac{1}{t(\tau)-z} \non\\
& \times & \left(-\frac{2}{h}\right)\frac{z-\tau}{(t_--z)(\tau-\tau(t_-)}
\frac{z-\tau}{(t_+-z)(\tau-\tau(t_+)}\bigg)\,,
\label{g2int}
\eena
where we used the coordinate transformation \eqref{tautrafo}. The contour
$D^-$ is the image of $C^+$ under this transformation. $D^-$ encircles $\gamma_2$ and runs clockwise,
not crossing the cuts $\gamma_1,\gamma_2$. We can deform $D^-$ into $C^-$ by which we mean the
contour $C^+$ with negative orientation. $\tau(t_+)$ is on the inside of $C^-$,
$\tau(t_-)$ on the outside. Thus the residue integral \eqref{g2int} is
%% identities: \tau(t_+)-\tau(t_-)=t_+-t_-
%%             (z-t_+)(z-t_-)=z^2-1
\bena
\label{g2contr}
\,&\,& \frac{\sin{\pi\delta_D}}{\pi}\,\frac{1}{\sqrt{1+h^2-2hz}}\,\frac{1-z^2}{2(z-t_+)}\non\\
&\times& B(1+\delta_D,1-\delta_D)F_1(1-\delta_D,1+\delta_D,1,2;(1-z)/2,(1-\tau(t_+))/2)\,.
\eena
Putting together
eqs.~\eqref{g1contr} and \eqref{g2contr}, we obtain
\bena
\sum_{l=0}^\infty h^l P(z,l+\delta_D,3)& = &\frac{\sin\delta_D\pi}{\pi}\frac{1}{\sqrt{1+h^2+2hz}}\non\\
& \times & \bigg(B(1+\delta_D,-\delta_D)F_1\left(-\delta_D,\delta_D,1,1;(1-z)/2,(1-t_-)/2\right)\non\\
& + & \frac{1-z^2}{2(z-t_+)}B(1+\delta_D,1-\delta_D)\non\\
& \times & F_1\left(1-\delta_D,1+\delta_D,1,2;(1-z)/2,(1-\tau(t_+))/2\right)\bigg)\,.
\label{gen_func2}
\eena
Above, we have used $\Re\delta_D<0$. However both sides of eq.~\eqref{gen_func2} are analytic
in $\delta_D$ throughout the complex plane for $|h|<1$ (with possible exceptions for $\delta_D \in \mathbb Z$).
This means that they are identical and
 eq.~\eqref{gen_func2} must hold for $\Re\delta\geq 0$ as well.
Eq.~\eqref{gen_func} follows from the first line of eq.~\eqref{gen_int}, the recurrence formula~\eqref{diffP},
standard identities for the gamma function and the relation
$(z-t_+)(z-t_-)=z^2-1$.\qed

We finally mention another representation of the Legendre functions used in the main text:

\begin{thm}
For $\nu \in \mc \setminus \mz$, $|z|<1$ we have the formula
\ben
\P(z,\nu,D) = \frac{\sin \pi \nu}{\pi} \, \frac{2^{-(D+1)/2}}{\Gamma(D/2)}
\sum_{n=0}^\infty (-2z)^n \frac{\Gamma(-\nu/2+n/2)\Gamma(\nu/2 + n/2+D/2-1)}{n!}  \, .
\een
{\em Proof:} We prove this first for $D=2$. Let $z=\cos \alpha$. We have the identities
\bena
\cos \nu \alpha &=& {}_2 F_1( -\nu/2, \nu/2; 1/2; \sin^2 \alpha) \\
&=& \frac{\pi}{\Gamma(-\nu/2+1/2)\Gamma(\nu/2 + 1/2)} \, {}_2 F_1 (-\nu/2,\nu/2; 1/2; \cos^2 \alpha)\non\\
&&-2 \cos \alpha \,
\frac{\pi}{\Gamma(-\nu/2)\Gamma(\nu/2)} \, {}_2 F_1 (-\nu/2+1/2,\nu/2+1/2; 3/2; \cos^2 \alpha) \, ,\non
\eena
where in the second line we have used a standard transformation formula for hypergeometric
functions. We now use $\P(\cos \alpha, \nu, 2) = \cos (\nu \alpha)/2\nu$, and we expand the
hypergeometric series in the second and third line, using the doubling identity of the Gamma function,
$\sqrt{\pi} \Gamma(2x) = 2^{2x-1} \Gamma(x) \Gamma(x+1/2)$, in various ways. Then we obtain the
statement of the theorem for $D=2$. The case $D=3$ is covered by formula~8.1.4 of~\cite{abramowitz},
together with the use of the doubling identity as above.

For general $D \in \mn$, we use the recurrence formula~\eqref{diffP}, combined with a standard formula for the derivatives of the hypergeometric function. This
then gives the formula for all even $D$ starting from $D=2$ and all odd $D$ starting from $D=3$.\qed
\end{thm}

\section{The free field vertex operators}
\label{Y0op}
The non-vanishing  partial derivatives of the basic field $\varphi$ in the theory defined
by the Schwinger functions eq.~\eqref{Schwinger} are
\ben
\label{philmdef}
\varphi^{l,m}=c_l^{-1}\bar t_{l,m;\alpha}\partial^\alpha\varphi
\een
where $c_l$ will be chosen later. The composite fields are labeled by multiindices
\ben
\mathcal{O}_a=(a!)^{-1/2}\prod_{l,m}\left(\varphi^{l,m}\right)^{a_{l,m}}\,.
\een
We use Latin letters for the multiindices denoting
composite fields and Greek letters for multiindices when dealing with polynomials in $x$ or $\partial$.
The basic field $\varphi$ is harmonic by the field equation \eqref{ffe}, so we may use eq.~\eqref{harmdecomp2},
\ben
\varphi(x)=\sum_{l,m}\frac{c_l}{k_l}\, h_{l,m}(x)\,\varphi^{l,m}(0)
\een
This has to be understood as an equation for insertions into Schwinger functions.
Now the OPE of $\varphi$ with a field $\mathcal{O}_a$ can easily be deduced from eq.~\eqref{Schwinger} and the definition
of $\varphi^{l,m}$, eq.~\eqref{philmdef}:
\bena
\varphi(x)\mathcal{O}_a(0)&=&\sum_{l,m}\frac{c_l}{k_l}\, h_{l,m}(x)\,\left(\varphi^{l,m}\,\mathcal{O}_a\right)(0)\non\\
&+&\sum_{l,m}\,c_l^{-1}\,\bar h_{l,m}(\partial)\, g(r)\, \frac{\partial\mathcal{O}_a}{\partial\varphi^{l,m}}(0)\,,
\label{OPE1}
\eena
where $r=|x|$ as always; see eq.~\eqref{Gr} for the definition of the ``Euclidean propagator'' $g$.
Again, this has to be understood as an equation for insertions:
\bena
\label{OPEinsert}
\left\langle \varphi(x)\mathcal{O}_a(0)\varphi(x_1) \dots \varphi(x_s)\right\rangle=
\left\langle\left( \text{RHS of eq.~\eqref{OPE1}}\right)\varphi(x_1) \dots \varphi(x_s)\right\rangle\,.
\eena
If we write down the left hand side of eq.~\eqref{OPEinsert} as in eq.~\eqref{Schwinger2},
where the points $x$ and $0$ are represented by vertices $v$ and $w$ respectively ($x_v=x, x_w=0$),
then the first term of the right hand side in eq.~\eqref{OPE1}
can be understood as a Taylor expansion (in $x$) of those
graphs without a line $(vw)$, and the second term as the Taylor expansion of graphs including that line.
In order for these Taylor expansions to converge, we must have $|x_1|,\dots,|x_s|>|x|$.
Using eqs.~\eqref{harmderiv} and ~\eqref{OPE1} we obtain
\bena
\varphi(x)\mathcal{O}_a(0)&=&\sum_{l,m}\sqrt{a_{l,m}+1}\,\frac{c_l}{k_l}\, h_{l,m}(x)\,\mathcal{O}_{a+e_{(l,m)}}(0)\non\\
&+&\sum_{l,m}\sqrt{a_{l,m}}\,\frac{q_l}{c_l}\,\bar h_{l,m}(x)r^{-2l-D+2}\, \mathcal{O}_{a-e_{(l,m)}}(0)
\label{freeOPE}
\eena
where by $e_{(l,m)}$, we mean the multiindex defined by $(e_{(l,m)})_{l',m'}=\delta_{l,l'}\delta_{m,m'}$,
and we define $(a-e_{(l,m)}):=0$ for $a_{l,m}=0$.
To obtain a symmetric form of the OPE, we choose
\ben
c_l=\sqrt{q_lk_l}=\begin{cases}2^l\Gamma(l)\sqrt{l/2} \quad                   & \text{for }D=2\\
                 2^l\Gamma(l+D/2-1)\sqrt{2(l+D/2-1)/(D-2)}\quad & \text{for }D>2\,.\end{cases}
\een
We introduce the abstract vector space $V$ spanned by the
field labels $a$ and creation and annihilation operators $\a^+_{l,m},\a_{l,m}$ on $V$ by
\bena
\a^+_{l,m} a&=&\sqrt{a_{l,m}+1} (a+e_{(l,m)})\non\\
\a_{l,m}a&=&\sqrt{a_{l,m}} (a-e_{(l,m)})\,.
\eena
We also introduce the vertex operator $Y_0(\varphi,x)$ that corresponds to a multiplication
of an insertion with the free field $\varphi(x)$. We rewrite the left-hand side of eq.~\eqref{freeOPE} in this notation by
\ben
Y_0(\varphi,x)a
\een
and we can read off the right hand side of eq.~\eqref{freeOPE} that
\ben
Y_0(\varphi,x)= K_D\sum_{l,m}\frac{1}{\sqrt{\omega(D,l)}}\left( h_{l,m}(x)\a^+_{l,m}+\bar h_{l,m}(x)r^{-2l-D+2}\a_{l,m}\right)
\een
with $K_D=1$ for $D=2$, $K_D=\sqrt{D-2}$ for $D>2$ and $\omega(D,l)=2l+D-2$.\\
That the vertex operators for composite fields $Y_0(a,x)$ are given by eq.~\eqref{freevertex} is evident from
the remarks on how to determine the Schwinger functions of composite operators from those
of the basic field below eq.~\eqref{Schwinger}.
\\
In the main text, we also used Wick's theorem, which we cite for completeness in the following form:
Let $\wick{1}{<1 \a_{l,m}^{\#} >1 \a_{l',m'}^{\#}}$
denote the ``contraction'' between two creation
and/or annihilation operators, defined by
$$\a_{l,m}^{\#} \a_{l',m'}^{\#}=: \a_{l,m}^{\#} \a_{l',m'}^{\#}:
+\wick{1}{<1 \a_{l,m}^{\#} >1 \a_{l',m'}^{\#}}\,.$$
Writing $\a_{i}^\#$ for $\a_{l_i,m_i}^\#$ and $\a_{i'}^\#$ for $\a_{l_i',m_i'}^\#$, we have the combinatoric identity
\bena
:\a_{1}^\#...\a_{n}^\#: \,\, :\a_{1'}^\#...\a_{p'}^\#:\, &=&
\sum_{s=0}^{{\rm min}(n,p)}\sum_{\substack{i_1<...<i_s\\j_1\neq...\neq j_s}}
\wick{1}{<1 \a_{i_1}^\# >1 \a_{j_1'}^\# }...
\wick{1}{<1 \a_{i_s}^\# >1 \a_{j_s'}^\#}\notag\\
&\times&:\a_{1}^\#...\a_{n}^\#\a_{1'}^\#...\a_{p'}^\#:_{(i_1,...,i_s;j_1,...,j_s)}\label{Wick}
\eena
where double dots mean normal ordering and the subscript $(i_1,...,i_s;j_1,...,j_s)$ means that
the respective creation or annihilation operators have been
removed from the normal ordered product.\\

\bibliography{submit}{}
\bibliographystyle{plain}

\end{document}